\documentclass[12pt]{article}
 \pdfoutput=1
\usepackage{amssymb}
\usepackage{amsmath}
\usepackage{amstext}
\usepackage{graphicx,epsfig}
\usepackage{epsfig}
\usepackage{verbatim} 
\usepackage{fancybox}
\usepackage{color}
\usepackage{ulem}
\usepackage{enumitem}
\usepackage{subfigure}
\usepackage{bbm}
\usepackage{parskip}
\usepackage{cite}
\usepackage{youngtab}
\usepackage{graphicx}
\usepackage{jheppub}
\usepackage{setspace}
\usepackage{longtable}
\usepackage{tabularx}
\usepackage{amsmath,amssymb,color,mathrsfs,verbatim,bbm,wasysym,pstricks,epsfig,colortbl}
\usepackage{slashed,mathtools,youngtab,xcolor,rotating,color}
\usepackage{ytableau}

\newcommand{\Comment}[1]{{}}
\definecolor{MyDarkBlue}{rgb}{0.15,0.15,0.45}
\usepackage[linktocpage=true]{hyperref}
\hypersetup{
colorlinks=true,
citecolor=MyDarkBlue,
linkcolor=MyDarkBlue,
urlcolor=MyDarkBlue,
pdfauthor={Kurt Hinterbichler},
pdftitle={},
pdfsubject={hep-th}
}

\newcommand{\be}{\begin{equation}}
\newcommand{\ee}{\end{equation}}
\newcommand{\bea}{\begin{eqnarray}}
\newcommand{\eea}{\end{eqnarray}}
\newcommand{\beas}{\begin{eqnarray*}}
\newcommand{\eeas}{\end{eqnarray*}}
\newcommand{\nn}{\nonumber}

\newcommand{\cO}{\mathcal{O}}

\newcommand{\cD}{\mathcal{D}}

\newcommand{\tr}{\mathrm{Tr}}
\newcommand{\R}{\mathbb{R}}

\newcommand{\dcft}{d}

\definecolor{Gray}{gray}{0.9}

\newcommand{\normord}[1]{:\mathrel{#1}:}

\def\({\left(}
\def\){\right)}
\newcommand{\la}{\langle}
\newcommand{\ra}{\rangle}

\numberwithin{equation}{section}

\begin{document}

%\maketitle

\begin{center}
{\LARGE { Free $\square^k$ Scalar Conformal Field Theory }}
\end{center} 
 \vspace{1truecm}
\thispagestyle{empty} \centerline{
{\large  { Christopher Brust$^{a,}$}}\footnote{E-mail: \Comment{\href{mailto:cbrust@perimeterinstitute.ca}}{\tt cbrust@perimeterinstitute.ca}},
{\large  { Kurt Hinterbichler$^{b,}$}}\footnote{E-mail: \Comment{\href{mailto:kurt.hinterbichler@case.edu}}{\tt kurt.hinterbichler@case.edu}}
                                                          }

\vspace{1cm}

\centerline{{\it 
${}^a$Perimeter Institute for Theoretical Physics,}}
 \centerline{{\it 31 Caroline St. N, Waterloo, Ontario, Canada, N2L 2Y5 }} 
 
  \vspace{.3cm}

\centerline{\it ${}^{\rm b}$CERCA, Department of Physics, Case Western Reserve University, }
\centerline{\it 10900 Euclid Ave, Cleveland, OH 44106, USA}

\begin{abstract}

We consider the generalizations of the free $U(N)$ and $O(N)$ scalar conformal field theories to actions with higher powers of the Laplacian $\square^k$, in general dimension $d$.  We study the spectra, Verma modules, anomalies and OPE of these theories.  We argue that in certain $d$ and $k$, the spectrum contains zero norm operators which are both primary and descendant, as well as extension operators which are neither primary nor descendant. In addition, we argue that in even dimensions $d \leq 2k$, there are well-defined operator algebras which are related to the $\square^k$ theories and are novel in that they have a finite number of single-trace states.

\end{abstract}

\newpage

\tableofcontents
\newpage

\section{Introduction\label{sec:intro}}
\parskip=5pt
\normalsize

It has proven to be a tall order to understand the details of quantum gravity. There has been great progress in explaining how quantum gravity works in anti-de Sitter space (AdS), thanks primarily to the AdS/CFT correspondence \cite{Maldacena:1997re} relating questions in quantum gravity on AdS to questions in conformal field theory (CFT). Nevertheless, how quantum gravity works with a vanishing or positive cosmological constant (such as our own universe) is still comparatively elusive.

There have been efforts to understand quantum gravity in de Sitter (dS) through a similar correspondence which would relate quantum gravity in dS to a CFT living at at least one of the past and future boundaries \cite{Strominger:2001pn}. However, it has proven difficult to fill out entries in a putative dS/CFT dictionary due to a lack of complete examples; most early examples of AdS/CFT arose from string theory, and there does not seem to be a way to obtain a stable dS vacuum in string theory.

Recently there was a proposal for the first example of this putative correspondence \cite{Anninos:2011ui}, owing to the fact that the bosonic Vasiliev higher-spin gauge theory \cite{Vasiliev:1990en, Vasiliev:1992av, Vasiliev:1999ba, Vasiliev:2003ev} (being a collection of classical equations of motion) can be constructed about a dS solution just as easily as about an AdS solution. The CFT dual of the bosonic CP-even Vasiliev theory on dS is conjectured to be the (non-unitary) Grassmann scalar version of the $U(N)$ or $O(N)$ models dual to the AdS theory. Further work has suggested that the de Sitter space in this example may possibly be unstable \cite{Anninos:2012ft}, but other work studying a related 5d critical version of the $Sp(N)$ model suggests that it defines a sensible CFT \cite{Fei:2015kta}. 

In order to make progress in understanding the details or veracity of a dS/CFT correspondence, we need additional examples. However, if (as proposed) there is to be matching between the isometry groups and representations of a Euclidean CFT and a Lorentzian dS theory, then the masses of {\it unitary} dS particles would generically correspond to {\it non-unitary} CFT operators. Indeed, the scaling dimension inferred from taking a scalar particle of mass $m$ to the future boundary of dS with radius $1/H$ would be given by $\Delta = {d\over 2} \pm \sqrt{{d^2\over 4} - {m^2\over H^2}}$ (with similar equations for spinning particles), and there would generally be one root below the CFT unitarity bound, or even complex roots. 

Supposing then that at least some non-unitary CFTs know something about quantum gravity in dS, it is fruitful to inquire about what non-unitary CFTs exist, in order to eventually gain a handle on universal properties of quantum gravity in de Sitter. Having conjectural examples of dS/CFT also paves the way for tests of a hypothetical correspondence.  It seems therefore prudent to find more examples of CFTs which are non-unitary, which might nevertheless have unitary duals on dS.

Some non-unitary CFTs are known to exist as well-defined theories; in two dimensions, the minimal models $\mathcal{M}_{p,q}$ with $|q-p| \neq 1$ are non-unitary and nevertheless integrable, allowing for a complete understanding of the spectrum, OPE coefficients, and correlators in the theory. The first example is the Yang-Lee edge singularity, $\mathcal{M}_{2,5}$ with $c=-\frac{22}{5}$.  In higher dimensions, there is the supergroup $\mathcal{N}=4$ theory of Vafa \cite{Vafa:2014iua}, and the Yang-Lee edge singularity in $2\leq d \leq 6$, for which substantial evidence was presented in \cite{Butera:2012tq}. We still have not yet classified unitary CFTs, and comparatively less is known about non-unitary CFTs. Nevertheless, signs point to a rich classification that may include interacting examples even in $d>6$, unlike unitary CFTs where interacting theories seem to be limited to $d\leq 6$ \cite{Stergiou:2015roa,Gracey:2015xmw}.

Besides the goal of de Sitter holography, there are other independent reasons to study non-unitary CFTs, for example to describe open quantum systems or the theory of elasticity (see e.g. \cite{Osborn:2016bev,Guerrieri:2016whh,Nakayama:2016dby,Peli:2016gio,Gwak:2016sma,Fujimori:2016udq,Gliozzi:2016ysv,Gliozzi:2017hni} for some recent work). Logarithmic CFTs are non-unitary in any dimension \cite{Hogervorst:2016itc}, and arise in studies of turbulence and percolation \cite{Polyakov:1992yw, Flohr:1996ik, Creutzig:2013hma}. Non-unitary CFTs may encode some data of unitary QFTs; for example the Yang-Lee edge singularity encodes the imaginary zeroes of the magnetic field in the Ising QFT \cite{PhysRevLett.40.1610}. Finally, dropping the restriction of unitarity can open the door to new techniques; for example in the conformal bootstrap, new algorithms can be developed \cite{Gliozzi:2013ysa, Gliozzi:2014jsa,Esterlis:2016psv} and can illuminate more efficient ways to obtain conformal blocks via recursion relations \cite{Kos:2013tga, Penedones:2015aga}. 

With these motivations in mind, we study non-unitary free field theories defined by higher-derivative actions. These theories are well-defined because they are free; correlators are all well defined via Wick contraction and there are no interactions which could cause e.g. runaway instabilities or infinities. Nevertheless, considering theories like the $U(N)$ or $O(N)$ models with higher-derivative actions have been argued to give rise to AdS or dS duals which generalize or extend the Vasiliev theory to include additional partially massless and massive states, corresponding to the appearance of additional operators in the spectrum \cite{Bekaert:2013zya,Basile:2014wua,Grigoriev:2014kpa,Alkalaev:2014nsa,Joung:2015jza}.   We study these AdS duals in a complementary paper \cite{Brust:2016zns}.

The plan of this paper is as follows. We study the singlet sector of free scalar $U(N)$ or $O(N)$ CFTs described by the flat-space Lagrangian ${\cal L}\sim \phi^\dagger_a\square^k\phi^a$, with $\phi^a$ a field in the fundamental representation.
 $k=1$ is the usual $U(N)$ or $O(N)$ model. We begin in section \ref{sec:hs2} with $k=2$ for illustrative purposes, paying close attention to the structure of the Verma modules and the OPE. We work out the structure of the theory explicitly in the $k=2$ case. We then move on in section \ref{sec:boxk} to general $k$.  In the appendices, we briefly discuss the higher spin symmetry algebra (which we label here $hs_k$) of these theories and the associated underlying geometric symmetries of the theory, although it does not play a crucial role in the main body of work in this paper. We explore these symmetries because they serve as the primary technical tool in the AdS dual of these theories, which we explore in \cite{Brust:2016zns}. In addition, in the appendix we compute the conformal anomaly in even dimensions and the free energy on a sphere in odd dimensions for a few of the lowest lying cases, and review the ``partition function'' construction counting of the number of operators in these theories.

We will see the appearance of so-called ``multiply conserved currents'', operators which vanish when not one but $c$ (for ``conservedness'') derivatives are dotted into them\footnote{In the higher-spin literature, such operators are colloquially referred to as ``partially conserved currents''. We deviate from this convention because in particle physics parlance, partially conserved currents are currents of spontaneously broken symmetries, such as the famous partially conserved axial current (PCAC). Here, these multiply conserved currents correspond to exact symmetries of the theory.} . In addition we will see that there are generally zero- and negative-norm states in the spectrum, reflecting the non-unitarity of the theory. As such, the space of states of the theory is not, strictly speaking, a Hilbert space, whose definition mandates positive-definite norm states, but we nevertheless refer to the space as a ``Hilbert'' space throughout the paper. These features causes peculiarities to arise in the corresponding Verma modules of the theory, such as the appearance of states which we refer to as ``zero-norm states'' (states which are both primary and descendant and therefore have vanishing norms, but unlike null states in 2d minimal models, are {\it not} orthogonal to every state in the theory) as well as ``extension states'' (states which are {\it neither primary nor descendant}) which are relevant to the OPE structure of the theory. These peculiarities arise in $d=3, 6$ for the $k=2$ theory, and for the general $k$ theory we provide a conjecture for when these states should appear.

In addition, we will see that when $d$ is even and $d\leq 2k$, there are two choices of operator algebra. The first is the solution to the $\square^k$ theory; $\phi$ is no longer a scaling operator in the theory, as its correlators have logs in them. This is familiar to us from the $d=2$, $k=1$ usual free field theory. We may construct a CFT by insisting on only having scaling operators in the spectrum; we may, for example, ``gauge away'' $\phi$ by insisting that it have a shift symmetry. Doing so allows things such as $\partial \phi$ in the $d=2$, $k=1$ theory to be called primary. The other choice, though, is to consider the operator algebra {\it defined} by saying that $\phi$ is a scaling operator with $\Delta_\phi = \frac{d}{2}-k$. This is clearly not a solution to the $\square^k$ theory as the two-point function is no longer a Green's function. Nevertheless, it produces a consistent CFT with a well-defined OPE. These theories are unique in that their correlation functions are all polynomial in the separations, and so they have a finite number of single-trace states in the theory, each with a finite number of descendants.

{\bf Notations and Conventions:} $\dcft$ refers to the dimension of the CFT, which we take to be $\geq 2$. Latin lowercase indices from the middle of the alphabet $i,j,\ldots$ are spacetime indices.  The letter $a$ is a color index. We work on flat Euclidean space, so the metric is $\delta_{ij}$ and $\square\equiv\delta^{ij}\partial_i\partial_j$.  Indices are symmetrized with unit weight, i.e. $T_{(ij)}\equiv {1\over 2}\left(T_{ij}+T_{ji}\right)$, and the notation ${(ij\ldots)_T}$ indicates to symmetrize and subtract all traces, i.e. $T_{(ij)_T}\equiv {1\over 2}\left(T_{ij}+T_{ji}\right)-{1\over d}T^l_{\ l}\delta_{ij}$.

We use $k$ to refer to which field theory we are considering; the $k^\mathrm{th}$ theory is the theory with a $\square^k$ in the action. The $k^\mathrm{th}$ theory has $k$ towers of operators in it, each containing spins $s=0,1,2,\ldots$ appearing once; which tower we are in is labelled by $b=0,1,\ldots,k-1$, for reasons which we explain in section \ref{sec:boxk}.

%%%%%%%%%%%%%%%%%%%%%%%%%%%%%%%%%%%%%%%%%%%%%%%%%%%%%%%%%%%%%%%%%%%%%%%%%%%%%%%%

\section{$\square^2$ Theory}
\label{sec:hs2}

We start in this section with the CFT described by the action

\begin{equation}S \propto \int d^\dcft x ~\phi_a^\dagger \square^2 \phi^a\label{box2Lagrangian}\end{equation}

\noindent on Euclidean flat space of dimension $d\geq 2$, with a ``gauge'' group $U(N)$ or $O(N)$ (i.e. in the limit where we decouple and throw away the gauge group). We explicitly consider the case $U(N)$ here, but all results for $O(N)$ can be obtained simply by replacing $\phi^\dagger \rightarrow \phi$. In practice, calling this group a ``gauge'' group means that we only consider operators which are singlets under the group, which is a consistent truncation.  Here $a$ is a fundamental index under this group, and we will suppress it, along with factors of $N$ (which are easily restored) in all that follows. The noncommittal normalization of the scalar field kinetic term (reflected by the $\propto$) allows us the freedom to fix the two-point function normalization in a convenient fashion later. 

The action \eqref{box2Lagrangian} is fully conformally invariant, i.e. invariant up to a total derivative under both dilations and special conformal transformations, if we choose $\phi$ to be a scaling operator with dimension 
\be \Delta_\phi = \frac{d-4}{2}.\ee

\noindent This choice is consistent with the equations of motion provided $d \neq 2, 4$. We will review below in subsection \ref{sectionfinitebox2} what happens if we do (or do not) choose $\phi$ to be a scaling operator in the $d=2,4$ cases.

The canonical quantization of this theory on ordinary time slices in flat space is subtle due to the Jordan form of the Hamiltonian (see e.g. \cite{Maldacena:2011mk}). However, it is straightforward to perform radial quantization, diagonalizing the dilatation operator (or equivalently, quantizing the theory on a cylinder $\mathbb{R}\times S^{d-1}$ by conformally coupling it to the background curvature. We review this coupling in appendix \ref{couplingtocurvapp}). We then have the usual operator-state correspondence, which pairs every local operator ${\cal O}(x)$ with a state $|{\cal O}\ra\equiv  {\cal O}(0)|0\ra$ (as the operator-state correspondence does not rely on unitarity), and so we may, as usual, freely pass between the language of operators and the language of states.

As we will see, the ``Hilbert'' space of states does not have a positive-definite inner product; there are zero- and negative-norm states in the spectrum. In this sense, the theory is non-unitary.  However, when a theory has such negative norm states (often called ghost states) there is usually a more physical quantization available in which the theory is perfectly unitary, i.e. the inner product is positive definite and the Hamiltonian is Hermitian, but has energies unbounded from below\footnote{For example, the wrong-sign harmonic oscillator can be quantized such that it has positive energies and negative norms, or negative energies and positive-definite norm.  The difference is reflected in the $i\epsilon$ prescription of the propagator \cite{vanTonder:2006ye}.} \cite{vanTonder:2006ye,Woodard:2006nt,Sbisa:2014pzo}.  Energies unbounded from below bring to mind instability, however in our case the theory is completely free; there can be no instabilities such as vacuum decay or finite-time runaway solutions.  In this paper, we will stick with the quantization involving negative norm states, since it is easier to apply standard CFT tools in this case.

The ``gauge invariant" operators, the spectrum of singlets under $U(N)$ or $O(N)$ which span our CFT ``Hilbert'' space, can be counted by the usual partition function and character arguments; this is the result of the ``generalized Flato-Fronsdal theorem'' \cite{Basile:2014wua} (which we review for $d=3$ in appendix \ref{sec:flatofronsdal}). The ``single-trace'' primary operators (which correspond to single-particle states in the bulk dual described in \cite{Brust:2016zns}) are the singlet primary operators which involve a single contraction of the ``gauge'' indices, and are therefore bilinear in the fields. An explicit construction of which bilinears are primary can be obtained by recursion relations or by brute force. We follow the recursive formula developed in \cite{Penedones:2010ue} and extended by \cite{Fitzpatrick:2011dm}, only we extend the procedure further to account for equations of motion. 

In this section, computations are done rather explicitly to exemplify the general structure of the $\square^k$ theory. The outline of this section is as follows: in subsection \ref{sec:primaries}, we compute by brute force the first few single-trace primary operators. In \ref{sec:boxsqsymms}, we discuss how these operators are multiply-conserved currents of the symmetry algebra of the theory, $hs_2$. In \ref{sec:correlators} we compute the two-point functions of these first few primaries. We see from this that in dimensions $2$ and $4$, the spectrum of single-trace operators collapses dramatically, which we explain in \ref{sectionfinitebox2}. We also see from the structure of these operators the appearance of operators which become both primary and descendant in dimensions $3$ and $6$. We study the structure of the associated Verma modules in \ref{sec:nuances}.

\subsection{$\square^2$ Spectrum}
\label{sec:primaries}

We would like to find all primary operators $j$ (spin indices suppressed) which are ``single-trace'' (bilinear in $\phi$) by demanding that $[K_i,j(0)] = 0$, where $K_i$ is the special conformal generator which acts as the lowering operator in the conformal algebra.  The fundamental field is primary, $[K_i, \phi^a(0)]=0$. We begin by expanding the most general linear combination of single trace operators $j$ in a complete basis:

\begin{equation}j=\sum_{k_1,k_2,m=0}^\infty \sum_{u_1, u_2=0}^1 a(k_1,k_2,u_1,u_2,m) T(k_1,k_2,u_1,u_2,m)\, ,\end{equation}

\noindent where $a$ are coefficients and $T$ is the basis operator

\begin{equation}T(k_1,k_2,u_1,u_2,m) \equiv \left(\partial_{i_1}\ldots \partial_{i_{k_1}}  \partial_{n_1}\ldots \partial_{n_m} \square^{u_1}\phi^\dagger \partial_{j_1}\ldots \partial_{j_{k_2}}  \partial^{n_1}\ldots \partial^{n_m} \square^{u_2}\phi\right)_{\mathrm{sym,T}}(0)\, . \label{basisopdef1}
\end{equation} 

\noindent Here sym,T refers to the symmetric traceless part of the operator, so that it is an irreducible representation. This basis operator has spin $k_1+k_2$ and total scaling dimension $2\Delta_\phi + k_1+k_2+m+2u_1+2u_2$, where $\Delta_\phi = \frac{1}{2}(d-4)$.

This basis operator vanishes if $u_1\geq 2$ or $u_2 \geq 2$ by the equations of motion $\square^2\phi=0$, and so we subject our coefficients $a$ to be zero if $u_1\geq 2$ or $u_2 \geq 2$. We of course subject $a$ to be zero if any argument is negative.

The general action of $K_i$ on $j$ was worked out in \cite{Penedones:2010ue,Fitzpatrick:2011dm,Bekaert:2015tva}; demanding that it be zero gives two different equations, one for the spin $k_1+k_2+1$ part and one for the spin $k_1+k_2-1$ part. They are
\begin{align}0=&2(u_1+1)(d-2u_1-2-2\Delta_\phi)a(k_1-1,k_2,u_1+1,u_2,m)\nonumber \\
&+2(u_2+1)(d-2u_2-2-2\Delta_\phi)a(k_1,k_2-1,u_1,u_2+1,m) \nonumber \\
&-2(m+1)(\Delta_\phi+m+2u_1)a(k_1,k_2-1,u_1,u_2,m+1)\nonumber \\
&-2(m+1)(\Delta_\phi+m+2u_2)a(k_1-1,k_2,u_1,u_2,m+1) \nonumber \\
&+(m+1)(m+2)a(k_1-1,k_2,u_1,u_2-1,m+2) \nonumber \\
&+(m+1)(m+2)a(k_1,k_2-1,u_1-1,u_2,m+2)\, , \label{topequationr} \\ \nonumber \\
0=&-2(k_1+1)(\Delta_\phi+m+k_1+2u_1)a(k_1+1,k_2,u_1,u_2,m) \nonumber \\
&-2(k_2+1)(\Delta_\phi+m+k_2+2u_2)a(k_1,k_2+1,u_1,u_2,m)\, . \label{bottomequationr} \end{align}
Solving these recursively for $a$ for all $k_1,k_2,u_1,u_2,m$ subject to the aforementioned boundary conditions yield all primary operators in the theory, up to normalization. The overall normalization of the primary operator would normally be fixed either by using the Ward identity of the associated symmetry or by unit normalizing the two-point functions, but for our later interests it is more convenient to leave the normalizations arbitrary for the moment.

As an example, let us consider $k_1=k_2=u_1=u_2=m=0$. Equation \eqref{topequationr} is trivial ($0=0$ because there are no operators with spin $-1$) but equation \eqref{bottomequationr} yields $0=a(0,1,0,0,0)+a(1,0,0,0,0)$. This primary operator is $\propto\partial_i \phi^\dagger \phi - \phi^\dagger \partial_i \phi$.

There turn out to be two primaries of each spin in the $U(N)$ theory (each even spin in the $O(N)$ theory).  These operators are organized into two towers, or ``Regge trajectories''.  Each tower contains all spins $s=0,1,2,\ldots$ occurring once.  We denote these operators $j_{i_1\ldots i_s}^{(b)}$, where $s$ is the spin and $b\in \{ 0,1\}$ labels the ``Regge trajectory'', and stands for ``boxes'', i.e. the number of pairs of contracted derivatives the primaries will have.  We will often use a shorthand notation for the spin indices, writing $j_{s}^{(b)}$, with $s$ standing in for the set of indices $i_1\ldots i_s$.  

The form of the operators in the first tower, $b=0$, is 
\be { j }^{(0)}_{s} \sim  \phi^\dag  \,\partial_{i_1}\ldots \partial_{i_s}\phi+\ldots,\ \ \ \ s=0,1,2,\ldots\ \ \ \  , \ \ \Delta=d+s-4\ \ ,\label{operatorsk20}\ee
where the ellipses denote other orderings of the derivatives with relative coefficients uniquely fixed by the requirement that the operators are symmetric, fully traceless, primary, and hermitian.
The second tower, $b=1$, consists of operators of the form
\be { j }^{(1)}_{s} \sim \phi^\dag  \,\square \partial_{i_1}\ldots \partial_{i_s}\phi+\ldots,\ \ \ \ s=0,1,2,\ldots\ \ \  , \ \ \Delta=d+s-2\ \ . \ \label{operatorsk202}\ee 

We briefly interject with definitions. Suppose a spin-$s$ operator $\cO_s$ vanishes when contracted with $c\leq s$ derivatives,

\begin{equation}\partial^{i_1}\ldots \partial^{i_c} \cO_{i_1 \ldots i_c \ldots i_s} = 0\ \ .\end{equation}

\noindent We refer to $\cO$ as a multiply-conserved operator of ``conservedness'' $c$, or a $c$-conserved operator for short.  (Multiply-conserved single-trace primary operators are dual to partially massless particles in AdS \cite{Dolan:2001ih}.)

The operators $ { j }^{(1)}_{s} $ have the same scaling dimension as the single-trace primaries (STP's) of the ordinary free scalar.  Those with $s\geq 1$ saturate the unitarity bound and are $c=1$ conserved currents, satisfying 
\be \partial^{i_1}{ j }^{(1)}_{i_1 \ldots i_s} =0, \ \ \ s\geq 1.  \ee
by virtue of the equations of motion $\square^2\phi=0$.  The scalar operator ${ j }^{(1)}_{0}$ does not satisfy any kind of conservation condition. 

The operators $ { j }^{(0)}_{s} $ are new to the $\square^2$ theory.  Those with $s\geq 3$ are ``triply conserved'' with $c=3$, satisfying 
\be \partial^{i_1}\partial^{i_2}\partial^{i_3}{ j }^{(0)}_{i_1 \ldots i_s} =0, \ \ \ s\geq 3. \label{consintos32} \ee
by virtue of the equations of motion $\square^2\phi=0$.  The first three operators in this tower, ${ j }^{(0)}_{0},\ { j }^{(0)}_{1},\ { j }^{(0)}_{2}$ satisfy no conservation condition.

The explicit form for the first few STPs in the $U(N)$ theory are (with arbitrary normalization, chosen only so that it does not vanish in any dimension $d\geq 2$) shown in table \ref{tab:primaries}.  The $O(N)$ primaries can be obtained simply by replacing $\phi^\dagger \rightarrow \phi$, which kills all primaries of odd spin.

There are Ward identities which determine a preferred normalization for the form of those operators that have conservation conditions associated to them. We do not concern ourselves with these Ward identities here, as they are irrelevant for the discussion of extended modules to follow. However, they would be very interesting to study; for specific operators, Ward identities (as well as explicit formulae for the primaries) have been studied previously \cite{Guerrieri:2016whh, Osborn:2016bev}.

{\renewcommand{\arraystretch}{1.0}
\begin{table}[h]
\centering
\begin{tabular}{|l|c|c|} \hline
{\bf Operator $j_s^{(b)}$} & {\bf $\Delta$} & {\bf $c$} \\ \hline
\rowcolor{Gray} $j_{0}^{(0)} ~=~ \phi^\dag \phi$ & $d-4$ & \\
$j_{0}^{(1)} ~=~ (d-4)\left(\phi^\dag\square \phi +\square\phi^\dag \,\phi \right)+4\, \partial\phi^\dag\partial\phi$ & $d-2$ & \\
\rowcolor{Gray} $j_{1}^{(0)} ~=~ i\left(\phi^\dag \partial_i\phi - \partial_i\phi^\dag \phi\right)$& $d-3$ & \\
$j_{1}^{(1)} ~=~ i \bigg( \partial_i\phi^\dag\square \phi -{d-4\over d}\phi^\dag\partial_i\square \phi -{4\over d} \partial^j\phi^\dag \partial_i\partial_j\phi-c.c.\bigg)$ & $d-1$ & $1$ \\
\rowcolor{Gray} $j_{2}^{(0)} ~=~ \left(-(d-4) \phi^\dag \partial_i\partial_j\phi+(d-2)\partial_i\phi^\dag \partial_j\phi+c.c.\right)_{sym,T}$ & $d-2$ & \\
$\begin{aligned}j_{2}^{(1)} ~=~ {1\over d-1}\bigg[&-4(d-2)\partial_k\phi^\dag\partial^k\partial_i\partial_j\phi-{d(d+2)}\square\phi^\dag\partial_i\partial_j\phi+{4d}\partial_k\partial_i\phi^\dag\partial^k\partial_j\phi \nonumber \\ &-(d-4)(d-2)\phi^\dag \partial_i\partial_j\square\phi+2(d+2)(d-2)\partial_i\phi^\dag\partial_j\square\phi\bigg]_{sym,T} \end{aligned}$ & $d$ & $1$ \\ 
\rowcolor{Gray} $j_{3}^{(0)} ~=~ i\left(-{d-4\over 3d}\phi^\dag\partial_i\partial_j\partial_k\phi+\,\partial_i\phi^\dag\partial_j\partial_k\phi-c.c.\right)_{sym,T}$ & $d-1$ & $3$ \\
$\begin{aligned}j_{3}^{(1)} ~=~ i\bigg[ &-{(d-4)(d-2)\over 12(d+2)}\phi^\dag \partial_i\partial_j\partial_k\square\phi+ {(d+4)(d-2)\over 4(d+2)}\partial_i\phi^\dag \partial_j\partial_k\square\phi \nonumber \\ &-{d-2\over 3(d+2)}\partial_l\phi^\dag\partial_i \partial_j\partial_k\partial^l\phi-{d+4\over 12} \square\phi^\dag\partial_i\partial_j\partial_k \phi +{d+4\over 4}\partial_i \square\phi^\dag\partial_j\partial_k \phi \nonumber \\ &+\partial_i\partial_l \phi^\dag \partial_j\partial_k\partial^l \phi -c.c. \bigg]_{sym,T}\end{aligned}$ & $d+1$ & $1$ \\
\cellcolor{Gray} $\begin{aligned}j_{4}^{(0)} ~=~ \bigg[&{(d-4)(d-2)\over 3d(d+2)}\phi^\dag \partial_i\partial_j\partial_k\partial_l\phi-{4(d-2)\over 3d}\partial_i\phi^\dag \partial_j\partial_k\partial_l\phi \nonumber \\ &+ \partial_i\partial_j\phi^\dag\partial_k\partial_l\phi+c.c.\bigg]_{sym,T}\end{aligned}$ & \cellcolor{Gray} $d$ & \cellcolor{Gray} $3$ \\
$\begin{aligned}j_{4}^{(1)}~=~ \bigg[ &-{(d-2)(d-4)\over 16(d+4)}\phi^\dag \partial_i\partial_j\partial_k\partial_l\square\phi+{(d+6)(d-2)\over 4(d+4)}\partial_i\phi^\dag \partial_j\partial_k\partial_l\square\phi \nonumber \\ &+{d-2\over 4(d+4)}\partial_m\phi^\dag \partial_i\partial_j\partial_k\partial_l\partial^m\phi -{1\over 16}(d+6)\square\phi^\dag \partial_i\partial_j\partial_k\partial_l\phi \nonumber \\ &+{(d+6)(d+2)\over 4d}\partial_i\square\phi^\dag \partial_j\partial_k\partial_l\phi-{3(d+2)\over 4d}\partial_i\partial_j\partial^m\phi^\dag \partial_k\partial_l\partial_m\phi \nonumber \\ &-{3\over 8}(d+6)\partial_i\partial_j\phi^\dag\partial_k\partial_l\square\phi+\partial_i\partial_m\phi^\dag\partial_j\partial_k\partial_l\partial^m\phi  +c.c. \bigg]_{sym,T}\end{aligned}$ & $d+2$ & $1$ \\ \hline
\end{tabular}
\caption{The first few primaries in the $\square^2$ theory in $d$ dimensions. There are two operators of each spin; we refer to them as the ``$b=0$ tower'' and the ``$b=1$'' tower. The operator is denoted $j_{s}^{(b)}$. We also show the scaling dimension, $\Delta$, and conservedness, $c$, of each operator.   The operator $j_{2}^{(1)}$ is proportional to the stress tensor $T_{ij}$ of the theory (in the cases where there is a stress tensor, as discussed later).  Color and spin indices are suppressed.}
\label{tab:primaries}
\end{table}
}

\subsection{$\square^2$ Conserved Currents and Symmetries}
\label{sec:boxsqsymms}

As is well-known in field theory, conserved currents correspond to global symmetries \cite{Noether:1918zz}.  In our case, we have a tower of higher-spin conserved currents ${ j }^{(1)}_{s}$ for $s\geq 1$.  As reviewed in appendix \ref{genktapp}, each current can be contracted with a spin-$s$ conformal Killing tensor\footnote{It is a slight abuse of notation to call this Killing tensor spin-$s$ when it is in fact has $s-1$ indices; regardless we label these Killing tensors by the spin $s$ of their associated currents.} $K_{(1)}^{i_1\ldots i_{s-1}}$ (which is a symmetric traceless tensor satisfying the conformal Killing equation $\partial^{(i_1}K_{(1)}^{i_2\ldots i_{s})_T}=0$) to form a Noether current,
\be J_{i_1}^{(s)}={ j }^{(1)}_{i_1\, i_2\ldots i_{s}}K_{(1)}^{i_2\ldots i_{s}}\, , \quad \partial^iJ_i^{(s)}=0. \ee
Each of these Noether currents is associated via Noether's theorem to a linearly realized global symmetry of the Lagrangian \eqref{box2Lagrangian}, with the leading-derivative part given by the conformal Killing tensor,
\be \delta_K\phi=K_{(1)}^{i_1\ldots i_{s-1}}\partial_{i_1}\ldots \partial_{i_{s-1}}\phi +\ldots \label{symgenbox2f}\ee
Here the ellipses are terms with the derivatives acting in all other possible ways on $K$ and $\phi$, with coefficients uniquely determined by the requirement that \eqref{symgenbox2f} leave the Lagrangian \eqref{box2Lagrangian} invariant up to a total derivative.
Each independent conformal Killing tensor gives an independent symmetry of the Lagrangian.  These are the higher spin symmetries familiar from the ordinary free scalar \cite{Eastwood:2002su}.  

In particular, there is only one spin-1 conformal Killing tensor; it is just a constant, and the Noether current is $\propto { j }^{(1)}_{1}$ which is associated with $U(1)$ charge rotation $\delta\phi=i\phi$.  The spin-2 conformal Killing tensors are the ordinary Killing vectors $K^i$ satisfying $\partial^{(i}K_{(1)}^{j)_T}=0$ and are associated with global conformal transformations.  The conserved current is the stress tensor (with the exception of the finite theory cases $d=2,4$ which have no true stress tensor, see section \ref{sectionfinitebox2} and Appendix \ref{couplingtocurvapp})
\be { j }^{(1)}_{2}\sim T_{ij}\, ,\ee
 which is the Noether current associated with the global conformal symmetries given by $\delta\phi=K_{(1)}^i\partial_i\phi+{\Delta_\phi\over d}\partial_iK_{(1)}^i\phi$.

We also have the triply-conserved currents ${ j }^{(0)}_{s}$, satisfying \eqref{consintos32}.
Even though they are not conserved in the ordinary sense, they nevertheless correspond to symmetries of the action which are affiliated with ``third-order'' conformal Killing tensors $K_{(3)}^{i_1\ldots i_{s-3}}$, extending the conformal Killing tensor symmetries of the usual two-derivative free theory.   Third-order conformal Killing tensors are symmetric traceless tensors satisfying the third-order conformal Killing equation 
\be \partial^{(i_1}\partial^{i_2}\partial^{i_3}K_{(3)}^{i_4\ldots i_{s})_T}=0\, .\label{thordcke2}\ee
Contracting with the ${ j }^{(0)}_{s}$ gives us new Noether currents,
\be \tilde J_{i_1}^{(s)} = { j }^{(0)}_{{i_1\ldots i_{s}}}\partial^{i_{2}}\partial^{i_3}K_{(3)}^{i_4\ldots i_{s}}-\partial^{i_{2}}{ j }^{(0)}_{{i_1\ldots i_{s}}}\partial^{i_3}K_{(3)}^{i_4\ldots i_{s}}+\partial^{i_{2}}\partial^{i_3}{ j }^{(0)}_{{i_1\ldots i_{s}}}K_{(3)}^{i_4\ldots i_{s}}\, . \ee
 These are conserved, $\partial^i\tilde J_{i}^{(s)}=0$, as can be seen by using \eqref{consintos32} and \eqref{thordcke2} along with the symmetry and tracelessness of ${ j }^{(0)}_{s}$.
Each of these conserved currents is the Noether current associated with a linearly realized global symmetry of the Lagrangian \eqref{box2Lagrangian}, with the leading derivative part given in terms of the third-order conformal Killing tensor and involving one power of the Laplacian,
\be \delta_K\phi=K_{(3)}^{i_1\ldots i_{s-3}}\partial_{i_1}\ldots \partial_{i_{s-3}}\square\phi +\ldots \label{symgenbox2}\ee
 We work out as an example the third-order conformal killing ``spin-3'' scalar symmetries in appendix \ref{genktapp}. 

These new symmetries combine with the old ones to form an algebra.  This symmetry algebra underlying the $\square^2$ theory was first studied by \cite{2006math.....10610E}, and its bilinear form was worked out in \cite{Joung:2015jza}. (In the latter work the algebra was referred to as $\mathfrak{p}_2$; we will refer to it in later contexts as $hs_2$.) We will not need details of the algebra here, but it plays a central role in the AdS dual story \cite{Brust:2016zns}.

\subsection{$\square^2$ Correlation Functions}
\label{sec:correlators}

Since we are dealing with a free field theory, all correlators can be worked out by knowing the basic two-point functions among $\phi,\phi^\dag$
and then extending to all other operators and correlators via Wick contraction.  If $\phi$ is itself a conformal field, the only form of the two point function consistent with conformal symmetry is
\begin{equation} \la \phi^\dag(x)\phi(0)\ra = \frac{1}{|x|^{d-4}}\, , \label{basiccork2}\end{equation}
with $\la \phi(x)\phi(0)\ra=\la \phi^\dag(x)\phi^\dag(0)\ra=0$.
 We will take this as the basic correlator in all dimensions.  For $d\neq 2,4$, this is consistent with the Green's equation for the theory
\begin{equation}\square^2 \la \phi^\dag(x)\phi(0)\ra \propto \delta^d(x)\, ,\end{equation}
whereas for $d=2,4$ there is no delta function on the right hand side (which raises issues, see section \ref{sectionfinitebox2}).

From here the singlet ``Hilbert'' space of the CFT can be constructed by computing various two point functions via Wick contraction.  The singlet ``Hilbert'' space is graded by particle number;  the vacuum is the state with no particles in it, the single trace operators are the singlet two-particle states, double-trace operators are singlet four-particle states, etc. We list the first few two-point functions for the STPs in table \ref{tab:correlators}.

{\renewcommand{\arraystretch}{1.8}
\begin{table}[h]
\centering
\begin{tabular}{|c|rcl|} \hline
{\bf Operator $j_s^{(b)}$} & \multicolumn{3}{|c|}{$\langle j_s^{(b)}(x)j_s^{(b)}(0)\rangle$} \\ \hline
\rowcolor{Gray}$j_{0}^{(0)}$ & & ${1\over x^{2(d-4)}}$ & \\
$j_{0}^{(1)}$ & $-8 d(d-6 ) (d-4)^2$ & ${1\over x^{2(d-2)}}$ & \\
\rowcolor{Gray}$j_{1}^{(0)}$ & $2(d-4)$ & ${1\over x^{2(d-3)}}$ & $I_{i_1}^{j_1}$ \\
$j_{1}^{(1)}$ & $-{16(d-4)^3(d-2)(d+2)\over d^2}$ & ${1\over x^{2(d-1)}}$ & $ I_{i_1}^{j_1}$ \\
\rowcolor{Gray}$j_{2}^{(0)}$ & $8(d-4)^2(d-3)( d-2)$ & ${1\over  x^{2(d-2)}}$ & $ I_{(i_1}^{(j_1} I_{i_2)_T}^{j_2)_T}$ \\
$j_{2}^{(1)}$ & $-{64 (d-4 )^2 d (d-2 )^3 (d+4 )\over d-1}$ & ${1\over  x^{2d}}$ & $I_{(i_1}^{(j_1} I_{i_2)_T}^{j_2)_T}$ \\
\rowcolor{Gray}$j_{3}^{(0)}$ & ${16 (d-4 )^2 (d-2 ) (d-1 )\over 3 d}$ & ${1\over  x^{2(d-1)}}$ & $ I_{(i_1}^{(j_1} I_{i_2}^{j_2}I_{i_3)_T}^{j_3)_T}$ \\
$j_{3}^{(1)}$ & $-\frac{8d^2  (d-4)^2 (d-2)^2 (d+1) (d+6)}{3 (d+2)}$ & ${1\over  x^{2(d+1)}} $ & $I_{(i_1}^{(j_1} I_{i_2}^{j_2}I_{i_3)_T}^{j_3)_T}$ \\
\rowcolor{Gray}$j_{4}^{(0)}$ & $\frac{32 (d-4)^2 (d-2)^2 (d-1) (d+1)}{3 d (d+2)}$ & ${1\over  x^{2d}}$ & $ I_{(i_1}^{(j_1} I_{i_2}^{j_2}I_{i_3}^{j_3}I_{i_4)_T}^{j_4)_T}$ \\
$j_{4}^{(1)}$ & $-\frac{12 (d-4)^2 (d-2)^2 (d+1) (d+2)^2 (d+3) (d+8)}{d+4}$ & ${1\over  x^{2(d+2)}}$ & $ I_{(i_1}^{(j_1} I_{i_2}^{j_2}I_{i_3}^{j_3}I_{i_4)_T}^{j_4)_T}$ \\ \hline
\end{tabular}
\caption{The two-point functions of the first few primaries of the $\square^2$ theory in $d$ dimensions.  The $i$ indices are associated with the left operator inserted at $x$ and the $j$ indices are associated with the right operator inserted at $0$, and we have defined $I_{ij}\equiv\delta_{ij}-2{x_i x_j\over x^2}$.  The correlators are all computed with the operators as normalized in table \ref{tab:primaries}.  As usual, correlators between two different primaries vanish. }
\label{tab:correlators}
\end{table}
}

Looking at these these two point functions, we see several features.  There are some critical dimensions in which certain two-point functions vanish.  In $d=4$, all the two-point functions except those of $ j_{0}^{(0)}$ vanish.  In $d=2$, all the two-point functions except those of $j_0^{(0)},\ j_0^{(1)}$ and $j_1^{(0)}$ vanish.  In $d=6$ the two-point function of $ j_{0}^{(1)}$ vanishes, and in $d=3$ the two-point function of $ j_{2}^{(0)}$ vanishes.
In all these dimensions, zero norm states are appearing, and compared to unitary CFTs, unfamiliar things are happening with the Verma modules in the ``Hilbert'' space.  In the cases $d=2,4$ the space of single-trace states is becoming finite-dimensional, as we will discuss in subsection \ref{sectionfinitebox2}, and in the cases $d=3,6$ two different Verma modules are being glued together, as we will discuss in subsection \ref{sec:nuances}.

\begin{figure}[h!]
\centering
\includegraphics[width=4in]{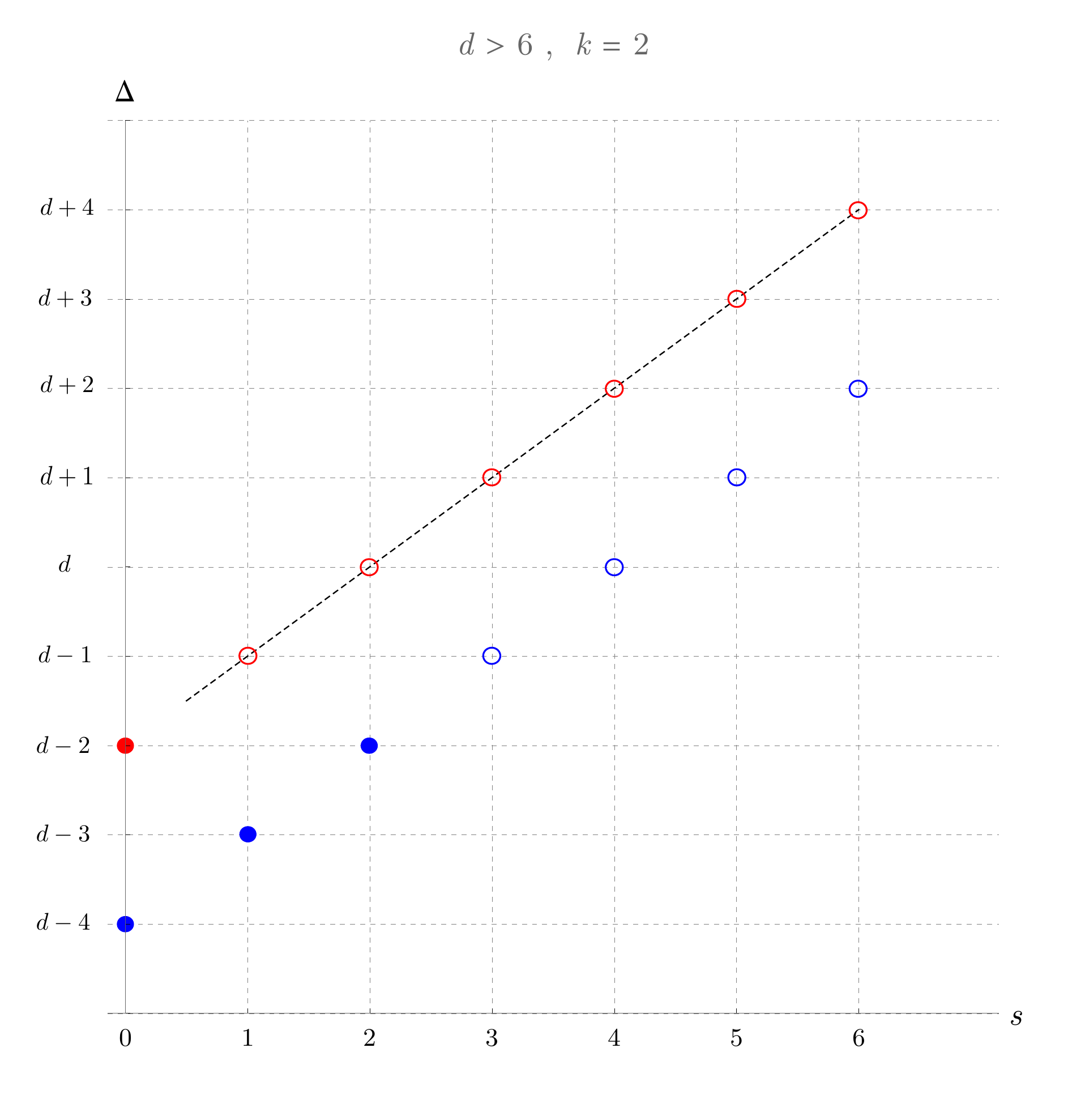}
\caption{Spectrum of single trace primaries in the case $d>6$ for the $\square^2$ theory.  The two Regge trajectories are clearly visible, with $b=0$ on the bottom, $b=1$ on the top.  Unfilled circles are the operators satisfying conservation conditions, with the top trajectory being singly conserved and the bottom trajectory being triply conserved. Filled circles are the operators satisfying no conservation condition.  Blue means the two point function has positive norm, red means it has negative norm.  The dotted line is the unitarity bound (not shown is the $s=0$ bound $\Delta\geq {d\over 2}-1$).   The case $d=5$ looks the same except that the $b=1$ scalar now has positive norm.  The cases $d=2,4$ are discussed in section \ref{sectionfinitebox2}, the $d=3,6$ cases in section \ref{sec:nuances}. }
\label{hs2d7}
\end{figure}

For the operators with $s\geq 3$, and $d\not=2,4$, the two-point functions are of opposite relative sign: positive for the triply-conserved $b=0$ tower and negative for the singly-conserved $b=1$ tower.  This is another symptom of non-unitarity; we can flip these signs by changing the overall sign of the basic correlator \eqref{basiccork2} (equivalent to changing the overall sign of the Lagrangian \eqref{box2Lagrangian}), but we cannot change the relative sign.  We illustrate the spectrum in the generic case $d>6$ in figure \ref{hs2d7}.

\subsection{$\square^2$ Finite Theories\label{sectionfinitebox2}}

For $d \neq 2,4$ the two-point function \eqref{basiccork2} is indeed a Green's function of the operator $\square^2\phi$, i.e. $\square^2  \la \phi^\dag (x)\phi (0)\ra \propto \delta^\dcft(x)$.  
However, in $d=2,4$ the correlator \eqref{basiccork2} becomes polynomial/analytic in the separation $x^i$, 
\bea  &&\la \phi^\dag (x)\phi (0)\ra = 1,\ \ \ d=4, \nn\\
&& \la \phi^\dag (x)\phi(0)\ra = x^2,\ \ \ d=2, \label{finitegfunck2}
\eea 
and so we have $\square^2  \la \phi^\dag(x)\phi(0)\ra=0$, with no delta function source.  Thus, in $d=2,4$ the correlator \eqref{finitegfunck2} does not properly describe the theory with Lagrangian \eqref{box2Lagrangian}. However, simply {\it defining} a CFT whose basic building block is  \eqref{finitegfunck2} serves to give a sensible operator algebra and OPE, which as we'll see shortly has a finite number of single-trace states.

The alternative in $d=2,4$ is to take the basic two point function to have logs in it, so that it becomes a true Green's functions of $\square^2$ with a delta function source,
\bea \la \phi^\dag (x)\phi (0)\ra \sim x^2 \ln |x| \, , \qquad d=2\, , \nn\\ 
\la \phi^\dag (x)\phi (0)\ra \sim  \ln |x|\, , \qquad d=4\, .
\eea
In this case, the field $\phi$ itself, and consequently many of the composite operators derived above, fail to be conformal fields and should be removed from the spectrum (which should be accomplished by imposing various ``gauged'' shift symmetries, perhaps along the lines of those considered in \cite{Griffin:2013dfa,Hinterbichler:2014cwa,Griffin:2014bta,Griffin:2015hxa}, and only considering ``gauge-invariant'' operators), leaving only operators whose correlators take the required conformally invariant form without logarithms\footnote{This is one way to treat the free scalar CFT in $d=2$.  $\phi$ itself is not a scaling operator, and the conformally invariant operators are those invariant under a shift symmetry $\phi\rightarrow \phi+c$. Of course, one could also consider vertex operators which transform covariantly under this symmetry instead as well.}.  This will give a different CFT,  where presumably the underlying symmetry algebra is different from $hs_2$.  We will not pursue this possibility in this paper, instead choosing the conformally invariant correlators \eqref{finitegfunck2} in the cases $d=2,4$. We choose to focus on the study of the finite theories rather than the log theories as the finite theories are what emerge in the partially massless higher-spin bulk dual, as we discuss in \cite{Brust:2016zns}. These two finite theories, $d=4$ and $d=2$, correspond to the two possible truncations of the $hs_2$ algebra as exhibited in \cite{Joung:2015jza}.

\subsubsection{$\square^2$ in $d=4$\label{box2d4subsecn}}

In $d=4$, our basic two point function becomes a constant,
\be \la \phi^\dag(x)\phi(0)\ra =1.\label{constant2pt1}\ee
As is apparent from table \ref{tab:correlators}, every two-point function except that of $j_0^{(0)}$ vanishes. This remains true of all higher-point functions; any correlator containing any operator other than $j_0^{(0)}$ vanishes, because the presence of any derivative kills the constant two-point function \eqref{constant2pt} occurring in the Wick contraction.  Furthermore, every descendant state of $j_0^{(0)}$ and every descendant of every other operator vanishes in any correlator, for example
\be\la \partial_ij_0^{(0)}(x)\partial_jj_0^{(0)}(0)\ra =0.\ee  

\begin{figure}[h!]
\centering
\includegraphics[width=6in]{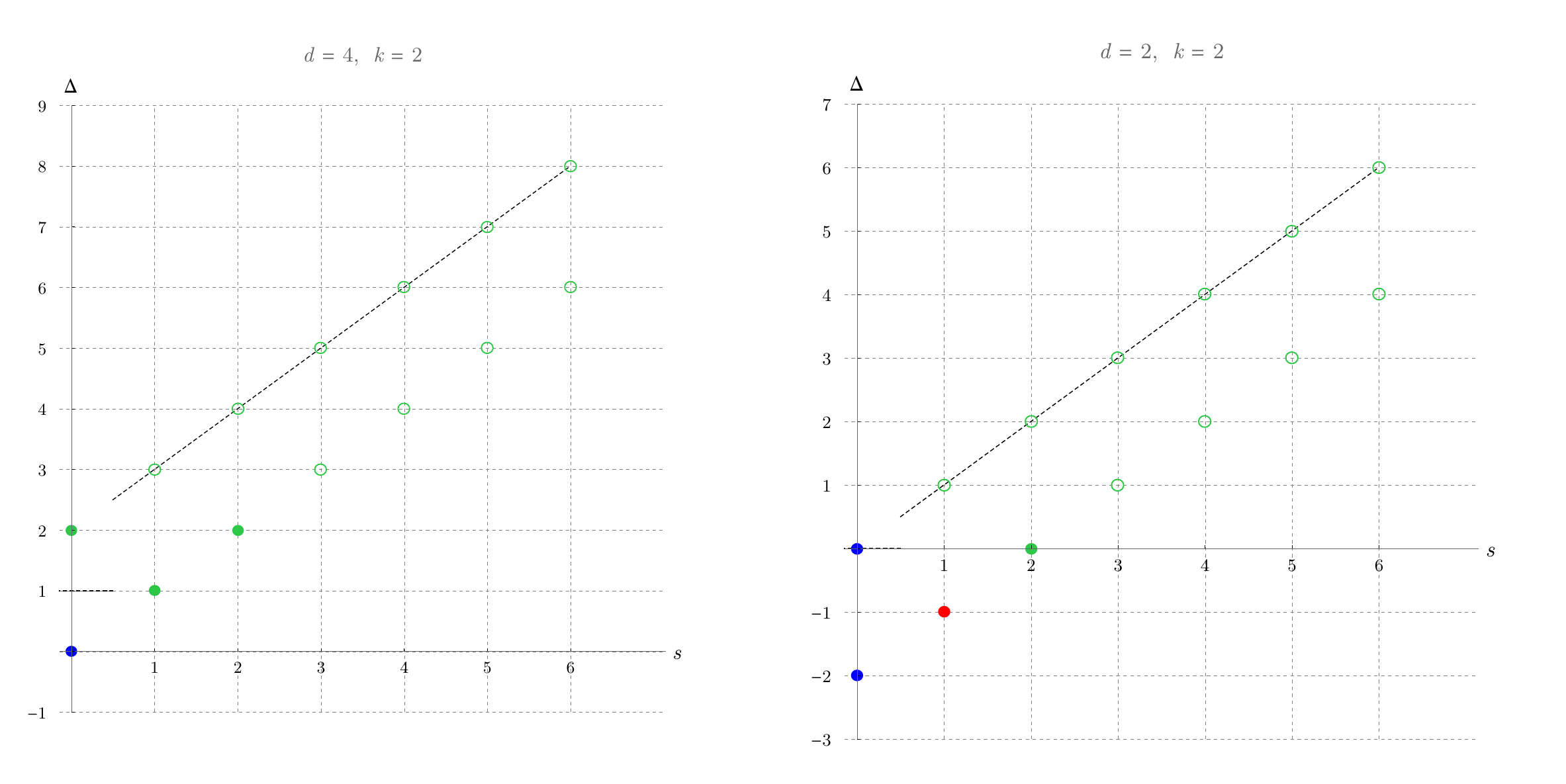}
\caption{Spectrum of single trace primaries in the finite case $d=4,2$ for the $\square^2$ theory.  Blue states have positive norm, red states have negative norm, and green states are zero norm null states which are projected out.  (Unfilled circles are the operators satisfying conservation conditions, filled circles are the operators satisfying no conservation condition, and the dotted line is the unitarity bound.)}
\label{fig:hs2dfinite}
\end{figure}

Therefore, every state except for $j_0^{(0)}$ is a null state and can be factored out of the theory.  What remains is perhaps the simplest possible non-empty ``CFT''; the single-trace Hilbert space is one-dimensional, spanned by the state $|j_0^{(0)}\ra \equiv j_0^{(0)}(0)|0\ra$ with the same quantum numbers ($\Delta=0,\ s=0$) as the vacuum. We use ``CFT'' rather liberally to mean a theory with conformal symmetry and an associative operator algebra. These theories would not be conformal field theories in the usual sense, though, due to the absence of a conserved stress tensor. (The would-be stress tensor and its descendants are all null and may be consistently set to zero.) This structure is indicated on the left of figure \ref{fig:hs2dfinite}. 
The $n$-trace singlet Hilbert space ($n=0,1,2,\ldots$) is just given by $n$ copies of $j_0^{(0)}$ acting on the vacuum at the origin, with no derivatives anywhere.

The alternative is, of course, to consider the logarithmic theory. We do not study this theory here, although it would be interesting to think about in the future. Of note is that as $\Delta_\phi=0$, we may construct ``vertex-like'' operators $e^{i\alpha \phi}$ in such a theory, leading one to wonder if there is some non-unitary Liouville-esque theory based around a $\square^2$ action.

\subsubsection{$\square^2$ in $d=2$}

In $d=2$, our basic two point function becomes,
\be \la \phi^\dag(x)\phi(0)\ra =x^2,\label{constant2pt}\ee
and we see from \ref{tab:correlators} that every correlator except those involving $j_0^{(0)},\ j_0^{(1)}$ and $j_1^{(0)}$ vanishes. 

The 2d story plays out similarly to the 4d story.  There are only three single-trace primary operators in the theory, $j_0^{(0)},\ j_0^{(1)}$ and $j_1^{(0)}$.  All the other single-trace primaries are null and can be factored away, including the stress tensor.  Furthermore, these three non-null primaries have a finite number of descendants; since the two point function \eqref{constant2pt} is a polynomial, once there are enough derivatives, the correlators of all of the higher descendants with anything all vanish, and so it is consistent to truncate the Verma module there.  The single-trace Hilbert space is again therefore finite-dimensional. We illustrate the single-trace spectrum of this finite theory on the right of figure \ref{fig:hs2dfinite}.

In order to call this a sensible ``CFT'', we must discuss the underlying operator algebra. Indeed, the OPE and conformal partial wave decomposition in these theories is perfectly well-defined.  This can be seen already at the level of the global conformal blocks. We use the conventions
\begin{align}x_{ij}^2&=(x_i-x_j)^2 \, ,\nonumber \\
u&=\frac{x_{12}^2x_{34}^2}{x_{13}^2x_{24}^2}=z\bar{z}\, , \nonumber \\
v&=\frac{x_{14}^2x_{23}^2}{x_{13}^2x_{24}^2}=(1-z)(1-\bar{z})\, , \nonumber \\
k_{\alpha}(x) &= x^{\frac{\alpha}{2}} ~\phantom{}_2F_1\left(\frac{\alpha}{2},\frac{\alpha}{2};\alpha;x\right)\, , \nonumber \\
g_{\Delta,\ell}(z,\bar{z}) &= k_{\Delta+\ell}(z)k_{\Delta-\ell}(\bar{z})+k_{\Delta-\ell}(z)k_{\Delta+\ell}(\bar{z})\, .\end{align}

For example, suppose we have four external $j_0^{(0)}$ ($\Delta = -2$, $s=0$), and exchange another $j_0^{(0)}$. This corresponds to the conformal block
\begin{equation}g_{-2,0}(z,\bar{z}) = \frac{2}{z\bar{z}} -\left(\frac{1}{z}+\frac{1}{\bar{z}}\right)+\frac{1}{2}\,.
\end{equation}
From this we see that the block associated with an exchanged $j_0^{(0)}$ only ``transmits''\footnote{The reader should be cautious that just because these states are the only ones which appear in the OPE does {\it not} imply that these are all of the non-null descendants. In this case, for example, the finite-dimensional module with highest weight state $j_0^{(0)}$ contains a total of 9 states, with the lowest weight state being $\square^2 j_0^{(0)}$.} the descendant states $j_0^{(0)}(0)$, $\partial_i j_0^{(0)}(0)$, and $\square j_0^{(0)}(0)$. Furthermore, the complete four-point function computed from Wick contractions again also decomposes into a finite number of partial waves:
\begin{equation}\langle j_0^{(0)}(x_1)j_0^{(0)}(x_2)j_0^{(0)}(x_3)j_0^{(0)}(x_4)\rangle \equiv x_{12}^4 x_{34}^4 F(u,v)\, ,\end{equation}
\begin{align}F(u,v)&=1+\frac{1}{u^2}+\left(\frac{v}{u}\right)^2 + \frac{2}{N}\left(\frac{1}{u}+\frac{v}{u}+\frac{v}{u^2}\right) \nonumber \\
&=1+\left(1+\frac{1}{N}\right)g_{-4,0}(z,\bar{z})+\left(\frac{2}{3}-\frac{1}{3N}\right)g_{-2,2}(z,\bar{z})\nonumber \\ &\qquad+\left(1+\frac{2}{N}\right)g_{-2,0}(z,\bar{z})+\left(\frac{1}{9}+\frac{19}{36N}\right)g_{0,0}(z,\bar{z})\, .\end{align}
Here we have restored the factors of $\frac{1}{N}$ so it can be seen explicitly what is and is not a contribution from a single-trace versus a double-trace state. Note that we have explicitly pulled out the vacuum contribution, a $1$ out front, but nevertheless there is a constant piece from the block $g_{0,0}=1$. These correspond to single-dimensional single-trace and double-trace modules which are not the identity.

This illustrates that only a finite number of primaries, each with only finite number of descendants, are exchanged.

\subsection{$\square^2$ Extended Modules\label{sec:nuances}}
In the $\square^2$ theory in $d=3$ and $d=6$, there are operators which are both primary and descendant. These are therefore zero-norm states, and have vanishing two-point functions with themselves, $\langle \cO(x) \cO(0)\rangle = 0$. These states, however, are {\it not} orthogonal to every other state in the theory, as we will see, and so we do not refer to them as null, a term which we reserve for states orthogonal to every other state in the theory.

However, due to this degeneration, another operator, $\tilde{\cO}$, necessarily appears in the spectrum.  This new operator is neither primary nor descendant, a phenomenon which cannot happen in a unitary CFT\footnote{This phenomenon has been seen to occur in nonrelativistic CFT Verma modules in \cite{Golkar:2014mwa}, where the authors referred to such bizarre operators $\tilde{\cO}$ as {\it alien operators}.  To our knowledge, this phenomenon has not been observed before in a relativistic CFT.}.  As we will show, this new operator forms its own module which is attached to the module which $\cO$ is present in, forming an ``extension'' of the Verma module associated with $\cO$, and intricately connected to it because $\langle \cO(x) \tilde{\cO}(0)\rangle\neq 0$.  As such, we will refer to operators like $\tilde{\cO}$ as ``extension operators.'' 

In the $\square^k$ theory, this phenomenon occurs in specific dimensions for various low-lying operators in the spectrum. This does not happen in $k=1$ (or any unitary relativistic CFT), and in $k=2$, it only happens in the module associated with $j_0^{(0)}$, in $d=3$ and in $d=6$.  We illustrate these two examples in detail here, and discuss the general $k$ case in section \ref{boxkextendedsecn}.

\subsubsection{$\square^2$ in $d=6$\label{6dspec}}

In the $\square^2$ theory away from $d=6$, there are two scalar primary operators which sit in independent Verma modules, $j_0^{(0)}$ and $j_0^{(1)}$. As we will show, as $d\rightarrow 6$, $j_0^{(1)} \propto \square j_0^{(0)}$, while maintaining the property that it is primary. The operator in $d=6$ is both primary and descendant, and therefore has zero norm, but is not orthogonal to every other state in the theory, and so is not null. The reason why is because a new operator $\tilde{j}_0^{(1)}$ is required to appear in the spectrum, which is an extension operator; it is neither primary nor descendant and has non-vanishing correlators with $j_0^{(1)}$. These two together form an extended Verma module. Let us see this in more detail.

Consider the space of states with $\Delta = \dcft-2$, $s=0$ which are bilinear in the fields and survive in the $O(N)$ theory, i.e. are symmetric under $\phi^\dag\leftrightarrow \phi$.  This space is spanned by two such basis states, $\phi^\dag\square \phi +\square\phi^\dag \,\phi$ and $\partial_i\phi^\dag\partial^i\phi$, and so this space is two-dimensional. 

\begin{figure}[h]
\centering
\includegraphics[width=\textwidth]{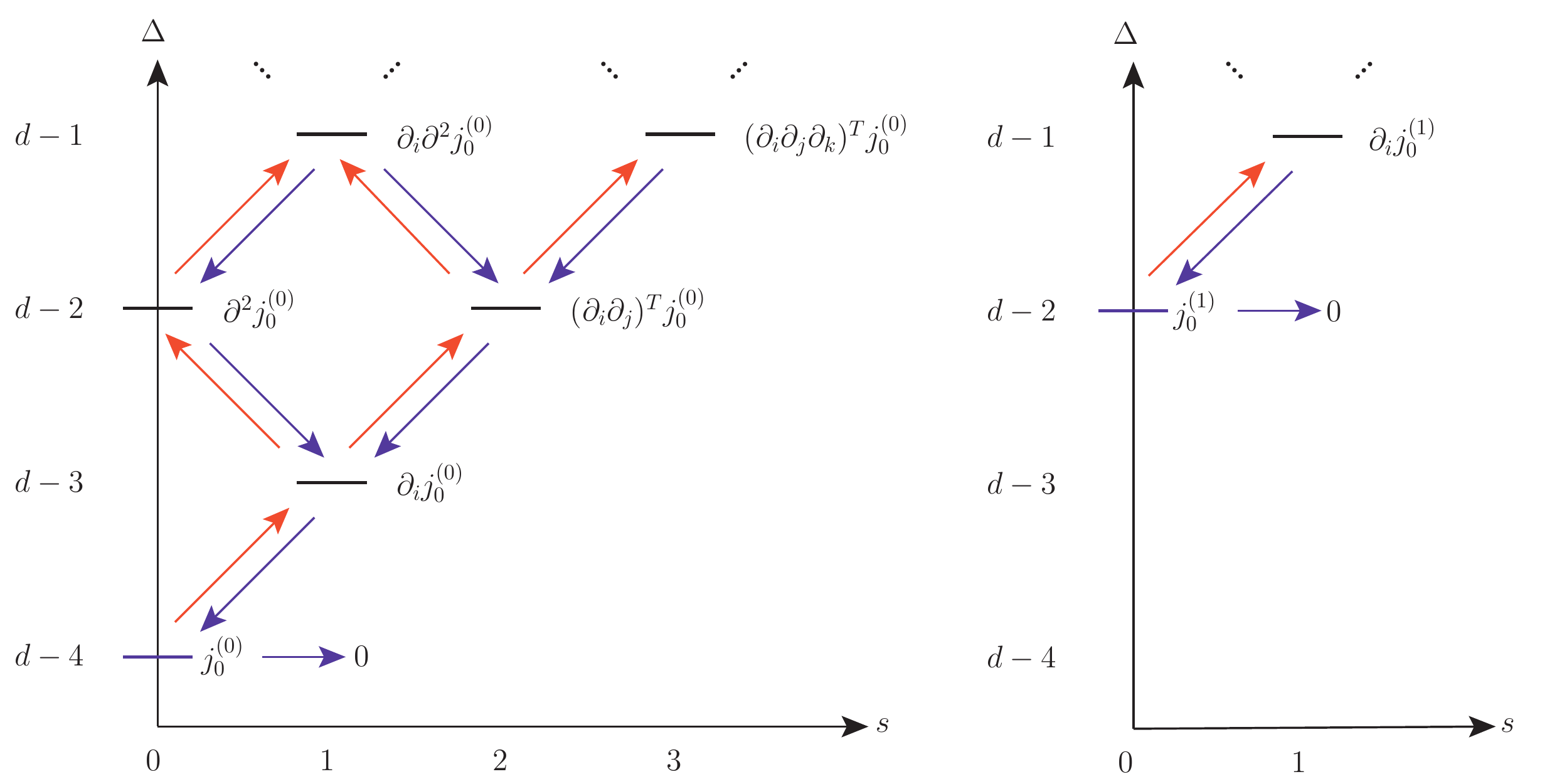}
\caption{A diagram of the generic scalar Verma modules in the $\square^2$ theory. There are two separate Verma modules corresponding to $j_{0}^{(0)}$ (left) and $j_{0}^{(1)}$ (right) in the theory. The blue states are the highest-weight states and are primary; they are annihilated by $K$. One can move around the module by raising with a $P$ (red arrow) or lowering with a $K$ (blue arrow). Of course, the module continues up and to the right.}
\label{fig:dVermaModule}
\end{figure}

When $d\not=6$, this space is spanned by the scalar primary $j_{0}^{(1)}$, and a scalar descendant operator arising from the primary $j_{0}^{(0)}=\phi^\dag \phi$,
\begin{align}
j_{0}^{(1)} &= (d-4)\left(\phi^\dag\square \phi +\square\phi^\dag \,\phi \right)+4\, \partial_i\phi^\dag\partial^i\phi ,\\
\square j_{0}^{(0)} &= \left(\square \phi^\dagger \phi  + \phi^\dagger \square \phi\right)+2 \partial_i \phi^\dagger \partial^i\phi,
\end{align}
and these two operators are orthogonal in the Hilbert space inner product.  This is business as usual in conformal field theories: each primary operator gives rise to an independent highest-weight Verma module whose states we get to by acting with the momentum generator $P_i\propto \partial_i$, and each state in one module is orthogonal to all states in any other module.  (This situation is shown in figure \ref{fig:dVermaModule}.)

However, when $\dcft=6$, these two operators are no longer independent:  
\begin{equation}j_{0}^{(1)} = 2\square j_{0}^{(0)}\, ,\quad d=6\, .\end{equation}
Thus, although $j_{0}^{(1)}$ is still a primary (as it is annihilated by $K$), it is also a descendant (as it's the derivative of another operator). The property of being both primary and descendant implies that the norm of the state is zero,
\be \la j_{0}^{(1)}|j_{0}^{(1)}\ra=\la P^2 j_{0}^{(0)} | P^2 j_{0}^{(0)}\ra =0\, ,\quad d=6. \ee

Usually, for example in the study of 2d minimal models, the presence of a zero-norm state indicates that the Verma module is shortening; the zero norm state also has zero inner product with every other state, and it is then known as a null state.  All the descendant of the null state are also null.  In such a module, one can reach the states of the null submodule by acting with $P$, but can not get back from the null module via $K$, i.e. access to the submodule is via a one-way gate.  In this situation, where the inner product of null states with every other state in the Hilbert space is 0, the null states can be consistently factored out of the Hilbert space.  In our case, as we'll see now, this is {\it not} what is happening. $ j_{0}^{(1)}$ in $d=6$ may be primary and descendant and have zero norm, but it is not a null state.

To see why, we first observe that when $d=6$ there are still two possible operators in the space of states $\Delta = \dcft-2$, $s=0$ which are bilinear in the fields and survive in the $O(N)$ theory: $\phi^\dag\square \phi +\square\phi^\dag \,\phi$ and $\partial_i\phi^\dag\partial^i\phi$, and so this space is still two dimensional.  The linear combinations $j_{0}^{(1)}$, $\square j_{0}^{(0)}$ spanned the space of such operators for $\dcft \neq 6$. However, in $\dcft=6$ where $j_{0}^{(1)}$ and $\square j_{0}^{(0)}$ degenerate, we
have a puzzle; what happened to the other linear combination? For example, in $\dcft=6$, the operator $\partial_i\phi^\dagger \partial^i \phi$ cannot be obtained as a linear combination of $j_{0}^{(1)}$, $\square j_{0}^{(0)}$ as these latter two point in the same direction in the operator-space.  

The resolution of this puzzle is instructive: the missing direction must be spanned by some operator which is {\it not descendant} as it cannot be obtained as the derivative of another operator (because the image of $P$ is spanned by $j_{0}^{(1)}$), but it is also {\it not a primary} (because the kernel of $K$ is also spanned by $j_{0}^{(1)}$). Instead, acting with $K$ on any other operator independent from $j_{0}^{(1)}$ returns us to the descendant state $\partial_i j_{0}^{(0)}$, e.g,
\begin{equation}0 \neq K_i |\partial_j\phi^\dagger \partial^j \phi\rangle \propto |\partial_i j_{0}^{(0)}\rangle \, .\end{equation}

In a unitary relativistic CFT, all states are a linear combination of primary or descendant states (see e.g. \cite{Rychkov:2016iqz}), but we see here, by counterexample, that this statement does not extend to non-unitary theories\footnote{The presence of an operator which is neither primary nor descendant may make us doubt whether the theory is really conformal, but one can check explicitly that the defining Lagrangian \eqref{box2Lagrangian} is invariant under all conformal transformations (including the special conformal transformations) in any dimension, and can be coupled (except in $d=2$) to a background metric in a Weyl-invariant fashion so that the stress tensor is automatically traceless (see appendix \ref{couplingtocurvapp}).}.
 
To learn more about this missing state, we consider the inner product matrix\footnote{Recall that the inner product of two states $|{\cal O}_1\ra={\cal O}_1(0)|0\ra$ and $|{\cal O}_2\ra={\cal O}_2(0)|0\ra$ associated with (not necessarily primary) operators ${\cal O}_1$ and ${\cal O}_2$ is computed from the two-point function by
\be \la {\cal O}_1|{\cal O}_2\ra =\lim_{x\rightarrow \infty,\, y\rightarrow 0}\left.\la { R}\left[{\cal O}_1(x')\right] {\cal O}_2(y) \ra\right|_{x'={\cal R}(x)}\, ,  \label{innerprodformfotne}\ee
 where ${\cal R}$ is the inversion operator which acts on the coordinate as ${\cal R}(x)^i={x^i\over x^2}$ and $R$ is the inversion which acts on the basic field $\phi$ as ${ R}\left[\phi(x)\right]={1\over x^{2\Delta_\phi}}\phi\left({\cal R}(x)\right)$.
 } in the generic basis spanned by
\bea && {\cal E}_1= \phi^\dag\square \phi +\square\phi^\dag \,\phi \, , \ \ \nn \\
&& {\cal E}_2= \partial_i\phi^\dagger \partial^i \phi\, , \\
&&  \left(\begin{array}{cc}\la {\cal E}_1|{\cal E}_1\ra & \la {\cal E}_1|{\cal E}_2\ra \\\la {\cal E}_2|{\cal E}_1\ra & \la {\cal E}_2|{\cal E}_2\ra\end{array}\right)=\left(\begin{array}{cc}-96 & 0 \\0 & 24\end{array}\right).  \label{innerprodm6dcase}
\eea

Note that this inner product matrix is the true Hilbert space inner product, which in this case is not the same as the coefficients of the matrix of finitely-separated two-point functions. If $\mathcal{E}_1$ and $\mathcal{E}_2$ were primary operators, then the inner product matrix would be the same as the matrix of two-point functions, up to an overall $\frac{1}{x^8}$. However, they are not primaries, and so the finite-separation correlators ``see'' the full structure of the generalized Verma module plus extension, and the two point functions are not equal to the inner products:
\be \left(\begin{array}{cc}\la {\cal E}_1(x) {\cal E}_1(0)\ra & \la {\cal E}_1(x) {\cal E}_2(0)\ra \\\la {\cal E}_2(x){\cal E}_1(0)\ra & \la {\cal E}_2(x) {\cal E}_2(0)\ra\end{array}\right)=\frac{1}{x^8}\left(\begin{array}{cc}32 & -64 \\-64 & 56\end{array}\right). 
\ee

From the inner product matrix \eqref{innerprodm6dcase} we see that there are no true null states, because the inner product matrix is non-degenerate, with no zero eigenvalues.
But there are two independent zero norm states, because the inner product matrix is Lorentzian and there are two ``light-like directions."  One of these zero-norm states is $j_0^{(1)}$, but there is a second, linearly independent zero norm state which we'll call $\tilde j_0^{(1)}$,
\bea && j_0^{(1)}= 2\left(\square \phi^\dagger\phi + \phi^\dagger \square \phi \right)+4 \partial_i \phi^\dagger \partial^i \phi\,  , \nn \\
&&  \tilde j_0^{(1)}= \partial_i \phi^\dagger \partial^i \phi-{1\over 2}\left(\square \phi^\dagger\phi + \phi^\dagger \square \phi \right)\, , \label{tildeopsdef6} \\
&&  \left(\begin{array}{cc}\la j_0^{(1)}|j_0^{(1)}\ra & \la j_0^{(1)}|\tilde j_0^{(1)} \ra \\\la \tilde j_0^{(1)} |j_0^{(1)}\ra & \la \tilde j_0^{(1)} | \tilde j_0^{(1)} \ra\end{array}\right)=\left(\begin{array}{cc}0 & 192 \\192  & 0\end{array}\right). 
\eea
Again, the inner product matrix is not the same as the matrix of two point functions:
\be \left(\begin{array}{cc}\la j_0^{(1)}(x) j_0^{(1)}(0)\ra & \la j_0^{(1)}(x)\tilde j_0^{(1)}(0) \ra \\\la \tilde j_0^{(1)}(x) j_0^{(1)}(0)\ra & \la \tilde j_0^{(1)}(x) \tilde j_0^{(1)}(0) \ra\end{array}\right)=\frac{1}{x^8}\left(\begin{array}{cc}0 & 192 \\192  & 128\end{array}\right). 
\ee

\begin{figure}[h]
\centering
\includegraphics[width=\textwidth]{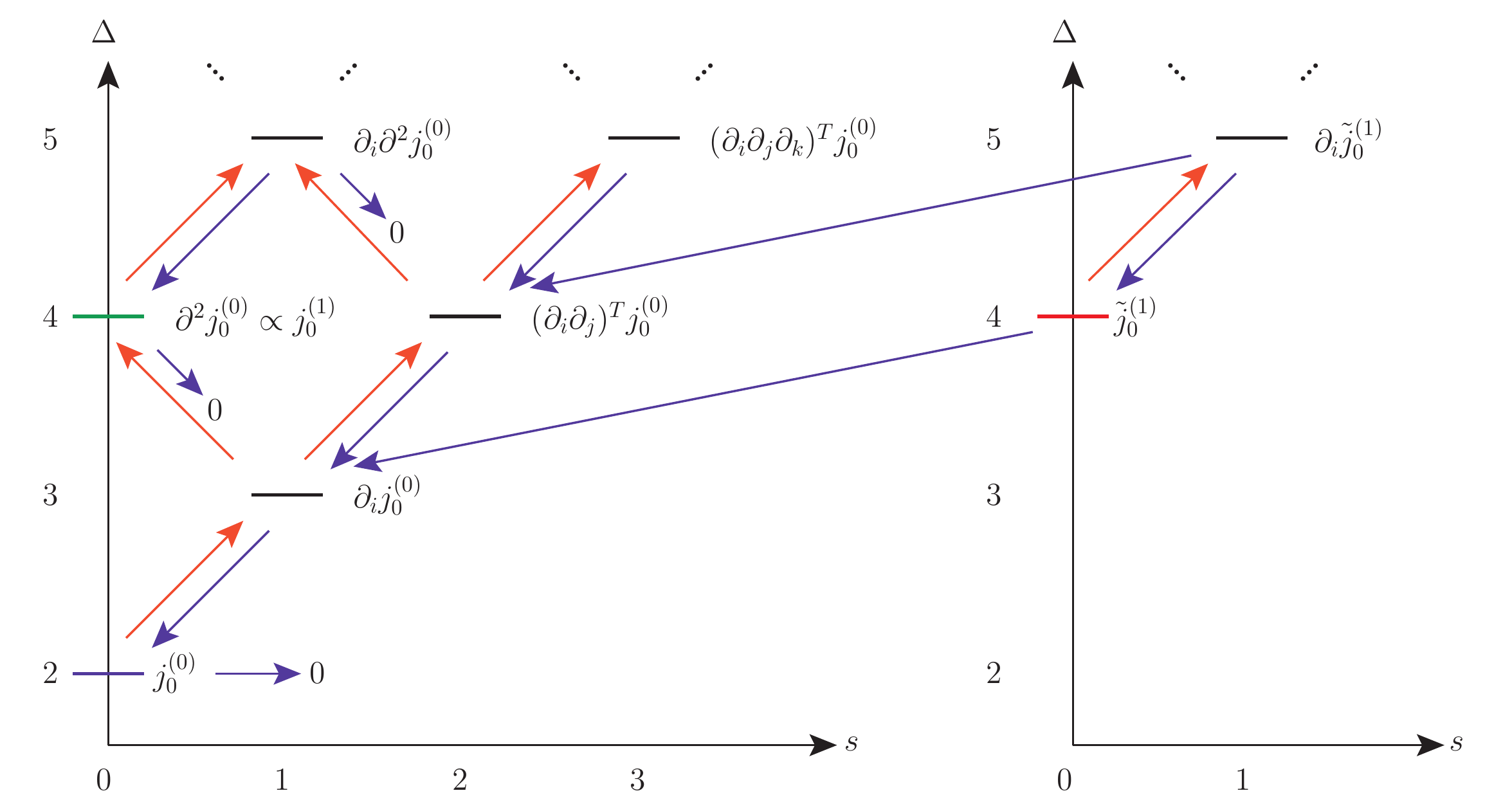}
\caption{A diagram of the scalar Verma modules in the $\dcft=6$ $\square^2$ theory. There is one generalized Verma module with an extension, the extension being a projective Verma module as shown on the right. The single Verma module corresponds to $j_{0}^{(0)}$ (on the left). The blue state is the highest-weight state and is primary; it is annihilated by $K$. One can move around the module by raising with a $P$ (red arrow) or lowering with a $K$ (blue arrow). The state in green is the null state $\square j_{0}^{(0)}$; it is both a primary and a descendant. The state in red is an extension state $\tilde{j}_{0}^{(1)}$; it is neither a primary nor a descendant. Of course, the modules continue up and to the right.}
\label{fig:6VermaModule}
\end{figure}

It is $j_0^{(1)}$ which is the primary, i.e. is the subspace annihilated by $K$, and the image of $P$ is this same subspace, so the other linear combination is neither primary nor descendant.
The presence of the new non-primary, non-descendant state $\tilde j_0^{(1)}$ tells us that some sort of new Verma module is arising to compensate for the degeneration of the original two Verma modules.  
We refer to $\tilde j_0^{(1)}$ as an {\it extension state} (for lack of a better word). The extension state spawns its own collection of descendants, forming a Verma module which extends the original Verma module.  We refer to the entire structure as a {\it Verma module with an extension}. We illustrate this structure in figure \ref{fig:6VermaModule}.  We can think of the extension module as the result of ``gluing together" the two modules associated with ${j}_{0}^{(0)}$ and ${j}_{0}^{(1)}$.  

We can get from the extension back to the original module through the action of $K$, since we have
\begin{equation}0 \neq K_i |\tilde {j}_{0}^{(1)} \rangle \propto |\partial_i j_{0}^{(0)}\rangle\, ,\end{equation}
but we cannot get to the extension module from the original module using either $K$ or $P$.
Thus the extension state is in some sense the ``dual'' of a null state; just as a null module is connected to the larger module through the action of $P$ but not $K$, the extension module is connected to the larger module through the action of $K$ but not $P$.  It is not a lowest or highest weight state in the module, in that there is no one state from which all the others can be reached through the action of only $P$, or through the action of only $K$.  It is a different kind of irreducible representation, which can only occur in non-unitary CFT's.  We describe the AdS dual of this structure in \cite{Brust:2016zns}.  The spectrum with the joining of modules is illustrated in the left panel of figure \ref{hs2d35}.

We may also consider the operator product expansion (OPE) and conformal partial wave decomposition in this theory. The theory is free, and so we may compute the OPE via Wick contractions (as reviewed for example in chapter 2 of \cite{Polchinski:1998rq}.) The answer in $d$ dimensions in the identity- and single-trace sectors for the $j^{(0)}_0\times j^{(0)}_0$ OPE is

\begin{align}j^{(0)}_0(x) j^{(0)}_0(0) &=\frac{I}{x^{2(d-4)}} + \frac{N^{-\frac{1}{2}}}{x^{d-4}}\left(\normord{\phi^\dagger(x) \phi(0)} + \normord{\phi(x) \phi^\dagger(0)} \right)+\ldots\nonumber \\ 
&=\frac{I}{x^{2(d-4)}} + \frac{N^{-\frac{1}{2}}}{x^{d-4}}\sum_{k=0}^\infty \frac{1}{k!}\left(\normord{(x\cdot\partial)^k \phi^\dagger \phi}(0) + \normord{(x\cdot\partial)^k\phi \phi^\dagger}(0) \right)+\ldots\nonumber \\
&=\frac{I}{x^{2(d-4)}} + \frac{2N^{-\frac{1}{2}}}{x^{d-4}}\left(\phi^\dagger \phi + \frac{1}{2}x\cdot \partial (\phi^\dagger \phi) +\frac{1}{4d} x^2 (\square \phi^\dagger \phi+\phi^\dagger\square \phi) + \ldots\right)(0)+ \ldots \label{eqn:ddimope}\end{align}
Here the $\ldots$ refer to both higher-order and double-trace terms, which are irrelevant to this discussion. Again, we've restored the factors of $\frac{1}{N}$ so that it's clear that these arise from single-trace terms.

What's clear from equation \eqref{eqn:ddimope} is that nothing singular happens as $d\rightarrow 6$. We normally characterize the OPE by contributions of primaries and descendants, and that characterization should break down as $d\rightarrow 6$. If we choose to continue insisting on breaking things into primaries and descendants then we encounter singularities as we approach $d\rightarrow 6$, having formally analytically continued in $d$. We may match the Wick-contracted OPE to the usual expressions from \cite{Rychkov:2016iqz} in $d$ dimensions, plugging in $\Delta=d-4$ for $j_0^{(0)}$, in order to read off OPE coefficients:

\begin{equation}j^{(0)}_0(x)j^{(0)}_0(0) = \frac{2N^{-\frac{1}{2}}}{x^2} \left(j^{(0)}_0+\frac{1}{2}x\cdot \partial j^{(0)}_0 - \frac{1}{2d(d-6)}x^2 \square j^{(0)}_0 + \ldots \right)(0) + \frac{N^{-\frac{1}{2}}}{2d(d-6)} j_0^{(1)}(0)+\ldots\end{equation}

From this we read off the $j^{(0)}_0$ OPE coefficient as $2N^{-\frac{1}{2}}$ and the $j^{(1)}_0$ OPE coefficient as $\frac{N^{-\frac{1}{2}}}{2d(d-6)}$, the latter of which is singular as $d\rightarrow 6$. Nevertheless, this singularity precisely cancels the singularity present in the higher-order $\square$ term in the $j^{(0)}_0$ channel.

Here, this divergence in the $j^{(0)}_0$ channel is arising because $d=6$ and $\Delta = 2$. This is precisely at the scalar unitarity bound, and so in unitary theories this forces $\square j^{(0)}_0=0$ upon us to cancel this divergence and make the OPE finite. Here we have no such constraint, hence the need for $j_0^{(1)}$ to come in to resolve the singular nature of the OPE.

Working directly in $d=6$, we see that instead what breaks down is our assumption that we may group terms in the OPE up in terms of primaries and descendants. Indeed, if we plug $d=6$ into \eqref{eqn:ddimope} and collect into the operators \eqref{tildeopsdef6}, there are no divergences of any sort,
\begin{equation}j^{(0)}_0(x)j^{(0)}_0(0) \supset \frac{2N^{-\frac{1}{2}}}{x^2} \left(j^{(0)}_0+\frac{1}{2}x\cdot \partial j^{(0)}_0 + \frac{1}{48}x^2 \square j^{(0)}_0 + \ldots \right)(0) - \frac{N^{-\frac{1}{2}}}{12} \tilde{j}^{(1)}_0(0) + \ldots\end{equation}

Thus we may make sense of the OPE in two ways; first as a limiting case where we always adjoin the $j^{(0)}_0$ and $j^{(1)}_0$ channels of the OPE so as to cancel singularities as we approach $d\rightarrow 6$, or the latter (and perhaps more natural) way is to work directly with the operators  \eqref{tildeopsdef6}, avoiding any and all discussions of singularities and subtleties.

A similar story plays out for the conformal partial wave decomposition. Indeed, we may compute the four-point function again with just Wick contractions, and nothing funny happens:
\begin{equation}\langle j^{(0)}_0(x_1)j^{(0)}_0(x_2)j^{(0)}_0(x_3)j^{(0)}_0(x_4)\rangle = \frac{1}{x_{12}^4x_{34}^4} F(u,v)\, ,\end{equation}
\begin{equation}F(u,v)=1+u^2+\left(\frac{u}{v}\right)^2 + \frac{2}{N}\left(u+\frac{u}{v}+\frac{u^2}{v}\right)\, . \end{equation}
If we attempt the usual conformal partial wave decomposition of this correlator, we end up with poles due to the singularity in the n\"aive block associated with $j_0^{(0)}$. However, as we have shown above, the OPE is perfectly finite, and so instead one is forced to consider an ``extended block'' $g_{\mathrm{ext}}$ associated with the exchange of not only the $j_0^{(0)}$ but also $\tilde{j}_0^{(1)}$. It is clear from the above that there will not be any singularity in the proper conformal partial wave decomposition of the above four-point function. Here we may easily understand what block to use by utilizing the limiting behavior approach above\footnote{We thank J. Penedones for discussions concerning this point.}. We simply take the $d\rightarrow 6$ limit of the blocks associated with $j^{(0)}_0$ and $j^{(1)}_0$, weighted by the squares of their OPE coefficients,
\begin{equation}C_{\mathrm{ext}}^2 g_{\mathrm{ext}}(z,\bar{z}) \equiv \frac{4}{N}g_{d-4,0}(z,\bar{z})+\frac{1}{4Nd^2(d-6)^2} g_{d-2,0}(z,\bar{z})\, ,\end{equation}
which again would be singular except for the precise ratio of the OPE coefficients between the two, allowing for a finite limit. We have not carried out this exercise explicitly, but it would be interesting to do.

\subsubsection{$\square^2$ in $d=3$\label{3dspecs}}

In $d=3$, a similar phenomenon occurs between the scalar primary $j_{0}^{(0)}$ and the spin-2 primary $j_{2}^{(0)}$.  Consider the space of states with $\Delta = \dcft-2$, $s=2$ which are bilinear in the fields and survive in the $O(N)$ theory, i.e. are symmetric under $\phi^\dag\leftrightarrow \phi$.  This space is spanned by two basis states, $\phi^\dag\partial_{(i}\partial_{j)_T} \phi +c.c.$ and $\partial_{(i}\phi^\dag\partial_{j)_T}\phi$, and so this space is two-dimensional. 

When $d\not=3$, this space is spanned by the spin-2 primary $j_{2}^{(0)}$, and a spin-2 descendant operator arising from the primary $j_{0}^{(0)}=\phi^\dag \phi$,
\begin{align}
j_{2}^{(0)} &=-(d-4)\left(\phi^\dag\partial_{(i}\partial_{j)_T} \phi +c.c.\right)+2(d-2)\partial_{(i}\phi^\dag\partial_{j)_T}\phi ,\\
\partial_{(i} \partial_{j)_T}j_{0}^{(0)}&=\left(\phi^\dag\partial_{(i}\partial_{j)_T} \phi +c.c.\right)+2\partial_{(i}\phi^\dag\partial_{j)_T}\phi ,
\end{align}
and these two operators are orthogonal in the Hilbert space inner product. 

When $\dcft=3$, these two operators are equal, and so are no longer independent:  
\begin{equation} j_{2}^{(0)} = \partial_{(i} \partial_{j)_T}j_{0}^{(0)} ,\quad d=3\, .\end{equation}
 $ j_{2}^{(0)}$ is now a primary and a descendant and its norm is zero
\be \la j_{i_1i_2}^{(0)}|j_{j_1j_2}^{(0)}\ra=\la\partial_{(i_i} \partial_{i_2)_T}j_{0}^{(0)} |\partial_{(j_1} \partial_{j_2)_T}j_{0}^{(0)}\ra =0\, ,\quad d=3\, . \ee

As in the spin-0 case, the missing direction in this two-dimensional space of operators will be spanned by an extension state.  To find it, we first look at the matrix of inner products 
in the generic basis spanned by
\bea && {\cal E}_{1,ij}= \phi^\dag\partial_{(i}\partial_{j)_T} \phi +c.c., \ \ \nn \\
&& {\cal E}_{2,ij}=\partial_{(i}\phi^\dag\partial_{j)_T}\phi , \\
&&  \left(\begin{array}{cc}\la {\cal E}_{1,i_1i_2}|{\cal E}_{1,j_1j_2}\ra &\la {\cal E}_{1,i_1i_2}|{\cal E}_{2,j_1j_2}\ra \\ \la {\cal E}_{2,i_1i_2}|{\cal E}_{1,j_1j_2}\ra &\la {\cal E}_{2,i_1i_2}|{\cal E}_{2,j_1j_2}\ra \end{array}\right)=\left(\begin{array}{cc}-4 & 0 \\0 & 1\end{array}\right){1\over 2}\left(\delta_{i_1j_1}\delta_{i_2j_2}+\delta_{i_2j_1}\delta_{i_1j_2}-{2\over 3}\delta_{i_1i_2}\delta_{j_1j_2}\right). \nn\\  
\eea
The matrix is again Lorentzian, with two zero-norm directions.  One zero-norm direction is the primary-descendant $ j_2^{(0)}$, and the second is the non-primary non-descendant extension state $\tilde j_2^{(0)}$
\bea 
&&  \tilde j_2^{(0)}=  {\cal E}_{2,ij}-{1\over 2}{\cal E}_{1,ij}= \partial_{(i}\phi^\dag\partial_{j)_T}\phi -{1\over 2}\left(\phi^\dag\partial_{(i}\partial_{j)_T} \phi +c.c\right), \\
&&  \left(\begin{array}{cc}\la j_{i_1i_2}^{(0)}|j_{i_1i_2}^{(0)}\ra  & \la j_{i_1i_2}^{(0)}|\tilde j_{i_1i_2}^{(0)}\ra  \\  \la \tilde j_{i_1i_2}^{(0)}|j_{i_1i_2}^{(0)}\ra  & \la \tilde j_{i_1i_2}^{(0)}|\tilde j_{i_1i_2}^{(0)}\ra  \end{array}\right)=\left(\begin{array}{cc}0 & 4 \\ 4  & 0\end{array}\right){1\over 2}\left(\delta_{i_1j_1}\delta_{i_2j_2}+\delta_{i_2j_1}\delta_{i_1j_2}-{2\over 3}\delta_{i_1i_2}\delta_{j_1j_2}\right). \nn\\
\eea
The correlators take the form
\bea  \la j_{i_1i_2}^{(0)}(x) j_{j_1j_2}^{(0)}(0)\ra =&&0 \, ,\nn\\
   \la j_{i_1i_2}^{(0)}(x) \tilde j^{(0)\, j_1j_2}(0)\ra =&&{4\over  x^2}  I_{(i_1}^{(i_1} I_{j_2)_T}^{j_2)_T} \, ,\nn\\
   \la \tilde j_{i_1i_2}^{(0)}(x) \tilde j^{(0)\, j_1j_2}(0)\ra =&&{1\over  x^2} \bigg[ -{4\over 9}\delta_{i_1i_2}\delta^{j_1j_2}+{8\over x^2}\delta_{(i_1}^{(j_1}x_{i_2)}x^{j_2)}+{4\over 3x^2}\left(\delta_{i_1i_2}x^{j_1}x^{j_2} +\delta^{j_1j_2}x_{i_1}x_{i_2}\right)  \nn\\
  &&-{12\over x^4} x_{i_1}x_{i_2}x^{j_1}x^{j_2} \bigg] \, .
  \eea

Note that the off-diagonal correlator takes a conformally invariant form with a coefficient matching that of the inner product, despite the fact that $\tilde j^{(0)}_2$ is not primary.

\begin{figure}[h!]
\centering
\includegraphics[width=7in]{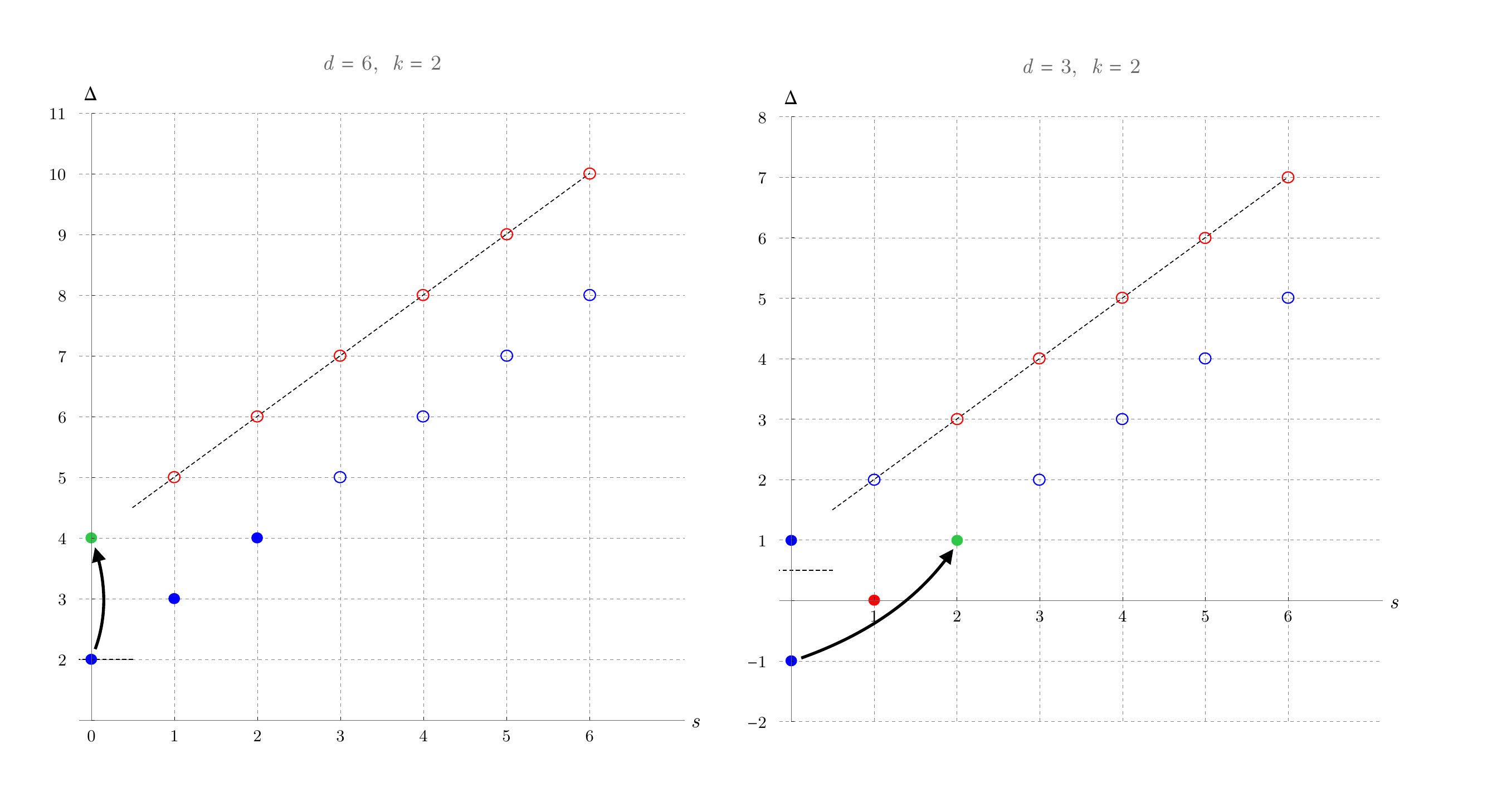}
\caption{Spectrum of primaries in the cases $d=6$ (left) and $d=3$ (right) where modules join in the $\square^2$ theory.  Blue states have positive norm, red states have negative norm, and green states are zero-norm states which are both primary and descendant and get paired with extension states. The mixing of two modules is denoted by an arrow. (The bottom row's unfilled circles are the operators satisfying triple conservation conditions, the top row's unfilled circles are the operators satisfying single conservation conditions, filled circles are the operators satisfying no conservation condition, and the dotted line is the unitarity bound.)}
\label{hs2d35}
\end{figure}

The extension state spawns its own extension module of descendants, which is glued to the $j_0^{(0)}$ module by the action of $K$,
\begin{equation}0 \neq K^j |\tilde {j}_{ij}^{(0)} \rangle \propto |\partial_i j_{0}^{(0)}\rangle\, . \end{equation}
As before, we cannot get to the extension module from the original module using either $K$ or $P$.  The spectrum in $d=3$ with the joining of modules is illustrated in the right panel of figure \ref{hs2d35}.
 
%%%%%%%%%%%%%%%%%%%%%%%%%%%%%%%%%%%%%%%%%%%%%%%%%%%%%%%%%%%%%%%%%%%%%%%%%%%%%%%%

\section{$\square^k$ Theory}
\label{sec:boxk}

We now extend the story to theories with higher powers of box.  We start from the action
\be S\propto \int d^dx ~\phi^\dag  \square^k\phi\, ,\label{CFTgenlag}\ee
with $k$ some integer $\geq 1$. (We again write $\propto$ to allow ourselves the freedom to normalize the two-point function.)

It is straightforward to show that this is indeed a CFT for any $d\geq 2$ and any $k$, i.e. the action \eqref{CFTgenlag} is invariant up to a total derivative under both dilations and special conformal transformations, if $\phi$ has the scaling dimension 
\be \Delta_\phi={d\over 2}-k\,. \label{scalingphiboxke}\ee
We define the CFT by taking the two-point function of $\phi$ to be the only one compatible with conformal symmetry,
\be \langle\phi^\dag (x)\phi(0)\rangle={1\over x^{2\Delta_\phi}},\ee
and determining all other correlators, and those of composite operators, through Wick contraction.

\subsection{$\square^k$ Spectrum}

The single trace primary operators of the CFT come in $k$ ``Regge trajectories''.   Each trajectory is labelled by an integer $b$ representing the number of boxes in the operator, $b\in \{ 0,1,2,\ldots,k-1\}$. Each trajectory has one operator of each spin $s=0,1,2,\ldots$ in the $U(N)$ case, and one of each even spin in the $O(N)$ case.  The operators are of the form
\be { j }^{(b)}_{i_1 \ldots i_s} \sim \phi^\dag  \,\partial_{i_1}\ldots \partial_{i_s}\square^{b}\phi+\ldots,\ \ \ \ s=0,1,2,\ldots\ \ \ \ b=0,\ldots,k-1 \quad , \label{operatorsg}\ee
where the ellipses denote other orderings of the derivatives with coefficients chosen such that on shell they are symmetric, fully traceless, primary, and hermitian.  (As before, we will often use a shorthand for the indices on the operators, writing only $ { j }^{(b)}_{s}$.)  Given the scaling dimension \eqref{scalingphiboxke}, the scaling dimension of $j^{(b)}_s$ is

\begin{equation}\Delta = d-2k+s+2b \, .\label{gencdimke}\end{equation}
The twist is $\tau=\Delta-s=d-2k+2b$, so the sets of operators of identical twist are precisely the Regge trajectories.

The operators come in two types.  Those with $\Delta\geq d-1$ (i.e. $s\geq 2(k-b)-1$) satisfy an on-shell conservation condition by virtue of the equations of motion $\square^k\phi=0$,
\be \partial^{i_1}\ldots \partial^{i_{c}}{ j }_{i_1\ldots i_c \ldots i_s}^{(b)}=0,\ \ \ \ \Delta\geq d-1, \ \ \  c\equiv 2(k-b)-1 \,.\label{consstps}\ee
These operators are $c$-conserved, meaning they vanish on the equations of motion when contracted with $c$ derivatives, but not when contracted by fewer than $c$ derivatives, where $c= 2(k-b)-1$.  The highest trajectory (meaning the single-trace primary with the largest scaling dimension of a given spin) is $b=k-1$, and these operators have $c=1$. They are therefore conserved in the usual sense, and are the usual higher-spin currents familiar from the ordinary free scalar. In particular, the stress tensor is (with the exception of the finite theory cases $d=2,4,6,\ldots,2k$ which have no true stress tensor, see section \ref{finitetheorysecks} and Appendix \ref{couplingtocurvapp})
\be T_{ij}\sim { j }_{ij}^{(k-1)}.\ee

The next highest trajectory with $b=k-2$ has $c=3$, and so on in steps of two, down to $b=0$ corresponding to $c=2k-1$.  These operators are the generalization of the triply-conserved higher-spin operators appearing in the $\square^2$ theory. The operators with $\Delta<d-1$ do not satisfy any such conservation condition.

The unitarity bound for spin $s\geq 1$ is $\Delta\geq d+s-2,$ so among all the $s\geq 1$ operators, those with $b< k-1$ (or $c>1$) violate it, whereas those with $b= k-1$ (or $c=1$) saturate it.  For the scalars $s=0$, the unitarity bound is $\Delta\geq {d\over 2}-1$, and the condition for unitarity is $2(k-b)\leq {d\over 2}+1$, so the $b=k-1$ scalar always satisfies it, whereas the $b< k-1$ scalars will violate it or not depending on whether the dimension is large enough.

\subsection{$\square^k$ Conserved Currents and Symmetries}

As reviewed in appendix \ref{genktapp}, when we have $c$-conserved currents such as \eqref{consstps}, we can contract with rank $s-c$ generalized conformal Killing tensors to get conserved Noether currents,
\be J_{i}^{(s,c)}={ j }_{i\,i_1\ldots i_{s-1}}^{(b)}\partial^{i_1}\ldots\partial^{i_{c}} K^{i_{c+1}\ldots i_{s-1}}_{(s-c)}+\ldots \ee 
Each of these Noether currents is associated with a linearly realized global symmetry of the action \eqref{CFTgenlag}, with the leading derivative part given by the generalized conformal Killing tensor,
\be \delta_K\phi=K^{i_1\ldots i_{s-c}}_{(s-c)} \partial_{i_1}\ldots \partial_{i_{s-c}}\square^{k-{ b}-1}\phi +\ldots \label{symgen}\ee
Here $\ldots$ are terms with the derivatives acting in all other possible ways on $K$ and $\phi$, with coefficients uniquely determined by the requirement that \eqref{symgen} leave the Lagrangian \eqref{CFTgenlag} invariant up to a total derivative.

These symmetries close to form an algebra.  By commuting two of these symmetries, with killing tensors $K$, $K'$, we get a third which is a linear combination of some Killing tensors, which we call $[K,K']$,
\be \left[ \delta_{K},\delta_{K'}\right]\phi=\delta_{[K,K']}\phi+{\rm on\ shell\ trivial}.\ee
This puts a Lie algebra structure on the space of generalized conformal Killing vectors of odd conservedness $1\leq c\leq 2k-1$, and this algebra is the higher spin algebra denoted $hs_k$. This algebra is the algebra of non-trivial symmetries of the equation $\square^k\phi=0$, as discussed in \cite{2009arXiv0911.5265G,Joung:2015jza}.

\subsection{$\square^k$ Correlation Functions}

The two-point functions of primary operators are completely fixed by conformal symmetry up to a constant, $C_{s,b}$,
\be \langle { j }_s^{(b)} (x) { j }_{s'}^{(b')} (0) \rangle = C_{s,b}{1\over x^{2\Delta}}\delta_{s,s'}\delta_{b,b'}I_{(i_1}^{(j_1}\ldots I_{i_s)_T}^{j_s)_T}.\ee

\noindent where $I_{ij}\equiv\delta_{ij}-2{x_i x_j\over x^2}$.

We are free to rescale the operators by any nonzero real constant (real because we prefer the operators to stay hermitian so that we may easily read off positivity).
By doing so, we may scale $C_{s,b}$ by any positive constant, so the only invariant data at the two-point level is the sign of $C_{s,b}$, or whether it vanishes\footnote{For the (multiply) conserved operators \eqref{consstps}, we could instead fix their normalization by looking beyond the two-point level and demanding that they generate via the Ward identity the global symmetries \eqref{symgen} with the appropriate normalization.  The two-point functions then become higher spin generalizations of the central charge \cite{Anselmi:1998bh}. However, as the symmetries are not our concern here, we do not do this in this paper. Instead, in principle we could recover these central charges from the computations of the three-point functions.}: ${\rm sgn}(C_{s,b})\in \{-1,0,1\}$.  This sign determines the norm of the normalized state $|j_s^{(b)}\rangle\equiv j_s^{(b)}(0)|0\rangle$ (corresponding to the operator $j_s^{(b)}$ through the operator state correspondence) in radial quantization,
\be \langle j_s^{(b)}|j_s^{(b)}\rangle ={\rm sgn}(C_{s,b}).\ee

By working out many cases and generalizing, we have arrived at a conjectural expression for these signs:
\be {\rm sgn}(C_{s,b})=\begin{cases} {\rm sgn}\bigg[ (-1)^b \prod_{i=1}^s \left(s + 2 (b - 1-k) + i + d \right) \prod_{i=1}^b \left(2 (s+i+b-1) + d - 4k\right) & \\
\qquad \times \prod_{i=0}^{\left\lfloor{{s\over 2} +  b - 1 }\right\rfloor} \left(d  + 2 (i-k)\right)^2 \bigg] \, , & s\geq 1, \\
 {\rm sgn}\left[ (-1)^b  \prod_{i=1}^b \left(2 (b + i - 1) + d - 4\right) \prod_{i=1}^{b} \left(2 (i -k- 1) + d \right)^2 \right]  \, , & s=0.
\end{cases} \label{normsconjecturek}
\ee
We will use this later to deduce when zero-norm states and module gluings occur in the general case.

\subsection{$\square^k$ Finite Theories\label{finitetheorysecks}}

The only conformally invariant form for the propagator of the basic fields is $\langle \phi^\dag (x) \phi(0)\rangle\sim x^{-2\Delta_\phi}$.  Paralleling the discussion in section \ref{sectionfinitebox2}, when $\Delta_\phi={d\over 2}-k$ is an integer $\leq 0$, the propagator is analytic, and satisfies the equations of motion, but not the Green's function equation with a delta function source.  Of course, we could also consider the log theories instead; however, the finite theories are what will emerge from the dual AdS partially massless higher-spin theory. Furthermore, these theories again all lack stress tensors, and thus are only ``CFTs''.
These analytic cases occur in even dimensions 
\be d=2,4,6,\ldots,2k.\label{finitedvalues}\ee
Away from $\Delta_\phi$ a negative integer, propagators are non-analytic.  
In the dimensions \eqref{finitedvalues} the correlator does not properly describe the theory with Lagrangian \eqref{CFTgenlag}, but nevertheless defines a CFT with a finite number of single trace states.
The alternative is to solve for the Green's function with a delta function source, giving logarithmic correlators.  We again focus only on the finite cases here, leaving studies of the alternative for future work. 

When the basic propagator is analytic, all the correlation functions of all the primary operators are also analytic, because they are built out of derivatives of Wick contractions of the basic propagator.  Conformal invariance demands that the two-point function of a spin-$s$ primary take the form 
\be \langle { j }_{i_1\ldots i_s}(x){ j }_{i_{s+1}\ldots i_{2s}}(0)\rangle \sim x^{-2(\Delta+s)}x_{i_1}\ldots x_{i_{2s}}+\eta {\rm \ terms}\, .\ee
For this to be analytic in the $x^i$, we must have 
\be \Delta+s=0,-1,-2,\ldots.\label{finitedeltavalues}\ee
Otherwise the correlator must vanish in order to be compatible with analyticity.
For the analytic values \eqref{finitedvalues}, there are only a finite number of primaries which satisfy the constraint \eqref{finitedeltavalues}, and these have non-vanishing correlators.  Furthermore, each of these primaries has in turn a finite number of descendants, because as we take more and more derivatives of an analytic correlation function eventually we reach zero.  Any correlator which includes a primary or descendant operator which is not a member of this finite set, or a null descendant of one of these finite primaries, vanishes.  Therefore, for these theories we may factor out all the null operators and descendants.  The resulting single-trace Hilbert space is finite-dimensional, furnishing a (non-unitary) finite-dimensional representation of the conformal group, which breaks up into a direct sum of the finite dimensional  irreducible representations corresponding to each single-trace primary.  

The case $d=2k$ is the critical case.  In this case, there is only one non-vanishing single-trace state, the lowest scalar ${ j }^{(0)}_0\sim\phi^\dag\phi$ which has $\Delta=0$.  Its two-point function is a constant and correlators involving any other state vanish.  Furthermore, all the descendants vanish of ${ j }^{(0)}_0$ vanish, so this furnishes a trivial one-dimensional representation, as discussed in section \ref{box2d4subsecn}.

\begin{figure}[h!]
\centering
\includegraphics[width=\textwidth]{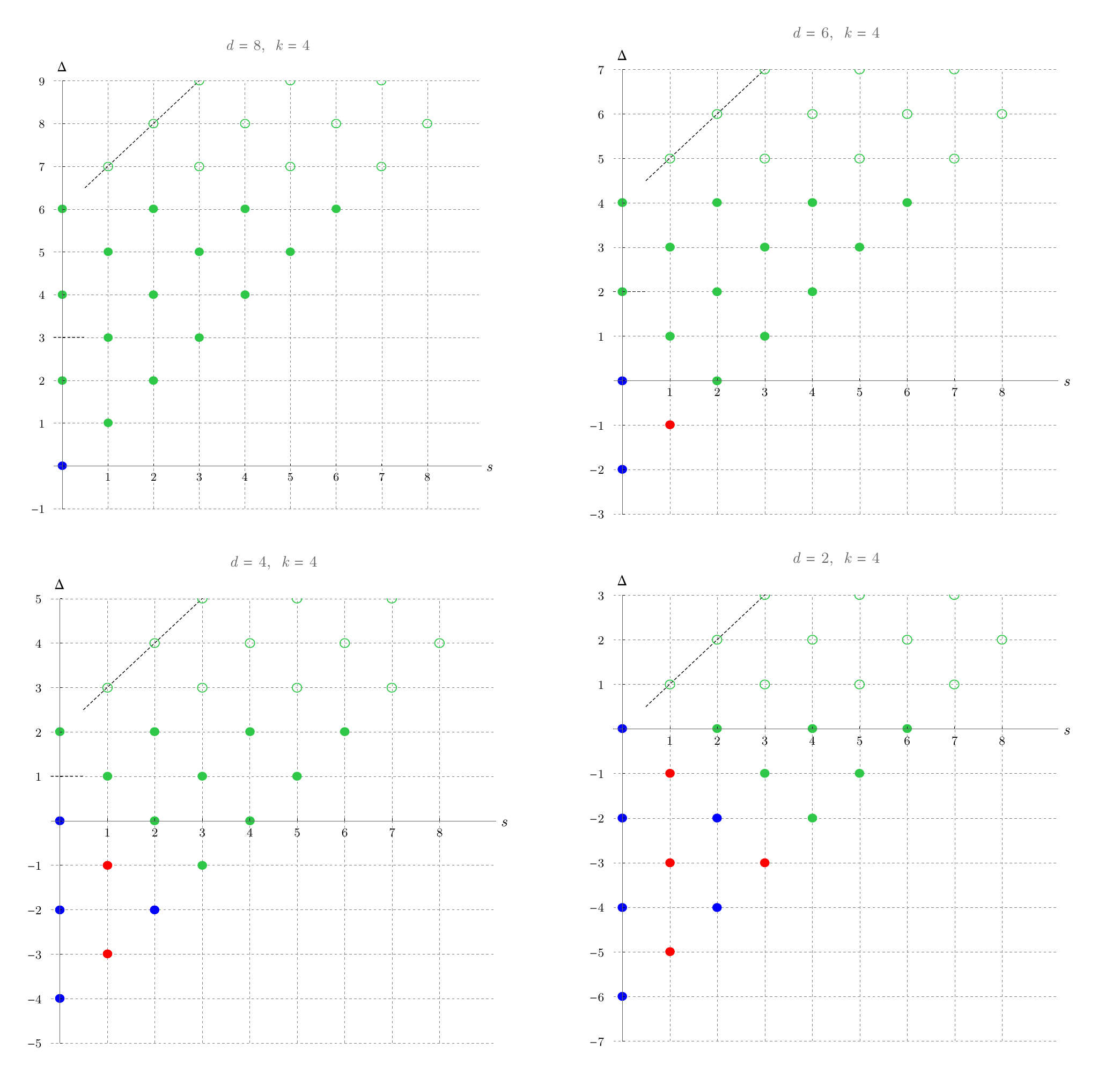}
\caption{Primaries in the case $k=4$ for all dimensions (even $d\leq 2k$) in which there exists a finite CFT, corresponding to the degeneration of the $hs_4$ algebra. The dotted line is the unitarity bound.  Blue states have positive norm, red states have negative norm, and green states are true null states which are projected out. }
\label{hs4finite}
\end{figure}

As we move down in dimension from the critical case, we find more and more non-zero primary states, in a pyramid-like pattern, as shown in table \ref{tab:finite}, and illustrated in figure \ref{hs4finite}.  These primary states are what would be obtained by starting with a $(d+2)$-dimensional symmetric traceless tensor of rank $k-{d\over 2}$ and dimensionally reducing $d+2\rightarrow d$.  In these finite theories, the representation of the higher spin symmetry algebra $hs_k$ degenerates, consistent with the degeneration of the entire algebra to the finite quotient algebras discussed in \cite{Joung:2015jza}.

{\renewcommand{\arraystretch}{1.5}
\begin{table}[h]
\footnotesize
\centering
\begin{tabular}{|c|c|c|c|c|c|c|}\hline
$d$ & $s=0$ & $s=1$ & $s=2$ & $s=3$ & $\ldots$ & $s= k -1$ \\ \hline
\rowcolor{Gray} $2 k $ &${ j }_0^{(0)}$ & & & & &\\
$2 k -2$ & ${ j }_0^{(0)},\ { j }_0^{(1)}$ & ${ j }_{1}^{(0)}$ & && & \\
\rowcolor{Gray} $2 k -4$ &${ j }_0^{(0)},\ { j }_0^{(1)}, \ { j }_0^{(2)}$ & ${ j }_{1}^{(0)},\  { j }_{1}^{(1)}$  & ${ j }_{2}^{(0)}$ & & &\\
$2 k -6$ &${ j }_0^{(0)},\ { j }_0^{(1)}, \ { j }_0^{(2)}, \ { j }_0^{(3)}$ & ${ j }_{1}^{(0)},\  { j }_{1}^{(1)}, \  { j }_{1}^{(2)}$  & ${ j }_{2}^{(0)},\ { j }_{2}^{(1)}$ &  ${ j }_{3}^{(1)}$ & & \\
\rowcolor{Gray} $\vdots$  & & & & & &\\ 
 2 & ${ j }_0^{(0)},\ \ldots,\ { j }_0^{( k-1 )}$ & ${ j }_{1}^{(0)},\ \ldots,\ { j }_{1}^{( k -2)}$ & ${ j }_{2}^{(0)},\ \ldots,\ { j }_{2}^{( k -3)}$ & ${ j }_{3}^{(0)},\ \ldots,\ { j }_{3}^{( k -4)}$ & $\ldots$ & ${ j }_{k -1}^{(0)}$ \\ \hline
\end{tabular}
\caption{All non-trivial single-trace primaries in the finite theories, which occur in even $d\leq 2k$.}
\label{tab:finite}
\end{table}
}

These cases \eqref{finitedvalues}, with the exception of the critical case $d=2k$, are precisely the cases for which the theory cannot be {conformally} coupled to a {general} background metric, as discussed in \cite{Karananas:2015ioa} (see appendix \ref{couplingtocurvapp}), and so are the cases where the theory does not contain a proper stress tensor.

Finally, we note in passing that the critical cases $d=2k$ are all ones for which $\Delta_\phi=0$, and so we could imagine constructing vertex operators in all of these dimensions, allowing for possible Liouville-esque theories. It would be very interesting to construct such theories.

\subsection{$\square^k$ Extended Modules\label{boxkextendedsecn}}

In the $\square^k$ theory, we will have ``extension states'' appearing in various specific dimensions, generalizing the discussion in section \ref{sec:nuances} for the $\square^2$ theory in $d=3,6$. In this subsection, we use \eqref{normsconjecturek} and known results about zero-norm descendant states in the conformal algebra to deduce the general pattern for when these linkings of modules occur.

We conjecture that modules link together whenever a primary has a symmetric tensor zero-norm descendant state which does not vanish by virtue of the $\square^k\phi=0$ equation of motion.  When this happens, there should also be a primary whose norm vanishes according to \eqref{normsconjecturek}, which degenerates with the descendant.
There should then be another operator, an extension operator of the same dimension and spin as the zero-norm descendant, which is neither primary nor descendant.  The extension operator then links to the original module via the $K$ operator, as described in section \ref{sec:nuances} for the $k=2$ case.  

When an operator has a zero-norm descendant is determined entirely by the conformal algebra, and thus depends only on the operator dimension and spin \cite{Shaynkman:2004vu,Kos:2013tga,Penedones:2015aga}.  In our case, we only need to deal with the symmetric tensor operators, and the dimensions at which zero-norm states can occur come in three types (using the terminology of \cite{Penedones:2015aga}),
\begin{equation}
\left\{
\begin{array}{llll}
\Delta= 1-s-n, ~~~&n=1,2,3,\ldots,    ~~~& \left|P_{(i_1}\ldots P_{i_n}|{\cal O}_{j_1\ldots j_s)_T}\rangle\right|= 0, ~~~&{\rm type\ I}\, , \\
\Delta= s+d-1-n, ~~~&n=1,2,\ldots,s-1,s, ~~~& \left|P^{i_1}\ldots P^{i_n}|{\cal O}_{i_1\ldots i_n \ldots i_s}\rangle\right|= 0, ~~~&{\rm type\ II} \, ,\\
\Delta= {d\over 2}-n, ~~~&n=1,2,3,\ldots, ~~~& \left|\left(P^2\right)^n |{\cal O}_{i_1\ldots i_s}\rangle\right|= 0, ~~~&{\rm type\ III}\, .
\end{array}
\right.
\end{equation}
Type I is the case where a symmetrized gradient of the state has zero norm, type II is the case where a divergence has zero norm, and type III is the case where some power of the Laplacian on the state has zero norm.

The extended modules occur in two different series: above and below the critical dimension $d= 2k$.  Above the critical dimension, they occur in the even dimensions $2k+2,2k+4,\ldots,4k-2$, and the degeneracies are all of type III.   Below the critical dimension, they occur in odd dimensions $d=3,5,\ldots, 2k-1$ and the degeneracies are all of type I.  In all these cases, the module gluings happen only for the non-conserved operators.  We leave aside the cases $d=2,4,\ldots,2k$; these are the finite cases discussed in section \ref{finitetheorysecks}.  { We do not find any degeneracies of type II.}

\begin{figure}[h]
\centering
\includegraphics[width=\textwidth]{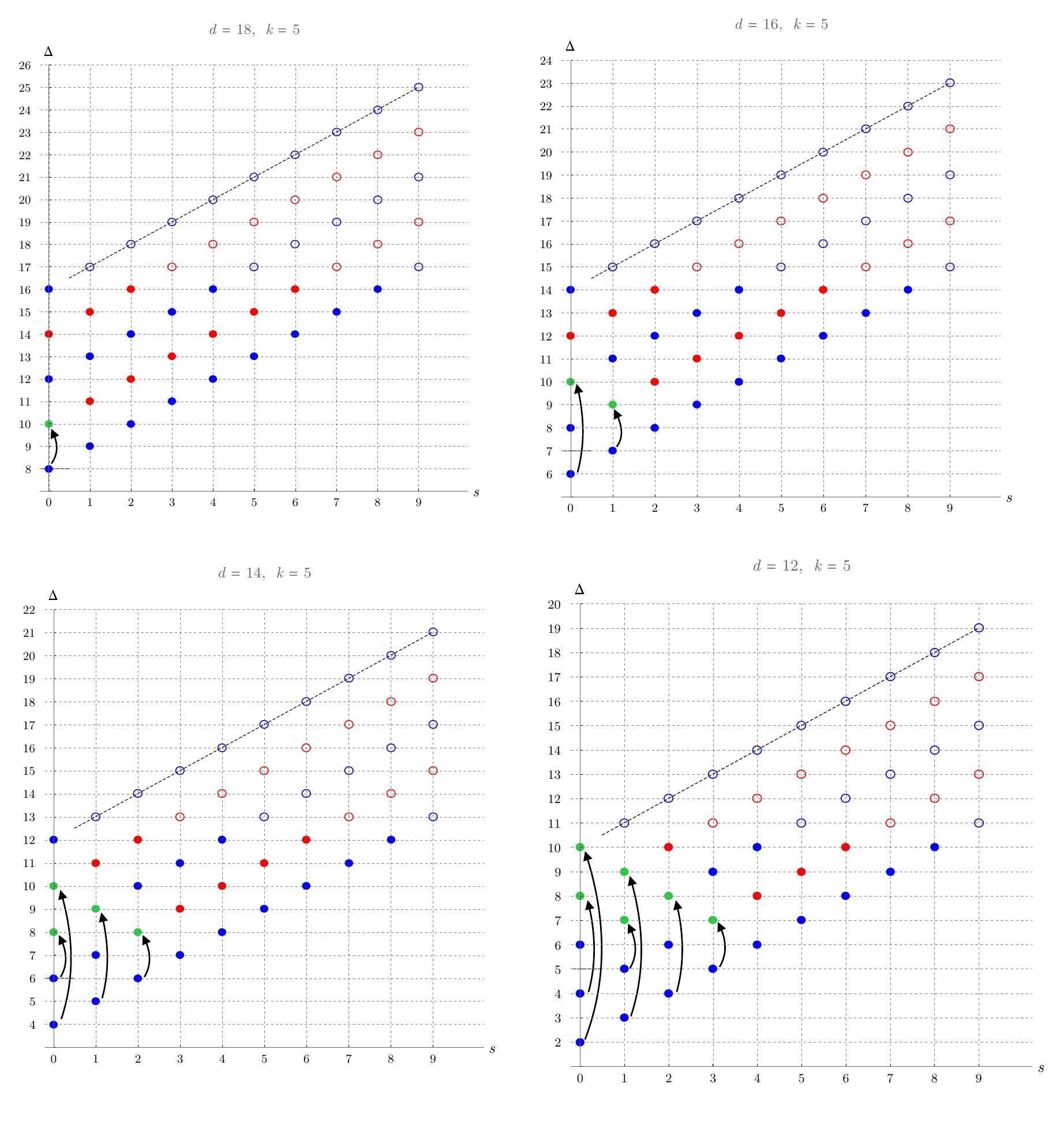}
\caption{Pattern of module gluings for even dimensions above the critical dimension $d=10$ in the case $ k =5$.  Unfilled circles are the (multiply) conserved  primaries, whereas filled circles are the non-conserved primaries. Blue states have positive norm, red states have negative norm, and green states are zero-norm states which are both primary and descendant and get paired with extension states. The mixing of two modules is denoted by an arrow. The Regge trajectories can be seen from $b=0$ on the bottom to $b =4$ at the top. The dotted line is the unitarity bound.}
\label{hs5deven}
\end{figure}

First we discuss the even dimensions $2k+2,2k+4,\ldots,4k-2$.  These generalize the $d=6$ case for $k=2$ of section \ref{6dspec}.  In these cases, all the operators link with other operators of the same spin, i.e. the degeneracies are all of type III.  First we describe the linking of the scalars.  The simplest case is $d=4k-2$.  In this case, the lightest scalar ${ j }_{0}^{(0)}$ is at the unitarity bound $\Delta=d-2k={d\over 2}-1$, and its descendant obtained by acting with $P^2$ has zero norm which links with the next lightest scalar ${ j }_{0}^{(1)}$ with $\Delta={d\over 2}+1$.  Moving down in dimensions, the pattern is as follows, 

\bea s=0\ \begin{cases}  d=4 k -2 \, , &  { j }_{0}^{(0)}\rightarrow { j }_{0}^{(1)} \\
d=4 k -4  \, ,& { j }_{0}^{(0)}\rightarrow { j }_{0}^{(2)} \\
d=4 k -6  \, ,& { j }_{0}^{(0)}\rightarrow { j }_{0}^{(3)},\ \ { j }_{0}^{(1)}\rightarrow { j }_{0}^{(2)}  \\
d=4 k -8  \, ,& { j }_{0}^{(0)}\rightarrow { j }_{0}^{(4)}, \ \ { j }_{0}^{(1)}\rightarrow { j }_{0}^{(3)}   \\
d=4 k -10  \, ,& { j }_{0}^{(0)}\rightarrow { j }_{0}^{(5)}, \ \ { j }_{0}^{(1)}\rightarrow { j }_{0}^{(4)}, \ \ { j }_{0}^{(2)}\rightarrow { j }_{0}^{(3)}   \\
d=4 k -12  \, , & { j }_{0}^{(0)}\rightarrow { j }_{0}^{(6)}, \ \ { j }_{0}^{(1)}\rightarrow { j }_{0}^{(5)}, \ \ { j }_{0}^{(2)}\rightarrow { j }_{0}^{(4)}   \\
 & \vdots \\
d=2 k +2  \, , & { j }_{0}^{(0)}\rightarrow { j }_{0}^{( k-1 )}, \ \ { j }_{0}^{(1)}\rightarrow { j }_{0}^{( k -2)}, \ \ { j }_{0}^{(2)}\rightarrow { j }_{0}^{( k -3)},\ \ \ldots 
\end{cases}\quad .
\eea
For the vectors, the pattern is similar but starts at dimension $d=4 k -4$,
\bea s=1\ \begin{cases}  d=4 k -4  \, ,&  { j }_{1}^{(0)}\rightarrow   { j }_{1}^{(1)} \\
d=4 k -6  \, ,&   { j }_{1}^{(0)}\rightarrow   { j }_{1}^{(2)} \\
d=4 k -8  \, ,&   { j }_{1}^{(0)}\rightarrow   { j }_{1}^{(3)},\ \   { j }_{1}^{(1)}\rightarrow   { j }_{1}^{(2)}  \\
d=4 k -10 \, , &   { j }_{1}^{(0)}\rightarrow   { j }_{1}^{(4)}, \ \   { j }_{1}^{(1)}\rightarrow   { j }_{1}^{(3)}   \\
 & \vdots \\
d=2 k +2  \, , &    { j }_{1}^{(0)}\rightarrow   { j }_{1}^{( k -2)}, \ \   { j }_{1}^{(1)}\rightarrow   { j }_{1}^{( k -3)}, \ \   { j }_{1}^{(2)}\rightarrow   { j }_{1}^{( k -4)},\ \ \ldots 
\end{cases} \quad ,
\eea
the tensors start at dimension $d=4 k -6$,
\bea s=2\ \begin{cases}  d=4 k -6  \, ,&  { j }_{2}^{(0)}\rightarrow   { j }_{2}^{(1)} \\
d=4 k -8 \, ,&   { j }_{2}^{(0)}\rightarrow   { j }_{2}^{(2)} \\
d=4 k -10 \, , &   { j }_{2}^{(0)}\rightarrow   { j }_{2}^{(3)},\ \   { j }_{2}^{(1)}\rightarrow   { j }_{2}^{(2)}  \\
d=4 k -12 \, , &   { j }_{2}^{(0)}\rightarrow   { j }_{2}^{(4)}, \ \   { j }_{2}^{(1)}\rightarrow   { j }_{2}^{(3)}   \\
 & \vdots \\
d=2 k +2  \, ,&   { j }_{2}^{(0)}\rightarrow   { j }_{2}^{( k -3)}, \ \   { j }_{2}^{(1)}\rightarrow   { j }_{2}^{( k -4)}, \ \   { j }_{2}^{(2)}\rightarrow   { j }_{2}^{( k -5)},\ \ \ldots 
\end{cases}\quad ,
\eea
and so on up the spins, until we get to spin $ k -2$, where we have the single linking 
\bea s= k -2\  \begin{cases} d=2 k +2  \, ,& { j }_{k -2}^{(0)}\rightarrow   { j }_{k -2}^{(1)}\quad .
\end{cases}
\eea
This pattern is illustrated for the case $ k =5$ in figure \ref{hs5deven}.

\begin{figure}[h!]
\centering
\includegraphics[width=\textwidth]{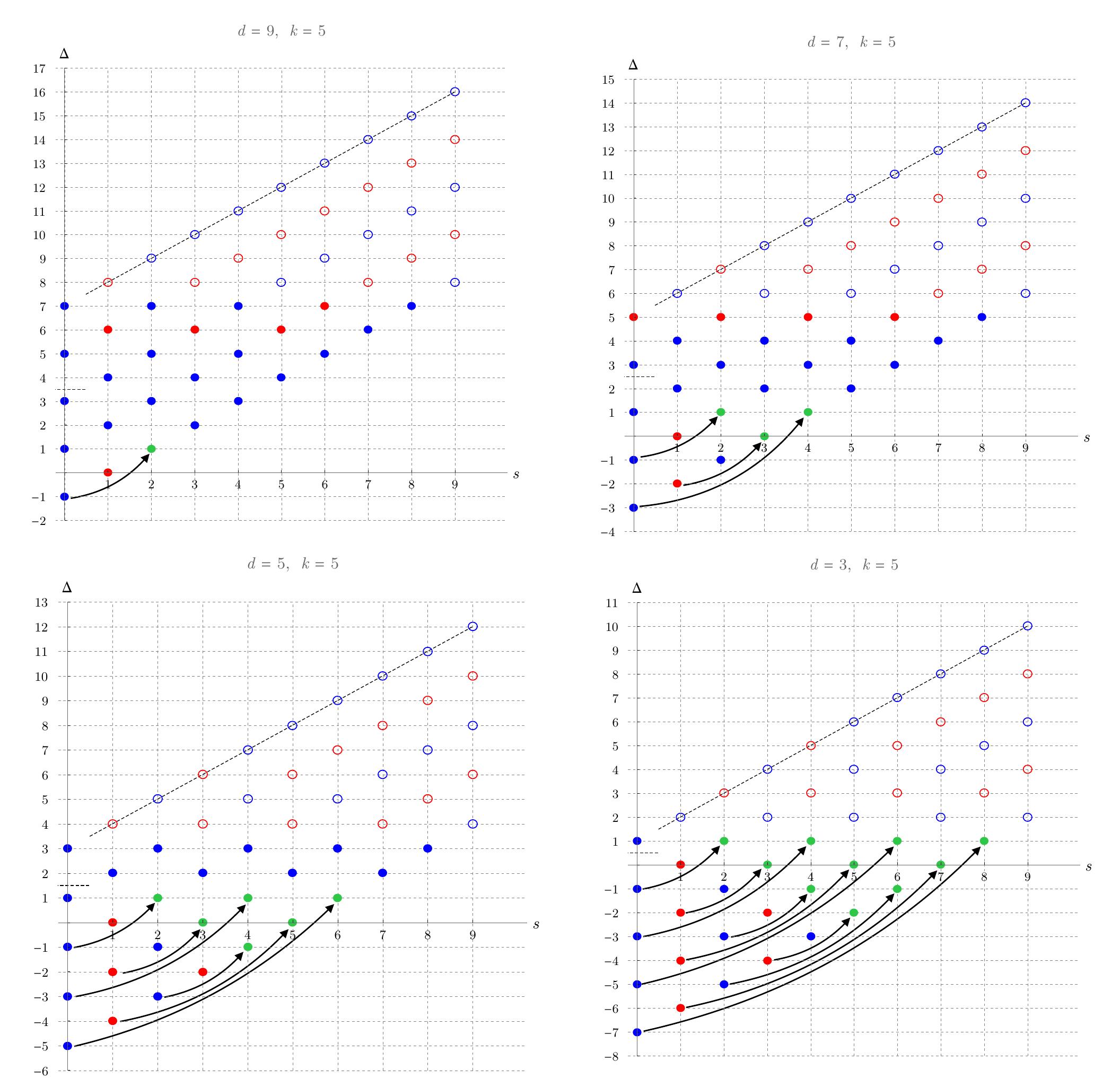}
\caption{Pattern of module gluing for odd dimensions below the critical dimension in the case $k=5$.  Unfilled circles are the (multiply) conserved  primaries, and filled circles are the non-conserved primaries.  The Regge trajectories can be seen from $b=0$ on the bottom to $b=4$ at the top.  The dotted line is the unitarity bound.}
\label{hs5dodd}
\end{figure}

Next we discuss the odd dimensions below the critical dimension, $d=3,5,\ldots, 2 k -1$.  These generalize the $d=3$ case for $k=2$ of section \ref{3dspecs}.  In these cases,  the operators link with other operators of higher spin along the same Regge trajectory, i.e. the degeneracies are all of type III.  The simplest case is $d=2 k -1$.  In this case, the lowest-$\Delta$ scalar ${ j }_0^{(0)}$ has scaling dimension $\Delta=d-2k=-1$, which is the scaling dimension where the descendant $P_{(i}P_{j)_T}{ j }_0^{(0)}$ has zero norm, linking ${ j }_0^{(0)}$ with the spin-2 primary ${ j }_2^{(0)}$ with $\Delta=1$.  The pattern of linking across the first Regge trajectory proceeds as follows,
\bea b=0\ \begin{cases}  
d=2 k -1  &  { j }_0^{(0)}\rightarrow   { j }_{2}^{(0)} \\
d=2 k -3  &  { j }_0^{(0)}\rightarrow   { j }_{4}^{(0)}, \ \  { j }_{1}^{(0)}\rightarrow   { j }_{3}^{(0)}  \\
d=2 k -5  &   { j }_0^{(0)}\rightarrow   { j }_{6}^{(0)}, \ \  { j }_{1}^{(0)}\rightarrow   { j }_{5}^{(0)},\ \  { j }_{2}^{(0)}\rightarrow   { j }_{4}^{(0)} \\
 & \vdots \\
d=3  &   { j }_0^{(0)}\rightarrow   { j }_{2 k -2}^{(0)},  \ \  { j }_{1}^{(0)}\rightarrow   { j }_{2 k -3}^{(0)},\ \ \ldots,  \ \  { j }_{k -2}^{(0)}\rightarrow   { j }_{ k }^{(0)}
\end{cases}\quad ,
\eea
the second Regge trajectory proceeds as,
\bea b=1\ \begin{cases}  d=2 k -3  &  { j }_0^{(1)}\rightarrow   { j }_{2}^{(1)} \\
d=2 k -5  &  { j }_0^{(1)}\rightarrow   { j }_{4}^{(1)}, \ \  { j }_{1}^{(1)}\rightarrow   { j }_{3}^{(1)}  \\
 & \vdots \\
d=3  &   { j }_0^{(1)}\rightarrow   { j }_{2 k -4}^{(1)},  \ \  { j }_{1}^{(1)}\rightarrow   { j }_{2 k -5}^{(1)},\ \ \ldots,  \ \  { j }_{k -3}^{(1)}\rightarrow   { j }_{k -1}^{(1)}
\end{cases}\quad ,
\eea
and so on up the Regge trajectories until we reach $b=k-2$, where we have
\bea b=k-2\ \begin{cases}  d=3  &   { j }_0^{(k-2)}\rightarrow   { j }_{2}^{(k-2)}\quad .
\end{cases}
\eea
This pattern is illustrated for the case $k=5$ in figure \ref{hs5dodd}.

Odd dimensions $d>2k$ and all dimensions $d>4k-2$ have no extended modules.  Note that the patterns of module gluing are all compatible with the truncation to $O(N)$ where we keep only even spins.
 
%%%%%%%%%%%%%%%%%%%%%%%%%%%%%%%%%%%%%%%%%%%%%%%%%%%%%%%%%%%%%%%%%%%%%%%%%%%%%%%%

\section{Conclusions and Future Directions}
\label{sec:conc}

In this paper, we have discussed aspects of the $\square^k$ free scalar field theories, restricted to the singlet sector of a $U(N)$ or $O(N)$ ``gauge'' group under which the scalar field transforms as a fundamental. We first worked through the $\square^2$ theory in detail as an example, and then moved on to the $\square^k$ theory, preferring to state results.

We began by working out the spectrum of single-trace primaries and computing their two-point functions. We used the vanishing of various two-point functions as a diagnostic tool for the appearance of unfamiliar representation theory, whereby the space of single-trace primaries could truncate and become finite-dimensional (for example in $d=2,4$ in the $\square^2$ theory). We investigated the consistency of the operator algebra in these theories. We also studied cases where two different representations become linked together due to the appearance of non-null but zero-norm states which are both primary and descendant. We argued that in order to continue spanning the space of operators, these zero-norm states imply the existence of other zero-norm states which are neither primary nor descendant, and we referred to them as extension states. The modules associated with these mix, forming one larger representation. We demonstrated this explicitly in $d=3,6$ in the $\square^2$ theory. We went on to discuss the general conditions under which these finite and extension patterns occur in the $\square^k$ theory.

There are many further avenues and questions which would be interesting to pursue.  For example:

\begin{itemize}

\item It would be interesting to study the AdS dual of these theories \cite{Bekaert:2013zya}. One approach is to use the unfolding structure in the Vasiliev equations, gauging the $hs_k$ algebra instead of the usual $hs$ algebra.  We plan to report on this for the $\square^2$ theory \cite{Brust:2016zns}.

\item Building upon this, one might conjecture that the ``$U(-N)$'' or $Sp(N)$ counterparts of these CFTs are dual to partially massless theories on dS, generalizing the conjecture for the usual Vasiliev theory \cite{Anninos:2011ui}. Can these additional examples of dS/CFT help us say anything about unitarity or the emergence of a timelike direction in the partially massless higher-spin theories in dS? As a first step, could we compute and learn anything from the Hartle-Hawking wavefunction of the dS theory, along the lines of \cite{Anninos:2012ft}?

\item In principle, there is nothing stopping us from gauging the finite truncations of $hs_k$ in the bulk. These bizarre gauge theories should reproduce at the boundaries the finite theories studied in this paper. What do these finite-dimensional theories look like in AdS or dS?

\item Do similar representation-theoretic phenomena occur in the fermionic counterpart theories described by the Lagrangian $\psi^\dagger_a\slashed{\partial}^k \psi^a$ and their associated algebras?

\item In the Vasiliev/$O(N)$ duality, a $\phi^4$ double-trace deformation at the boundary can be implemented by changing the scalar boundary condition in the bulk. Is it possible to turn on some double-trace deformation with $\Delta < \frac{d}{2}$ in these theories in such a way that the resulting theory is still sensible/doesn't suffer from vacuum runaways or other such pathologies? Do the $\frac{1}{N}$ or $\epsilon$ expansions offer us any control over the nearby non-unitary QFTs, and might it be possible to flow to a new strongly-coupled higher-derivative critical point?

\item Can this theory be supersymmetrized, and if so does the supersymmetric version exhibit similar phenomena? For example, in 3d with $\mathcal{N}=2$ supersymmetry, we could study the theory with the K\"ahler potential $\mathcal{K} = \bar{D}^2 \Phi_a^\dagger D^2 \Phi^a$. This n\"aively imparts a single-box kinetic term to the formerly auxiliary scalar field $F$ in the chiral multiplet. Does this supersymmetry offer us any control to allow for deformations, using the power of localization or other such techniques?

\item What do these 3d theories look like with Chern-Simons couplings turned on? What happens to the module mixings? Can we couple to suitable topological gauge theories to carry out something similar in $d>3$?

\item What is the structure of the log-CFTs? What is their underlying algebra and representation theory? Can the $d=2k$ cases be turned into Liouville-like theories exploiting $\Delta_\phi = 0$? Can we construct bulk duals of these?

\item Are there any useful lessons to be learned from adjoint-valued equivalents of this theory? E.g. Tr$(\square \phi \square \phi)$?

\item Might the extended modules and their associated extended conformal blocks play any role in the recursion relations of \cite{Penedones:2015aga}, or more generally in the conformal bootstrap program?

\end{itemize}

Progress in understanding dS/CFT has suffered from a lack of examples. If we believe that unitarity in de Sitter is reflected in some property of non-unitary CFTs, then we must greatly expand how much we know about non-unitary CFTs in order to make progress in understanding quantum gravity in dS.  One might speculate that the theories discussed here, along with their de Sitter higher-spin duals, form only the tip of an iceberg of new examples of dS/CFT.  There could be all sorts of ways of deforming or changing these theories, none of which has been explored in de Sitter, but all of which are tractable, by analogy with the Vasiliev/$O(N)$ duality. We might hope that by pursuing these avenues, we might finally gain a better handle into what quantum gravity in de Sitter truly is.

%%%%%%%%%%%%%%%%%%%%%%%%%%%%%%%%%%%%%%%%%%%%%%%%%%%%%%%%%%%%%%%%%%%%%%%%%%%%%%%%

{\bf Acknowledgements:} We thank Xavier Bekaert, Frederik Denef, Tudor Dimofte, Ethan Dyer, Davide Gaiotto, Simone Giombi, Bob Holdom, Euihun Joung, Jared Kaplan, Igor Klebanov, Shota Komatsu, Hee-Cheol Kim, Jo\~ao Penedones, Emilio Trevisani, Matt Walters, and Yuan Wan for helpful discussions and comments. We also thank Masha Baryakhtar and Robert Lasenby for helpful comments on the presentation of the material. Research at Perimeter Institute is supported by the Government of Canada through Industry Canada and by the Province of Ontario through the Ministry of Economic Development and Innovation.

\appendix

\section{Coupling to Background Curvature\label{couplingtocurvapp}}

Here we describe how to conformally couple the $\square^k$ CFTs to a background metric $g_{ij}$.  This is necessary to compute partition functions on curved spaces such as spheres, and also yields the stress tensor via functional differentiation.  The conformal coupling to the metric can be achieved by replacing $\square^k$ with covariant derivatives and then adding lower derivative terms proportional to the background curvatures,
\be  \int d^dx \,  \phi^\dag \square^k \phi\rightarrow  \int d^dx \sqrt{g}\left[ \phi^\dag  \square^k\phi+{\rm curvature\ terms}\right],\ee
 in such a way that the action is Weyl invariant with the appropriate conformal weight,
\be g_{ij}\rightarrow e^{2\sigma(x)}g_{ij},\ \ \ \phi\rightarrow e^{-{\Delta}\sigma(x)}\phi,\ \ \ \Delta={d\over 2}-k,\ee
for an arbitrary scalar function $\sigma(x)$. 
This Weyl invariance uniquely fixes the lower-derivative curvature terms, and can be achieved in all cases except the low even dimensions $d=2,4,\ldots,2k-2$.

The scalar equations of motion give the conformally covariant operators known as the GJMS operators \cite{Graham01121992}, 
$ P_k[g]$,
which are the unique $2k$-th order operators which reduce to $\square^k$ in flat space, $P_k[g]=\square^k +{\rm curvature\ terms}$, and are covariant under the above Weyl transformations,
\be P_k[e^{2\sigma} g]\left(e^{-{\Delta}\sigma}\phi\right)= e^{{-(\Delta+2k)}\sigma}  P_k[ g]\phi  .\ee

There are however exceptions in the cases of low even dimensions $d=2,4,\ldots,2k-2$.  In these cases, the curvature terms are singular, the GJMS operators do not exist, and the action cannot be coupled to gravity in a Weyl invariant manner.  The gravitational stress tensor doesn't exist, because finding it requires varying with respect to the metric, which requires being able to couple to an arbitrary metric.  These are precisely the cases of the (non-critical) finite or log theories, discussed in sections \ref{sectionfinitebox2} and \ref{finitetheorysecks}.  In these cases the theory is still globally conformally invariant, despite the fact that it cannot be made Weyl invariant \cite{Karananas:2015ioa,Farnsworth:2017tbz}.

\subsection{ $\square$ Theory}

In this case, we have the usual conformal coupling of the free scalar,
\be \int d^dx \,  \phi^\dag \square\phi\rightarrow  \int d^dx \sqrt{g}\left[ \phi^\dag  \square\phi -\frac{d-2}{4 (d-1)}R|\phi|^2\right],\ee
with $R$ the background Ricci curvature.  The coupled action is invariant under a Weyl transformation where the scalar transforms with the proper conformal weight,
\be g_{ij}\rightarrow e^{2\sigma(x)}g_{ij},\ \ \ \phi\rightarrow e^{-{\Delta}\sigma(x)}\phi,\ \ \ \Delta={d\over 2 }-1.\label{weylbox1eq}\ee

The scalar equations of motion give the first GJMS operator, 
\be P_1[g]\phi\equiv\left( \square -\frac{d-2}{4 (d-1)}R\right)\phi=0,\ee
which is known as the Yamabe operator \cite{yamabe1960}.  The Yamabe operator is the unique second-order operator which is covariant under the Weyl transformation \eqref{weylbox1eq} and which reduces to the Laplacian in flat space.

The on-shell flat space stress tensor is
\be T_{ij}=-\left.{2\over \sqrt{g}}{\delta S\over \delta g^{ij}}\right|_{g=\eta,\square\phi=0}={1\over 2(d-1)}\left[-(d-2) \phi^\dagger\partial_i\partial_j\phi+d\, \partial_i\phi^\dag \partial_j\phi+c.c\right]_{sym,T,\square\phi=0},\ee
and is conserved and traceless on-shell,
\be\left. \partial^j T_{ij}\right|_{\square\phi=0}= 0,\ \ \ \ \left. T^i_{\ i}\right|_{\square\phi=0}= 0.\ee

\subsection{ $\square^2$ Theory}

For the $\square^2$ theory, the curvature couplings take the form
\bea && \int d^dx \, \phi^\dag  \square^2\phi\rightarrow  \int d^dx \sqrt{g}\bigg[ \phi^\dag  \square^2\phi+\left({2\over d-2}R^{ij}\phi^\dag \nabla_i\nabla_j\phi-{d^2-4d+8\over 4(d-2)(d-1)}R\phi^\dag \square\phi +c.c.\right) \nn\\
&&+\left(-{d-4\over (d-2)^2}R_{ij}^2+{(d-4)(d^3-4d^2+16d-16)\over 16(d-2)^2(d-1)^2}R^2-{1\over 2(d-1)}\square R\right)|\phi|^2\bigg]. \label{paneitzact2}\eea
This action is invariant under the Weyl transformation
\be g_{ij}\rightarrow e^{2\sigma(x)}g_{ij},\ \ \ \phi\rightarrow e^{-{\Delta}\sigma(x)}\phi,\ \ \ \Delta={d\over 2}-2. \label{weylbox2eq}\ee

The scalar equations of motion give the second GJMS operator, 
\bea P_2[g]\phi & \equiv & \square^2\phi+\frac{4}{d-2}R_{ij}\nabla^i\nabla^j\phi-\frac{d^2-4 d+8}{2 (d-2) (d-1)}R\square\phi \nn\\ && +\left(-\frac{d-4}{4 (d-1)}R_{ij}R^{ij}+\frac{(d-4) \left(d^3-4 d^2+16 d-16\right)}{16 (d-2)^2 (d-1)^2}R^2-\frac{d-4}{4 (d-1)}\square R\right)\phi=0, \nn\\ 
\eea
which is known as the Paneitz operator \cite{2008SIGMA...4..036P} (who obtained it first in general dimensions), which made earlier appearances in \cite{Fradkin:1981jc,Fradkin:1981iu,Fradkin:1982xc,Riegert:1984kt}.  It is the unique fourth-order operator which is covariant under the Weyl transformation \eqref{weylbox2eq} which reduces to $\square^2$ in flat space. 

In $d=2$, the action \eqref{paneitzact2} is singular, and the Paneitz operator does not exist.  Thus the $\square^2$ theory in $d=2$  cannot be coupled to {an arbitrary background} metric, and does not have a proper stress tensor.

The on-shell flat space stress tensor is computed as
\bea && T_{ij}=-\left.{2\over \sqrt{g}}{\delta S\over \delta g^{ij}}\right|_{g=\eta,\square^2\phi\,=0}= \nn\\
&& {1\over d-1}\bigg[-(d-4)\phi^\dag  \partial_i\partial_j\square\phi+2(d+2)\partial_i\phi^\dag \partial_j\square\phi-4\partial_k\phi^\dag \partial^k\partial_i\partial_j\phi \nn\\ &&-{d(d+2)\over d-2}\square\phi^\dag \partial_i\partial_j\phi+{4d\over d-2}\partial_k\partial_i\phi^\dag \partial^k\partial_j\phi+c.c.\bigg]_{sym,T,\square^2\phi=0},
\eea
and one can check that it is indeed conserved and traceless on-shell,
\be\left. \partial^j T_{ij}\right|_{\square^2\phi=0}= 0,\ \ \ \ \left. T^i_{\ i}\right|_{\square^2\phi=0}= 0.\ee

\subsection{ $\square^k$ Theory\label{boxkmetricsubs}}

For $k\geq 3$, general expressions are quite involved \cite{2009arXiv0905.3992J,Juhl:2011ua,2012arXiv1203.0360F}, but in the case where the background metric is Einstein, so that $R_{ij}={R\over d}g_{ij}$ with constant $R$, the action can be written rather simply in a factorized form \cite{Beccaria:2015vaa} 
\be \int d^dx \,  \phi^\dag\, \square^k\phi\rightarrow  \int d^dx \sqrt{g}\, \phi^\dag  \prod_{l=1}^k \left[ \square-{R\over d(d-1)}\left({d\over 2}-l\right)\left({d\over 2}+l-1\right)\right]\phi .\label{generaleinsts}\ee
Some general properties are discussed in \cite{Gover:2005mn,Manvelyan:2006bk,2009arXiv0905.3992J,Juhl:2011ua,2012arXiv1203.0360F}.  Note that in low even dimensions $d=2,4,\ldots, 2k-2$ the action \eqref{generaleinsts} is not singular.  The general expression is singular in these dimensions, but the singularities cancel upon using the Einstein condition.  The partition function of the theory on Einstein spaces, such as spheres, may therefore be well-defined, even though the theory cannot be coupled to a general background metric.

\section{Higher-Order Conformal Killing Tensors\label{genktapp}}

A generalized conformal Killing tensor $K^{i_1\ldots i_{s-c}}_{(s,c)}$ of ``spin" $s$ and conservedness $c$ on ${\mathbb R}^d$ is a symmetric traceless tensor field satisfying
\be \partial^{(i_1}\ldots\partial^{ i_c}K^{i_{c+1}\ldots i_{s})_T}_{(s,c)}=0.\label{gcke}\ee
For example, the ordinary conformal Killing tensors are the case $c=1$, satisfying $\partial^{(i_1}K^{i_2\ldots i_{s})_T}_{(s,1)}=0$, and the ordinary conformal Killing vectors are the case $c=1$, $s=2$, satisfying $\partial^{i}K^{j}_{(2,1)}+\partial^{i}K^{j}_{(2,1)}={2\over d}\partial_l K^l_{(2,1)}\delta^{ij}.$

Given a symmetric traceless spin-$s$ operator satisfying a $c$-fold conservation condition,
\be \partial^{i_1}\ldots \partial^{i_{c}}{\cal O}_{i_1\ldots i_c \ldots i_s}^{(s,c)}=0,\ \label{consstpsc}\ee
 we can contract with a generalized conformal Killing tensor to form a conserved Noether current,
\begin{align}  J_{i}^{(s,c)}=&{\cal O}_{ii_1\ldots i_{s-1}}^{(s,c)}\partial^{i_1}\ldots\partial^{i_{c-1}}K^{i_c\ldots i_{s-1}}_{(s,c)}\nonumber\\
&-\partial^{i_1}{\cal O}_{ii_1\ldots i_{s-1}}^{(s,c)}\partial^{i_2}\ldots\partial^{i_{c-1}}K^{i_c\ldots i_{s-1}}_{(s,c)}, \nonumber \\
&+\partial^{i_1}\partial^{i_2}{\cal O}_{ii_1\ldots i_{s-1}}^{(s,c)}\partial^{i_3}\ldots\partial^{i_{c-1}}K^{i_c\ldots i_{s-1}}_{(s,c)}, \nonumber \\
&\vdots\nonumber \\
&+(-1)^{c-1}\partial^{i_1}\ldots\partial^{i_{c-1}}{\cal O}_{ii_1\ldots i_{s-1}}^{(s,c)}K^{i_c\ldots i_{s-1}}_{(s,c)}\, .\end{align}
The Noether current is conserved
\be \partial^i  J_{i}^{(s,c)}=0,\ee
upon use of the conservation condition \eqref{consstpsc}, the generalized conformal Killing equation \eqref{gcke}, and the symmetry and tracelessness of the current ${\cal O}$.

Let $X^{A}(x)$ be the following embedding of ${\mathbb R}^d$ into $\mathbb{R}^{d+1,1}$ with metric $\eta_{AB}={\rm diag}\left(-1,1,\delta_{ij}\right)$,
\be X^{A}(x)=\left(\begin{array}{c}{1\over 2} (1+x^2) \\ {1\over 2} (1-x^2) \\x^i\end{array}\right)\, .\ee
The most general solution to \eqref{gcke} is given by 
\be K_{i_1\ldots i_{s-c}}^{(s,c)}=X^{A_1}\ldots X^{A_{s-1}}\partial_{i_1}X^{B_1}\ldots \partial_{i_{s-c}}X^{B_{s-c}}K_{A_1B_1,\ldots, A_{s-c}B_{s-c},A_{s-c+1}\ldots A_{s-1}} ,\label{killingdesce}\ee
where $K_{A_1B_1,\ldots, A_{s-c}B_{s-c},A_{s-c+1}\ldots A_{s-1}}$ is a constant completely traceless tensor with the symmetries of a tableaux with one row of length $s-1$ (the $A$ indices) and a second row of length $s-c$ (the $B$ indices),
\be
K_{A_1B_1,\ldots, A_{s-c}B_{s-c},A_{s-c+1}\ldots A_{s-1}}~\in~
\begin{array}{|c c c c c|}\hline
&s-1\!\!\!\!\!\!&\!\!\!&&\\
\hline
\!\!\!  &\!\!\! ~~~s-c~~~  \vline\!\!\\
\cline{1-2}
\end{array}^{\ T}~, \label{tableaxeq}
\ee
where the superscript $T$ indicates total tracelessness.
Thus the space of generalized conformal Killing tensors is isomorphic to the vector space of constant ($d+2$)-dimensional traceless tensors of the symmetry type \eqref{tableaxeq}.

As an example, we describe the first set of new elements of $hs_2$. These are third-order conformal killing tensors $K_{(3,3)}$, which are scalar functions which solve
\begin{equation}\partial_{(i}\partial_j \partial_{k)}^{\ T} K_{(3,3)} = 0\, .\label{triplekillinge}\end{equation}
The most general solution to this can be inferred from the traceless symmetric tensor $K_{A_1A_2}\in\yng(2)^{\ T}$ on $\mathbb{R}^{d+1,1}$, which descends to $\mathbb{R}^{d}$ according to
\be \yng(2)^{\ T}\rightarrow \yng(2)^{\ T}+\yng(1)+\yng(1)+\bullet+\bullet+\bullet\, \ , \ee
which we can parametrize as 
\be K_{A_1A_2} =\left(\begin{array}{ccc}a^{(1)}+a^{(3)}+{d\over d+2}a^{(2)} & a^{(1)}-a^{(3)} & {1\over 2}(b_j^{(1)}+b_j^{(1)} )\\  a^{(1)}-a^{(3)} & a^{(1)}+a^{(3)}-{d\over d+2}a^{(2)} & {1\over 2}(b_j^{(1)}-b_j^{(2)} ) \\ {1\over 2}(b_i^{(1)}+b_i^{(2)} )  & {1\over 2}(b_i^{(1)}-b_i^{(1)} ) & c_{ij}+{2\over d+2}g_{ij}a^{(2)}\end{array}\right)\, ,\ee
with $d$ dimensional constant tensors: $a^{(1)}$, $a^{(2)}$, $a^{(3)}$, $b_i^{(1)}$, $b_i^{(2)}$ and symmetric-traceless $c_{ij}$.
Using \eqref{killingdesce} it descends to
\begin{equation}K_{(3,3)}=a^{(1)} + b_i^{(1)} x^i + c_{ij}x^i x^j+a^{(2)}x^2 + b_i^{(2)} x^i x^2 + a^{(3)} x^4\, \label{Kexprsp} .\end{equation}

In the $\square^2$ theory we have the spin-3 triply conserved operator $j^{(0)}_3$, 
\be \partial^{i}\partial^{j}\partial^{k}j^{(0)}_{ijk}=0\, . \label{triplyconsa}\ee
The Noether current reads
\be J_{i}^{(3,3)}\propto j^{(0)}_{ijk} \partial^j\partial^k K_{(3,3)}- \partial^j j^{(0)}_{ijk}\partial^k K_{(3,3)}+\partial^j\partial^k j^{(0)}_{ijk}  K_{(3,3)} \, ,\label{noethercappex}\ee
and one can check directly that it satisfies $\partial^i  J_{i}^{(s,c)}=0$ upon using \eqref{triplekillinge} and \eqref{triplyconsa} and the fact that $j^{(0)}_{ijk}$ is symmetric and traceless.  The symmetry responsible for this Noether current reads
\be \delta\phi=i\left[K_{(3,3)}\square\phi-\partial^iK_{(3,3)}\partial_i\phi-{d-4\over 2(d+2)}\square K_{(3,3)}\phi\right]\, .  \ee
Using this and \eqref{Kexprsp}, we can write the transformation in terms of $d$ dimensional differential operators,
\be \delta\phi=i\left[ a^{(1)}\mathcal{D}^1+a^{(2)}\mathcal{D}^2+a^{(3)}\mathcal{D}^3+b^{(1)i}\mathcal{D}^1_i +b^{(2)i}\mathcal{D}^2_i +c^{ij}\mathcal{D}_{ij} \right]\phi,\ee  defined as 

\begin{align}\mathcal{D}^1 &= \square\quad, \nn \\
\mathcal{D}^{2} &= x^2 \square - 2x\cdot \partial -\frac{d(d-4)}{d+2} \quad , \nn\\
\mathcal{D}^3 &= x^4\square - 4 x^2 x\cdot \partial -2(d-4)x^2 \quad, \nn\\
\mathcal{D}^1_i &= x_i \square - \partial_i \quad, \nn\\
\mathcal{D}^2_i &= x_i x^2 \square - 2 x_i x\cdot \partial - x^2 \partial_i -(d-4) x_i \quad , \nn\\
\mathcal{D}_{ij} &= (x_i x_j - \frac{1}{d} \eta_{ij} x^2) \square - x_i \partial_j - x_j \partial_i + \frac{2}{d}\eta_{ij} x\cdot \partial \quad .
\end{align}
These symmetries do not commute with conformal transformations, which permute them amongst themselves, and do not commute with each other, closing instead into the whole of $hs_2$.

\section{Conformal Anomalies and Free Energies}

In this section, we work out the $a$-type conformal anomalies and the free energies $F$ for the some of the lowest lying values of $d$ and $k$, specifically for the theory with one real scalar in it.   Some of these numbers have been computed before, for example\footnote{We thank Simone Giombi and Igor Klebanov for sharing preliminary results which helped us.} \cite{Dowker:2010qy,Dowker:2013oqa,Dowker:2013ysa,Dowker:2015xya,Beccaria:2015uta}.  For the $\square^2$ theory, we match to their AdS counterparts in \cite{Brust:2016xif} (see also \cite{Gunaydin:2016amv}).

\subsection{Even Dimensions: Conformal Anomalies}

In even dimensions, the $a$-type conformal anomaly can be extracted from the effective action, $W[r]$, of the theory on a $d$-sphere $S^d$ of radius $r$.
The effective action can be related to the vacuum expectation value of the trace of the stress tensor,
\begin{equation}\langle T_i^i\rangle = \frac{1}{2(4\pi)^{\frac{d}{2}} \left(\frac{d}{2}\right)!} a ~\mathrm{Euler}\, ,\end{equation}
where we have normalized $a$ in such a way that it is the coefficient of the $\ln r$ term of the effective action,
\begin{equation}\frac{dW}{d\ln r}=a\, .\end{equation}
The effective action can be written as
\begin{equation}W = \frac{1}{2} \ln \det \cD = \frac{1}{2} \tr \ln \cD  \, ,\end{equation}
where $\cD$ is the differential operator appearing in the quadratic Lagrangian. We may formally define the determinant in terms of a zeta function 
\begin{equation}\zeta_\cD(z) = \sum_n \frac{1}{\lambda_n^z} \, ,\end{equation}
where $\lambda_n$ are the (not necessarily distinct) eigenvalues of $\cD$. It is important that we arrange for the zeta function to be dimensionless by including an appropriate power of the cutoff. This zeta function converges at large $z$, and we may differentiate it then analytically continue it back to $z=0$ to define $W$,
\begin{equation}W = -\frac{1}{2}\zeta^\prime_\cD(0)\, .\end{equation}

In the case of a $\square^k$ theory, since we are on a sphere, we may define the operator $\cD$ via equation \ref{generaleinsts}.  We introduce a factor $(-1)^k$ to ensure that the corresponding Euclidean action is positive-definite. In doing so we may write
\begin{equation}\cD_{S^d} = (-1)^k \prod_{j=1}^k\left( \square -  \frac{1}{r^2}\left(\frac{d}{2}-j\right)\left(\frac{d}{2}+j-1\right)\right)\, .\end{equation}
The mass dimensions of $\cD$ and its eigenvalues $\lambda$ are $2k$. We therefore define the dimensionless zeta function as
\begin{equation}\zeta_{\cD_{S^d}}(z) = \Lambda^{2kz} \sum_{\ell=0}^\infty d_\ell \lambda_\ell^{-z}\end{equation}
where $\Lambda$ is a cutoff, $\lambda_\ell$ are the distinct eigenvalues of $\cD$ on $S^d$ and $d_\ell$ are their degeneracies (recall that the eigenvalues of $\square$ are $-\frac{1}{r^2}\ell(\ell+d-1)$),
\begin{align} \lambda_\ell &= r^{-2k}\prod_{j=1}^{2k} \left(\ell+\frac{d}{2}+j-k-1\right)\, , \\
d_\ell &= \left(\begin{matrix} d+\ell \\ d\end{matrix}\right) - \left(\begin{matrix}d+\ell-2 \\ d \end{matrix}\right) = \frac{(d+2\ell-1)\Gamma(d+\ell-1)}{\Gamma(d)\Gamma(1+\ell)}\, .\end{align}

The sum over $\ell$ converges at large $z$, and the log-divergent term in the effective action can be computed by analytic continuation to $z=0$,
\begin{equation}W = -k\ln(\Lambda r) \lim_{z\rightarrow 0} \zeta_{\tilde{\cD}_{S^d}}(z)\, ,\end{equation}
where the tilde on the $\tilde{\cD}$ indicates a dimensionless equivalent of the zeta function,
\begin{equation}\zeta_{\tilde{\cD}_{S^d}}(z) = \sum_{\ell=0}^\infty d_\ell \left(r^{2k}\lambda_\ell\right)^{-z}\, ,\end{equation}
\begin{equation}a_{CFT} = -k \lim_{z\rightarrow 0} \zeta_{\tilde{\cD}_{S^d}}(z)\, .\end{equation}

The way to evaluate this sum is to split up the sum into functions that wouldn't converge at $z=0$ (but do converge at large $z$) and a remainder that does converge at $z=0$ via extensive partial fractioning. We can then evaluate the sum assuming large $z$, then take the limit. In any even $d$ and for any $k$, we have observed that the remainder evaluates to $0$ at $z=0$; as the sum of the remainder converges, we may simply set $z=0$ at each term in the sum and see it vanish.

As an example, we demonstrate the computation for the standard $\square$ theory in 4d, and then cite answers for the $\square^k$ theory up to $k=8$ and $d=20$. This technique continues to work for the log theories, and so we state results for those theories as well. 

For $d=4$, $k=1$, we must evaluate
\begin{equation}\zeta_{\tilde{\mathcal{D}}_{S^4}}(z)=\sum_{\ell=0}^\infty \frac{1}{6} (\ell+1) (\ell+2) (2 \ell+3) (\ell (\ell+3)+2)^{-z} \, .\end{equation}
We first break off the $\ell=0$ term (which contributes 1). Then, we may partial-fraction to obtain those parts which do not converge at $z=0$ (dropping the remainder):
\begin{align}\sum_{\ell=1}^\infty &\frac{1}{6} (\ell+1) (\ell+2) (2 \ell+3) \ell^{-2 z}  \nonumber \\
&\times \left(\frac{27 z^4+90 z^3+97 z^2+34 z}{8 \ell^4}+\frac{-\frac{9 z^3}{2}-\frac{15 z^2}{2}-3 z}{\ell^3}+\frac{\frac{9 z^2}{2}+\frac{5 z}{2}}{\ell^2}-\frac{3 z}{\ell}+1\right)\, .\end{align}
This may be evaluated straightforwardly, e.g. in Mathematica:
\begin{align}f(z)=\frac{1}{48} \Big(&243 z^4 \zeta (2 (z+1))+162 z^4 \zeta (2 (z+2))+54 z^4 \zeta (2 z+1)+351 z^4 \zeta (2 z+3)-72 z^3 \zeta (2 z) \nonumber \\
&+342 z^3 \zeta (2 (z+1))+540 z^3 \zeta (2 (z+2))-144 z^3 \zeta (2 z+1)+954 z^3 \zeta (2 z+3)+204 z^2 \zeta (2 z)\nonumber \\
&+309 z^2 \zeta (2 (z+1)) +582 z^2 \zeta (2 (z+2))+72 z^2 \zeta (2 z-1)+122 z^2 \zeta (2 z+1)+901 z^2 \zeta (2 z+3) \nonumber \\
&-48 z \zeta (2 (z-1))-180 z \zeta (2 z)+114 z \zeta (2 (z+1)) +204 z \zeta (2 (z+2))-176 z \zeta (2 z-1) \nonumber \\
&-32 z \zeta (2 z+1)+298 z \zeta (2 z+3)+72 \zeta (2 (z-1))+48 \zeta (2 z)+16 \zeta (2 z-3)+104 \zeta (2 z-1)\Big)\, .\end{align}
The anomaly is therefore
\begin{equation}a_{CFT} = -\left(1 + \lim_{z\rightarrow 0}f(z)\right) = -\left(1 + \left(-\frac{91}{90}\right)\right) = \frac{1}{90}\, .\end{equation}
The results of this computation for the $\square^k$ theory in $d$ dimensions up to $k=8$ and $d=20$ (skipping cases with $d \leq 2k$) are shown in table $\ref{tab:anomalies}$. 

There is a caveat in the cases $d=2,4,\cdots,2k$.
As discussed in section \ref{couplingtocurvapp}, in the cases $d=2,4,\cdots,2k-2$, the theory cannot be coupled to an arbitrary background metric and has no true stress tensor, and so it is unclear what the underlying meaning of the $a$-anomaly we have computed actually is.  However, there is, curiously, no obstruction to computing the $a$-anomaly by the method here, because the coupling to the sphere happens to be well-defined (as discussed in section \ref{boxkmetricsubs}), and it is these values we have presented (colored in red) in  table \ref{tab:anomalies}.   Also, one should keep in mind that the choice we made for the propagator in section \ref{finitetheorysecks} in the cases $d=2,4,\cdots,2k$ gives the finite theories which do not properly correspond to the Lagrangian \eqref{CFTgenlag}.  Thus we expect the $a$-anomalies we have computed in these cases to correspond instead to the theory given by choosing logarithmic correlators for the basic fields.

{\renewcommand{\arraystretch}{1.5}
\begin{table}[h]
\centering
\begin{tabular}{|c|c|c|c|c|}\hline
$d$ & $\square$ & $\square^2$ & $\square^3$ & $\square^4$  \\ \hline
\rowcolor{Gray} 2 & \color{red} $-\frac{1}{3}$ & \color{red} $-\frac{8}{3}$ & \color{red} $-9$ & \color{red} $-\frac{64}{3}$ \\
4 &  $\frac{1}{90}$ & \color{red} $-\frac{14}{45}$ & \color{red} $-\frac{33}{10}$ & \color{red} $-\frac{688}{45}$ \\
\rowcolor{Gray} 6 & $-\frac{1}{756}$ & $\frac{8}{945}$ & \color{red} $-\frac{41}{140}$ & \color{red} $-\frac{736}{189}$ \\
8 & $\frac{23}{113400}$ & $-\frac{13}{14175}$ & $\frac{9}{1400}$ & \color{red} $-\frac{3956}{14175}$ \\
\rowcolor{Gray} 10 & $-\frac{263}{7484400}$ & $\frac{62}{467775}$ & $-\frac{19}{30800}$ & $\frac{2368}{467775}$ \\
12 & $\frac{133787}{20432412000}$ & $-\frac{28151}{1277025750}$ & $\frac{6887}{84084000}$ & $-\frac{275216}{638512875}$ \\
\rowcolor{Gray} 14 & $-\frac{157009}{122594472000}$ & $\frac{7636}{1915538625}$ & $-\frac{717}{56056000}$ & $\frac{7712}{147349125}$ \\
16 & $\frac{16215071}{62523180720000}$ & $-\frac{1488889}{1953849397500}$ & $\frac{2999}{1361360000}$ & $-\frac{531926}{69780335625}$ \\
\rowcolor{Gray} 18 & $-\frac{2689453969}{49893498214560000}$ & $\frac{694919}{4585799468250}$ & $-\frac{1848137}{4562734176000}$ & $\frac{241926224}{194896477400625}$ \\
20 & $\frac{26893118531}{2352122058686400000}$ & $-\frac{7984627349}{257263350168825000}$ & $\frac{65272217}{836501265600000}$ & $-\frac{7008930716}{32157918771103125}$ \\ \hline \multicolumn{5}{c}{}  \\ \hline
$d$ & $\square^5$ & $\square^6$ & $\square^7$ & $\square^8$\\ \hline 
\rowcolor{Gray} 2 & \color{red} $-\frac{125}{3}$ & \color{red} $-72$ & \color{red} $-\frac{343}{3}$ & \color{red} $-\frac{512}{3}$ \\
4 &  \color{red} $-\frac{875}{18}$ & \color{red} $-\frac{618}{5}$ & \color{red} $-\frac{24353}{90}$ & \color{red} $-\frac{23936}{45}$ \\
\rowcolor{Gray} 6 & \color{red} $-\frac{17225}{756}$ & \color{red} $-\frac{632}{7}$ & \color{red} $-\frac{151949}{540}$ & \color{red} $-\frac{702208}{945}$ \\
8 & \color{red} $-\frac{20225}{4536}$ & \color{red} $-\frac{5499}{175}$ & \color{red} $-\frac{2408497}{16200}$ & \color{red} $-\frac{7758112}{14175}$ \\
\rowcolor{Gray} 10 & \color{red} $-\frac{80335}{299376}$ & \color{red} $-\frac{9626}{1925}$ & \color{red} $-\frac{44003519}{1069200}$ & \color{red} $-\frac{105709312}{467775}$ \\
12 & $\frac{673175}{163459296}$ & \color{red} $-\frac{1364651}{5255250}$ & \color{red} $-\frac{2303435659}{416988000}$ & \color{red} $-\frac{33166818112}{638512875}$ \\
\rowcolor{Gray} 14 & $-\frac{307525}{980755776}$ & $\frac{3012}{875875}$ & \color{red} $-\frac{631693279}{2501928000}$ & \color{red} $-\frac{11555275136}{1915538625}$ \\
16 & $\frac{3490255}{100037089152}$ & $-\frac{70327}{297797500}$ & $\frac{3740727473}{1275983280000}$ & \color{red} $-\frac{120348894184}{488462349375}$ \\
\rowcolor{Gray} 18 & $-\frac{378009185}{79829597143296}$ & $\frac{4309271}{178231803750}$ & $-\frac{5325836117}{29092418784000}$ & $\frac{99059365376}{38979295480125}$ \\
20 & $\frac{3832329377}{5268753411457536}$ & $-\frac{5718287}{1867190325000}$ & $\frac{48921638269}{2823675940800000}$ & $-\frac{667748058496}{4593988395871875}$ \\ \hline
\end{tabular}
\caption{Anomalies for the first few $\square^k$ theories in the first few even dimensions. Red indicates that the theory is one of the log-theories.}
\label{tab:anomalies}
\end{table}
}

\subsection{Odd Dimensions: Free Energies}

In odd dimensions, the unambiguous computable quantity is instead the finite part of the effective action. We may define the exact same zeta function for the same operator $\tilde{\cD}$, borrowing all the expressions in the previous subsection, but now we are instead interested in computing the derivative of the zeta function at 0,
\begin{equation}F = -\frac{1}{2}\zeta_{\tilde{\mathcal{D}}_{S^d}}^\prime(0)\, .\end{equation}
Although the remainder evaluates to zero, its derivative does not, and must be computed more carefully. We are able to treat the remainder properly only in small-$d$, small-$k$ examples. For other cases, we must resort to splitting the sum in a way which is in general sensitive to multiplicative anomalies\footnote{A multiplicative anomaly in this context means having $\det(AB) \neq \det(A) \det(B)$ \cite{Chodos:1985hk,Wodzicki1987,GUILLEMIN1985131,zbMATH04149222,Kontsevich:1994nc,Kontsevich:1994xe}. There is of course no such anomaly for operators which act on finite-dimensional spaces, but such an anomaly can occur for infinite-dimensional spaces such as the spaces on which the Laplacians here are acting.  Since we are considering logs of determinants, the multiplicative anomaly shows up here as an additive anomaly.}. However, there is a particular choice of how precisely we split up the sum which agrees with the more careful computation in all cases we've been able to explicitly check. Therefore, we proceed under the assumption that this is a correct handling of the multiplicative anomaly.

Using the identity
\begin{equation}\lim_{z\rightarrow 0} \frac{d}{dz} x^{-z} = -\ln(x)\, , \end{equation}
we may convert the computation of the free energy into 
\begin{equation}F = \frac{1}{2}\left(\sum_\ell d_\ell \ln\lambda_\ell\right)\, .\end{equation}
 But because $\lambda_\ell$ is itself a product, we may exploit that the log of a product is the sum of the logs, and then pass back to a regularized function of $z$,
\begin{align}F &= \frac{1}{2}\left(\sum_\ell d_\ell\left(\sum_{j=1}^{2k} \ln\left(\ell+\frac{d}{2}+j-k-1\right)\right)\right) \\
&= -\frac{1}{2} \frac{d}{dz}\left( \sum_{j=1}^{2k}\sum_{\ell=0}^\infty d_\ell \left(\ell+\frac{d}{2}+j-k-1\right)^{-z}\right)\Bigg|_{z=0}\ \ \ \label{eqn:correctF}\, .\end{align}
This particular choice of passing through logs seems to lead to the correct free energy, with no multiplicative anomaly\footnote{Let us demonstrate also an erroneous computation that follows from the same logic, arising due to the multiplicative anomaly. We choose the case of $d=3$, $k=1$. In this case our prescription dictates that we should compute
\begin{align}F &= \frac{1}{2} \sum_{\ell=0}^\infty (\ell+1)^2 \ln\left(\left(\ell+\frac{1}{2}\right)\left(\ell+\frac{3}{2}\right)\right) \\
&\rightarrow \mathrm{Reg}\left(\frac{1}{2} \sum_{\ell=0}^\infty (\ell+1)^2 \left(\ln\left(\ell+\frac{1}{2}\right)+\ln\left(\ell+\frac{3}{2}\right)\right) \right)\\
&= \frac{\ln(2)}{8} - \frac{3\zeta(3)}{16\pi^2}\, .\end{align}
which is the correct free energy on $S^3$. If, however, we had split the logarithm by multiplying the first term by $2$ and the second term by $\frac{1}{2}$, we would have obtained different answers:
\begin{align}\mathrm{Wrong~}F &= \frac{1}{2} \sum_{\ell=0}^\infty (\ell+1)^2 \ln\left(\left(2\ell+1\right)\left(\frac{\ell}{2}+\frac{3}{4}\right)\right) \\
&\rightarrow \mathrm{Reg}\left(\frac{1}{2} \sum_{\ell=0}^\infty (\ell+1)^2 \left(\ln\left(2\ell+1\right)+\ln\left(\frac{\ell}{2}+\frac{3}{4}\right)\right)\right) \\
&= \frac{\ln(2)}{6} - \frac{3\zeta(3)}{16\pi^2}\, .\end{align}
}. 
The results of computing this with the scheme \eqref{eqn:correctF} are shown for $k=1,2,3$ for odd $d$ from $3$ to $13$ in table \ref{tab:freeenergies}. In this table, $A$ is Glaisher's constant $1.28243\ldots$. Zeta functions of two arguments are the Hurwitz zeta function, defined as the analytic continuation of
\begin{equation}\zeta(z,a) = \sum_{n=0}^\infty \frac{1}{(n+a)^z}\, .\end{equation}
Note that when $d<2k$, there are low-$\ell$ negative eigenvalues. Upon taking logarithms of such negative eigenvalues, giving these answers imaginary parts which depend on the choice of a branch for the logarithm.  For example, in the case $d=3$, $k=2$, we are interested in evaluating the following sums:
\begin{equation}F=-\frac{1}{2}\frac{d}{dz}\sum_{\ell=0}^\infty d_{\ell} \left(\left(\ell-\frac{1}{2}\right)^{-z}+\ldots\right)\, .\end{equation}
This sum is convergent and so we may freely look at individual terms inside the sum. In this case, focusing on just the $\ell=0$ part of the sum and noting that $d_0=1$, we have a contribution to $F$ which looks like
\begin{equation}F_{\ell=0} = -\frac{1}{2}\frac{d}{dz} \left(-\frac{1}{2}\right)^{-z} = -\frac{1}{2} \frac{d}{dz}e^{-z\ln(-\frac{1}{2})} = \frac{1}{2}\left(i\pi - \ln{2}\right)\, .\end{equation}
We may transition from $i\pi$ inside the parentheses to $ni\pi$, where $n$ is an odd integer, by changing which branch we evaluate the log on.

{\renewcommand{\arraystretch}{1.5}
\begin{table}[!h]
\centering
\begin{tabular}{|c|c|}\hline
$d$ & $\square$ \\ \hline
\rowcolor{Gray}$3$ & $\frac{\log (64)}{48}-\frac{3 \zeta (3)}{16 \pi ^2}$ \\
$5$ & $ -\frac{\log (4)}{256}+\frac{15 \zeta (5)}{256 \pi ^4}-\frac{\zeta (3)}{128 \pi ^2}$ \\
\rowcolor{Gray}$7$ & $ \frac{\log (2)}{1024}+\frac{41 \zeta (3)}{30720 \pi ^2}-\frac{5 \zeta (5)}{2048 \pi ^4}-\frac{63 \zeta (7)}{4096 \pi ^6}$ \\
$9$ & $ -\frac{5 \log (2)}{32768}+\frac{\zeta (5)}{32768 \pi ^4}+\frac{63 \zeta (7)}{32768 \pi ^6}+\frac{255 \zeta (9)}{65536 \pi ^8}-\frac{397 \zeta (3)}{1720320 \pi ^2}$\\
\rowcolor{Gray}$11$ & $ \frac{7 \log (2)}{262144}+\frac{3897 \zeta (3)}{91750400 \pi ^2}+\frac{485 \zeta (5)}{16515072 \pi ^4}-\frac{609 \zeta (7)}{2621440 \pi ^6}-\frac{425 \zeta (9)}{524288 \pi ^8}-\frac{1023 \zeta (11)}{1048576 \pi ^{10}}$ \\
$13$ & $ -\frac{21 \log (2)}{4194304}+\frac{1733 \zeta (7)}{62914560 \pi ^6}+\frac{3587 \zeta (9)}{25165824 \pi ^8}+\frac{2387 \zeta (11)}{8388608 \pi ^{10}}+\frac{4095 \zeta (13)}{16777216 \pi ^{12}}-\frac{596467 \zeta (3)}{72666316800 \pi ^2}-\frac{10957 \zeta (5)}{1132462080 \pi ^4}$ \\ \hline \multicolumn{2}{c}{} \\ \hline
$d$ & $\square^2$ \\ \hline
\rowcolor{Gray}$3$ & $ -\frac{1}{16}+\frac{9 i \pi }{8}+\frac{3 \log (8)}{16}+\frac{3 \log (A)}{4}-\frac{9 \zeta (3)}{32 \pi ^2}-\frac{1}{2} \zeta^{(1,0)}\left(-2,-\frac{1}{2}\right)-\frac{3}{2} \zeta^{(1,0)}\left(-1,-\frac{1}{2}\right)$ \\
$5$ & $ \frac{7 \log (2)}{64}+\frac{15 \zeta (5)}{128 \pi ^4}-\frac{13 \zeta (3)}{64 \pi ^2}$ \\
\rowcolor{Gray}$7$ & $ -\frac{1}{512} 3 \log (2)+\frac{55 \zeta (5)}{1024 \pi ^4}-\frac{79 \zeta (3)}{15360 \pi ^2}-\frac{63 \zeta (7)}{2048 \pi ^6}$ \\
$9$ & $ \frac{11 \log (2)}{16384}+\frac{751 \zeta (3)}{860160 \pi ^2}+\frac{255 \zeta (9)}{32768 \pi ^8}-\frac{39 \zeta (5)}{16384 \pi ^4}-\frac{189 \zeta (7)}{16384 \pi ^6} $\\
\rowcolor{Gray}$11$ & $ -\frac{13 \log (2)}{131072}+\frac{737 \zeta (5)}{8257536 \pi ^4}+\frac{1911 \zeta (7)}{1310720 \pi ^6}+\frac{595 \zeta (9)}{262144 \pi ^8}-\frac{2867 \zeta (3)}{19660800 \pi ^2}-\frac{1023 \zeta (11)}{524288 \pi ^{10}}$ \\
$13$ & $ \frac{35 \log (2)}{2097152}+\frac{189349 \zeta (3)}{7266631680 \pi ^2}+\frac{39701 \zeta (5)}{3963617280 \pi ^4}+\frac{4095 \zeta (13)}{8388608 \pi ^{12}}-\frac{1115 \zeta (7)}{6291456 \pi ^6}-\frac{6613 \zeta (9)}{12582912 \pi ^8}-\frac{1705 \zeta (11)}{4194304 \pi ^{10}}$ \\ \hline \multicolumn{2}{c}{} \\ \hline
$d$ & $\square^3$ \\ \hline
\rowcolor{Gray}$3$ & $ -\frac{1}{6}+\frac{59 i \pi }{8}-\frac{\log (2)}{6}+\frac{21 \log (3)}{8}+2 \log (A)-\frac{3 \zeta (3)}{8 \pi ^2}-\frac{1}{2} \zeta^{(1,0)}\left(-2,-\frac{3}{2}\right)$ \\
\rowcolor{Gray}&$-\frac{1}{2} \zeta^{(1,0)}\left(-2,-\frac{1}{2}\right)-\frac{5}{2} \zeta^{(1,0)}\left(-1,-\frac{3}{2}\right)-\frac{3}{2} \zeta^{(1,0)}\left(-1,-\frac{1}{2}\right) $\\
$5$ & $ -\frac{115}{1152}+\frac{175 i \pi }{128}+\frac{1175 \log (2)}{2304}+\frac{115 \log (A)}{96}-\frac{125 \zeta (3)}{256 \pi ^2}+\frac{75 \zeta (5)}{512 \pi ^4}-\frac{35}{96} \zeta '(-3)$ \\
&$-\frac{1}{24} \zeta^{(1,0)}\left(-4,-\frac{1}{2}\right)-\frac{5}{12} \zeta^{(1,0)}\left(-3,-\frac{1}{2}\right)-\frac{73}{48} \zeta^{(1,0)}\left(-2,-\frac{1}{2}\right)-\frac{115}{48} \zeta^{(1,0)}\left(-1,-\frac{1}{2}\right) $\\
\rowcolor{Gray}$7$ & $ \frac{99 \log (2)}{1024}+\frac{465 \zeta (5)}{2048 \pi ^4}-\frac{2199 \zeta (3)}{10240 \pi ^2}-\frac{189 \zeta (7)}{4096 \pi ^6}$ \\
$9$ & $ -\frac{143 \log (2)}{32768}+\frac{1603 \zeta (5)}{32768 \pi ^4}+\frac{765 \zeta (9)}{65536 \pi ^8}-\frac{5447 \zeta (3)}{1720320 \pi ^2}-\frac{1827 \zeta (7)}{32768 \pi ^6}$ \\
\rowcolor{Gray}$11$ & $ \frac{117 \log (2)}{262144}+\frac{49451 \zeta (3)}{91750400 \pi ^2}+\frac{6885 \zeta (9)}{524288 \pi ^8}-\frac{12283 \zeta (5)}{5505024 \pi ^4}-\frac{21987 \zeta (7)}{2621440 \pi ^6}-\frac{3069 \zeta (11)}{1048576 \pi ^{10}} $\\
$13$ & $ -\frac{255 \log (2)}{4194304}+\frac{314341 \zeta (5)}{2642411520 \pi ^4}+\frac{4513 \zeta (7)}{4194304 \pi ^6}+\frac{9027 \zeta (9)}{8388608 \pi ^8}+\frac{12285 \zeta (13)}{16777216 \pi ^{12}}-\frac{414199 \zeta (3)}{4844421120 \pi ^2}-\frac{25575 \zeta (11)}{8388608 \pi ^{10}}$ \\ \hline
\end{tabular}
\caption{Free energies for the first few $\square^k$ theories in the first few odd dimensions}
\label{tab:freeenergies}
\end{table}
}

\section{$\square^2$ Character Decomposition in $d=3$}
\label{sec:flatofronsdal}

In this section, we prove that the two towers of operators discussed in subsection \ref{sec:primaries} are indeed all of the single-trace primary operators of the $\square^2$ theory in $d=3$.  This has been proved before in \cite{Basile:2014wua}; we review the proof here for completeness' sake.  This proof generalizes straightforwardly to higher $d$ and $k$. The case $k=1$ in any $d$ was considered previously; see e.g. \cite{Dolan:2005wy}.

The idea is to compute the ``single-trace partition function'' $Z_1$, and decompose it into characters $\chi(\Delta,s)$ or $\chi(\Delta,s,c)$ labelling irreducible representations of $so(4,1)$, which are non-conserved or $c$-conserved, respectively. Strictly speaking, $Z_1$ is {\it not} the single-trace partition function, and its plethystic exponential does not give the true singlet partition function. The reason why is because the trace involved in computing the true partition function is not positive-definite, but rather weighted by the norms of states:
\begin{equation}Z_{true} = \tr(q^D y^{J_3}) = \sum_i \langle \cO_i|q^D y^{J_3}|\cO_i \rangle = \sum_i \langle \cO_i | \cO_i \rangle q^\Delta y^{j_3}\, .\end{equation}

In unitary CFTs, all states can be made to have norm $1$, and we may simply compute instead $\sum_i q^\Delta y^{j_3}$. However, the computation of the true partition function (as well as the associated characters) in a non-unitary CFT is complicated by the fact that there are zero- and negative-norm states.
Nevertheless, a n\"aive computation of the single-trace partition function where we simply ignore this issue and treat all states as if they had norm $1$ is still enough information to learn about what states are in the spectrum, and by using similarly ``blinded'' characters, we may still learn about the collection of single-trace primaries in the theory.
We therefore proceed with this ``blinded'' computation. We would like to compute the single-trace partition function
\begin{equation}Z_1(q,y) = \sum_{\mathrm{ST~}\cO} q^\Delta y^{j_3}\, ,\end{equation}
where $q = e^{-\beta}$ is the activity associated with the scaling dimension of the operator on $\R^\dcft$ (equivalently, the energy of the state on the cylinder) and $y$ is the activity associated with the $J_3$ eigenvalue of the rotations of the $S^{2}$. In this CFT, all single-trace operators can be written in the form
\begin{equation}\partial_0^{\bar{n}_0} \partial_+^{\bar{n}_+} \partial_-^{\bar{n}_-} \phi^\dagger_a \partial_0^{n_0} \partial_+^{n_+} \partial_-^{n_-} \phi^a\, ,\end{equation}
where as usual $\partial_\pm = \partial_1 \pm i \partial_2$ are eigenoperators of $J_3$ with eigenvalue $\pm 1$.

By employing the equations of motion, we may remove any operators with $n_0,\bar{n}_0> 3$, because we can replace $\partial_0^4\phi$ with other operators already counted in our sum. The single-trace ``partition function'' is therefore given by
\begin{align}Z_1(q,y) &= \sum_{n_+,n_-,\bar{n}_+,\bar{n}_-=0}^\infty \sum_{n_0,\bar{n}_0=0}^3 q^{-1+n_0+n_+ +n_-+\bar{n}_0+\bar{n}_++\bar{n}_-} y^{n_+-n_-+\bar{n}_+-\bar{n}_-} \nonumber \\
&= \frac{y^2 (1+q+q^2+q^3)^2}{q(q-y)^2(qy-1)^2}\, .
\end{align}

This may be decomposed into ``characters'' of $so(3,2)$ telling us which representations are present in $Z_1$. We first write down the ``character'' of a generic or long module with quantum numbers $\Delta$ and $s$, denoting this representation simply by $(\Delta,s)$. The generic state in the module can be written
\begin{equation}\partial_0^{n_0} \partial_+^{n_+} \partial_-^{n_-} |\Delta, s\rangle \, .\end{equation}
Therefore the associated ``character'' may be written as
\begin{align}\chi(\Delta, s)&=\sum_{n_0,n_+,n_-=0}^\infty \sum_{m=-s}^s q^{\Delta +n_0 + n_+ + n_-} y^{m+n_+-n_-} \nonumber \\
&= \frac{q^\Delta y^{1-s}(1-y^{1+2s})}{(q-1)(q-y)(y-1)(qy-1)}\, .
\end{align}

Suppose that our spin-$s$ operator vanishes when contracted with $c$ derivatives:
\begin{equation}\partial^{i_1} \ldots \partial^{i_c} \cO_{i_1 \ldots i_c \ldots  i_{s}} = 0\, .\end{equation}
As before, we refer to operators satisfying higher-$c$ conservation conditions as multiply conserved currents. Were this the case, the associated character would have fewer states than expected due to this null condition beginning at $\Delta_{\mathrm{null}} = \Delta_\cO + c$ and with $s_{\mathrm{null}} = s_{\cO} - c$. We denote this representation by $(\Delta,s,c)$. Therefore the associated character is
\begin{align}\chi(\Delta, s, c) &= \chi(\Delta, s) - \chi(\Delta+c, s-c) \nonumber \\
&=\frac{q^\Delta y^{1-c-s}(y^c-q^c y^{2c} + q^c y^{1+2s} - y^{1+c+2s})}{(q-1)(q-y)(y-1)(qy-1)} \, .
\end{align}

With these tools in hand we may readily verify that the character decomposition of the single-trace ``partition function'' written down above is
\begin{align}Z_1(q,y) &= \chi(-1,0) + \chi(0,1) + \chi(1,2) + \sum_{s=3}^\infty \chi(s-1,s,3) \nonumber \\
&\qquad + \chi(1,0) + \sum_{s=1}^\infty \chi(s+1,s,1)\, ,\end{align}
consistent with the results presented in subsection \ref{sec:primaries}.

 \bibliographystyle{utphys}
\addcontentsline{toc}{section}{References}
\bibliography{cft}

\providecommand{\href}[2]{#2}\begingroup\raggedright\begin{thebibliography}{10}

\bibitem{Maldacena:1997re}
J.~M. Maldacena, ``{The Large N limit of superconformal field theories and
  supergravity},'' \href{http://dx.doi.org/10.1023/A:1026654312961}{{\em Int.
  J. Theor. Phys.} {\bf 38} (1999)  1113--1133},
  \href{http://arxiv.org/abs/hep-th/9711200}{{\tt arXiv:hep-th/9711200
  [hep-th]}}.
[Adv. Theor. Math. Phys.2,231(1998)].
%%CITATION = HEP-TH/9711200;%%.

\bibitem{Strominger:2001pn}
A.~Strominger, ``{The dS / CFT correspondence},''
  \href{http://dx.doi.org/10.1088/1126-6708/2001/10/034}{{\em JHEP} {\bf 10}
  (2001)  034},
\href{http://arxiv.org/abs/hep-th/0106113}{{\tt arXiv:hep-th/0106113
  [hep-th]}}.
%%CITATION = HEP-TH/0106113;%%.

\bibitem{Anninos:2011ui}
D.~Anninos, T.~Hartman, and A.~Strominger, ``{Higher Spin Realization of the
  dS/CFT Correspondence},''
\href{http://arxiv.org/abs/1108.5735}{{\tt arXiv:1108.5735 [hep-th]}}.
%%CITATION = ARXIV:1108.5735;%%.

\bibitem{Vasiliev:1990en}
M.~A. Vasiliev, ``{Consistent equation for interacting gauge fields of all
  spins in (3+1)-dimensions},''
\href{http://dx.doi.org/10.1016/0370-2693(90)91400-6}{{\em Phys. Lett.} {\bf
  B243} (1990)  378--382}.
%%CITATION = PHLTA,B243,378;%%.

\bibitem{Vasiliev:1992av}
M.~A. Vasiliev, ``{More on equations of motion for interacting massless fields
  of all spins in (3+1)-dimensions},''
\href{http://dx.doi.org/10.1016/0370-2693(92)91457-K}{{\em Phys. Lett.} {\bf
  B285} (1992)  225--234}.
%%CITATION = PHLTA,B285,225;%%.

\bibitem{Vasiliev:1999ba}
M.~A. Vasiliev, ``{Higher spin gauge theories: Star product and AdS space},''
\href{http://arxiv.org/abs/hep-th/9910096}{{\tt arXiv:hep-th/9910096
  [hep-th]}}.
%%CITATION = HEP-TH/9910096;%%.

\bibitem{Vasiliev:2003ev}
M.~A. Vasiliev, ``{Nonlinear equations for symmetric massless higher spin
  fields in (A)dS(d)},''
  \href{http://dx.doi.org/10.1016/S0370-2693(03)00872-4}{{\em Phys. Lett.} {\bf
  B567} (2003)  139--151},
\href{http://arxiv.org/abs/hep-th/0304049}{{\tt arXiv:hep-th/0304049
  [hep-th]}}.
%%CITATION = HEP-TH/0304049;%%.

\bibitem{Anninos:2012ft}
D.~Anninos, F.~Denef, and D.~Harlow, ``{Wave function of Vasiliev�s universe:
  A few slices thereof},''
  \href{http://dx.doi.org/10.1103/PhysRevD.88.084049}{{\em Phys. Rev.} {\bf
  D88} (2013) no.~8, 084049},
\href{http://arxiv.org/abs/1207.5517}{{\tt arXiv:1207.5517 [hep-th]}}.
%%CITATION = ARXIV:1207.5517;%%.

\bibitem{Fei:2015kta}
L.~Fei, S.~Giombi, I.~R. Klebanov, and G.~Tarnopolsky, ``{Critical Sp(N) models
  in 6 dimensions and higher spin dS/CFT},''
  \href{http://dx.doi.org/10.1007/JHEP09(2015)076}{{\em JHEP} {\bf 09} (2015)
  076},
\href{http://arxiv.org/abs/1502.07271}{{\tt arXiv:1502.07271 [hep-th]}}.
%%CITATION = ARXIV:1502.07271;%%.

\bibitem{Vafa:2014iua}
C.~Vafa, ``{Non-Unitary Holography},''
\href{http://arxiv.org/abs/1409.1603}{{\tt arXiv:1409.1603 [hep-th]}}.
%%CITATION = ARXIV:1409.1603;%%.

\bibitem{Butera:2012tq}
P.~Butera and M.~Pernici, ``{Yang-Lee edge singularities from extended activity
  expansions of the dimer density for bipartite lattices of dimensionality two
  through seven},'' \href{http://dx.doi.org/10.1103/PhysRevE.86.011104}{{\em
  Phys. Rev.} {\bf E86} (2012)  011104},
\href{http://arxiv.org/abs/1206.0872}{{\tt arXiv:1206.0872
  [cond-mat.stat-mech]}}.
%%CITATION = ARXIV:1206.0872;%%.

\bibitem{Stergiou:2015roa}
A.~Stergiou, ``{Symplectic critical models in 6+epsilon dimensions},''
  \href{http://dx.doi.org/10.1016/j.physletb.2015.10.044}{{\em Phys. Lett.}
  {\bf B751} (2015)  184--187},
\href{http://arxiv.org/abs/1508.03639}{{\tt arXiv:1508.03639 [hep-th]}}.
%%CITATION = ARXIV:1508.03639;%%.

\bibitem{Gracey:2015xmw}
J.~A. Gracey, ``{Six dimensional QCD at two loops},''
  \href{http://dx.doi.org/10.1103/PhysRevD.93.025025}{{\em Phys. Rev.} {\bf
  D93} (2016) no.~2, 025025},
\href{http://arxiv.org/abs/1512.04443}{{\tt arXiv:1512.04443 [hep-th]}}.
%%CITATION = ARXIV:1512.04443;%%.

\bibitem{Osborn:2016bev}
H.~Osborn and A.~Stergiou, ``{C$_{T}$ for non-unitary CFTs in higher
  dimensions},'' \href{http://dx.doi.org/10.1007/JHEP06(2016)079}{{\em JHEP}
  {\bf 06} (2016)  079},
\href{http://arxiv.org/abs/1603.07307}{{\tt arXiv:1603.07307 [hep-th]}}.
%%CITATION = ARXIV:1603.07307;%%.

\bibitem{Guerrieri:2016whh}
A.~Guerrieri, A.~C. Petkou, and C.~Wen, ``{The free $\sigma$CFTs},''
\href{http://arxiv.org/abs/1604.07310}{{\tt arXiv:1604.07310 [hep-th]}}.
%%CITATION = ARXIV:1604.07310;%%.

\bibitem{Nakayama:2016dby}
Y.~Nakayama, ``{Hidden global conformal symmetry without Virasoro extension in
  theory of elasticity},''
\href{http://arxiv.org/abs/1604.00810}{{\tt arXiv:1604.00810 [hep-th]}}.
%%CITATION = ARXIV:1604.00810;%%.

\bibitem{Peli:2016gio}
Z.~P�li, S.~Nagy, and K.~Sailer, ``{Phase structure of the $O(2)$ ghost model
  with higher-order gradient term},''
\href{http://arxiv.org/abs/1605.07836}{{\tt arXiv:1605.07836 [hep-th]}}.
%%CITATION = ARXIV:1605.07836;%%.

\bibitem{Gwak:2016sma}
S.~Gwak, J.~Kim, and S.-J. Rey, ``{Massless and Massive Higher Spins from
  Anti-de Sitter Space Waveguide},''
\href{http://arxiv.org/abs/1605.06526}{{\tt arXiv:1605.06526 [hep-th]}}.
%%CITATION = ARXIV:1605.06526;%%.

\bibitem{Fujimori:2016udq}
T.~Fujimori, M.~Nitta, and Y.~Yamada, ``{Ghostbusters in higher derivative
  supersymmetric theories: who is afraid of propagating auxiliary fields?},''
\href{http://arxiv.org/abs/1608.01843}{{\tt arXiv:1608.01843 [hep-th]}}.
%%CITATION = ARXIV:1608.01843;%%.

\bibitem{Gliozzi:2016ysv}
F.~Gliozzi, A.~Guerrieri, A.~C. Petkou, and C.~Wen, ``{Generalized
  Wilson-Fisher critical points from the conformal OPE},''
\href{http://arxiv.org/abs/1611.10344}{{\tt arXiv:1611.10344 [hep-th]}}.
%%CITATION = ARXIV:1611.10344;%%.

\bibitem{Gliozzi:2017hni}
F.~Gliozzi, A.~L. Guerrieri, A.~C. Petkou, and C.~Wen, ``{The analytic
  structure of conformal blocks and the generalized Wilson-Fisher fixed
  points},''
\href{http://arxiv.org/abs/1702.03938}{{\tt arXiv:1702.03938 [hep-th]}}.
%%CITATION = ARXIV:1702.03938;%%.

\bibitem{Hogervorst:2016itc}
M.~Hogervorst, M.~Paulos, and A.~Vichi, ``{The ABC (in any D) of Logarithmic
  CFT},''
\href{http://arxiv.org/abs/1605.03959}{{\tt arXiv:1605.03959 [hep-th]}}.
%%CITATION = ARXIV:1605.03959;%%.

\bibitem{Polyakov:1992yw}
A.~M. Polyakov, ``{Conformal turbulence},''
\href{http://arxiv.org/abs/hep-th/9209046}{{\tt arXiv:hep-th/9209046
  [hep-th]}}.
%%CITATION = HEP-TH/9209046;%%.

\bibitem{Flohr:1996ik}
M.~A.~I. Flohr, ``{Two-dimensional turbulence: Yet another conformal field
  theory solution},''
  \href{http://dx.doi.org/10.1016/S0550-3213(96)00563-9}{{\em Nucl. Phys.} {\bf
  B482} (1996)  567--578},
\href{http://arxiv.org/abs/hep-th/9606130}{{\tt arXiv:hep-th/9606130
  [hep-th]}}.
%%CITATION = HEP-TH/9606130;%%.

\bibitem{Creutzig:2013hma}
T.~Creutzig and D.~Ridout, ``{Logarithmic Conformal Field Theory: Beyond an
  Introduction},'' \href{http://dx.doi.org/10.1088/1751-8113/46/49/494006}{{\em
  J. Phys.} {\bf A46} (2013)  4006},
\href{http://arxiv.org/abs/1303.0847}{{\tt arXiv:1303.0847 [hep-th]}}.
%%CITATION = ARXIV:1303.0847;%%.

\bibitem{PhysRevLett.40.1610}
M.~E. Fisher, \href{http://dx.doi.org/10.1103/PhysRevLett.40.1610}{``Yang-lee
  edge singularity and ${\ensuremath{\phi}}^{3}$ field theory,''{\em Phys. Rev.
  Lett.} {\bf 40} (Jun, 1978)  1610--1613}.
  \url{http://link.aps.org/doi/10.1103/PhysRevLett.40.1610}.

\bibitem{Gliozzi:2013ysa}
F.~Gliozzi, ``{More constraining conformal bootstrap},''
  \href{http://dx.doi.org/10.1103/PhysRevLett.111.161602}{{\em Phys. Rev.
  Lett.} {\bf 111} (2013)  161602},
\href{http://arxiv.org/abs/1307.3111}{{\tt arXiv:1307.3111}}.
%%CITATION = ARXIV:1307.3111;%%.

\bibitem{Gliozzi:2014jsa}
F.~Gliozzi and A.~Rago, ``{Critical exponents of the 3d Ising and related
  models from Conformal Bootstrap},''
  \href{http://dx.doi.org/10.1007/JHEP10(2014)042}{{\em JHEP} {\bf 10} (2014)
  042},
\href{http://arxiv.org/abs/1403.6003}{{\tt arXiv:1403.6003 [hep-th]}}.
%%CITATION = ARXIV:1403.6003;%%.

\bibitem{Esterlis:2016psv}
I.~Esterlis, A.~L. Fitzpatrick, and D.~Ramirez, ``{Closure of the Operator
  Product Expansion in the Non-Unitary Bootstrap},''
\href{http://arxiv.org/abs/1606.07458}{{\tt arXiv:1606.07458 [hep-th]}}.
%%CITATION = ARXIV:1606.07458;%%.

\bibitem{Kos:2013tga}
F.~Kos, D.~Poland, and D.~Simmons-Duffin, ``{Bootstrapping the $O(N)$ vector
  models},'' \href{http://dx.doi.org/10.1007/JHEP06(2014)091}{{\em JHEP} {\bf
  06} (2014)  091},
\href{http://arxiv.org/abs/1307.6856}{{\tt arXiv:1307.6856 [hep-th]}}.
%%CITATION = ARXIV:1307.6856;%%.

\bibitem{Penedones:2015aga}
J.~Penedones, E.~Trevisani, and M.~Yamazaki, ``{Recursion Relations for
  Conformal Blocks},''
\href{http://arxiv.org/abs/1509.00428}{{\tt arXiv:1509.00428 [hep-th]}}.
%%CITATION = ARXIV:1509.00428;%%.

\bibitem{Bekaert:2013zya}
X.~Bekaert and M.~Grigoriev, ``{Higher order singletons, partially massless
  fields and their boundary values in the ambient approach},''
  \href{http://dx.doi.org/10.1016/j.nuclphysb.2013.08.015}{{\em Nucl. Phys.}
  {\bf B876} (2013)  667--714}.
arXiv:1305.0162 [hep-th].
%%CITATION = ARXIV:1305.0162;%%.

\bibitem{Basile:2014wua}
T.~Basile, X.~Bekaert, and N.~Boulanger, ``{Flato-Fronsdal theorem for
  higher-order singletons},''
  \href{http://dx.doi.org/10.1007/JHEP11(2014)131}{{\em JHEP} {\bf 11} (2014)
  131}.
arXiv:1410.7668 [hep-th].
%%CITATION = ARXIV:1410.7668;%%.

\bibitem{Grigoriev:2014kpa}
X.~Bekaert and M.~Grigoriev, ``{Higher-Order Singletons and Partially Massless
  Fields},'' {\em Bulg. J. Phys.} {\bf 41} (2014)  172--179.
{\tt hal-01077511}.
%%CITATION = BJPHD,41,172;%%.

\bibitem{Alkalaev:2014nsa}
K.~B. Alkalaev, M.~Grigoriev, and E.~D. Skvortsov, ``{Uniformizing higher-spin
  equations},'' \href{http://dx.doi.org/10.1088/1751-8113/48/1/015401}{{\em J.
  Phys.} {\bf A48} (2015) no.~1, 015401}.
arXiv:1409.6507 [hep-th].
%%CITATION = ARXIV:1409.6507;%%.

\bibitem{Joung:2015jza}
E.~Joung and K.~Mkrtchyan, ``{Partially-massless higher-spin algebras and their
  finite-dimensional truncations},''.
arXiv:1508.07332 [hep-th].
%%CITATION = ARXIV:1508.07332;%%.

\bibitem{Brust:2016zns}
C.~Brust and K.~Hinterbichler, ``{Partially Massless Higher-Spin Theory},''
\href{http://arxiv.org/abs/1610.08510}{{\tt arXiv:1610.08510 [hep-th]}}.
%%CITATION = ARXIV:1610.08510;%%.

\bibitem{Maldacena:2011mk}
J.~Maldacena, ``{Einstein Gravity from Conformal Gravity},''
\href{http://arxiv.org/abs/1105.5632}{{\tt arXiv:1105.5632 [hep-th]}}.
%%CITATION = ARXIV:1105.5632;%%.

\bibitem{vanTonder:2006ye}
A.~van Tonder, ``{Non-perturbative quantization of phantom and ghost theories:
  Relating definite and indefinite representations},''
  \href{http://dx.doi.org/10.1142/S0217751X07036580}{{\em Int. J. Mod. Phys.}
  {\bf A22} (2007)  2563--2608},
\href{http://arxiv.org/abs/hep-th/0610185}{{\tt arXiv:hep-th/0610185
  [hep-th]}}.
%%CITATION = HEP-TH/0610185;%%.

\bibitem{Woodard:2006nt}
R.~P. Woodard, ``{Avoiding dark energy with 1/r modifications of gravity},''
  \href{http://dx.doi.org/10.1007/978-3-540-71013-4_14}{{\em Lect. Notes Phys.}
  {\bf 720} (2007)  403--433},
\href{http://arxiv.org/abs/astro-ph/0601672}{{\tt arXiv:astro-ph/0601672
  [astro-ph]}}.
%%CITATION = ASTRO-PH/0601672;%%.

\bibitem{Sbisa:2014pzo}
F.~Sbis�, ``{Classical and quantum ghosts},''
  \href{http://dx.doi.org/10.1088/0143-0807/36/1/015009}{{\em Eur. J. Phys.}
  {\bf 36} (2015)  015009},
\href{http://arxiv.org/abs/1406.4550}{{\tt arXiv:1406.4550 [hep-th]}}.
%%CITATION = ARXIV:1406.4550;%%.

\bibitem{Penedones:2010ue}
J.~Penedones, ``{Writing CFT correlation functions as AdS scattering
  amplitudes},'' \href{http://dx.doi.org/10.1007/JHEP03(2011)025}{{\em JHEP}
  {\bf 03} (2011)  025}.
arXiv:1011.1485 [hep-th].
%%CITATION = ARXIV:1011.1485;%%.

\bibitem{Fitzpatrick:2011dm}
A.~L. Fitzpatrick and J.~Kaplan, ``{Unitarity and the Holographic S-Matrix},''
  \href{http://dx.doi.org/10.1007/JHEP10(2012)032}{{\em JHEP} {\bf 10} (2012)
  032}.
arXiv:1112.4845 [hep-th].
%%CITATION = ARXIV:1112.4845;%%.

\bibitem{Bekaert:2015tva}
X.~Bekaert, J.~Erdmenger, D.~Ponomarev, and C.~Sleight, ``{Quartic AdS
  Interactions in Higher-Spin Gravity from Conformal Field Theory},''
  \href{http://dx.doi.org/10.1007/JHEP11(2015)149}{{\em JHEP} {\bf 11} (2015)
  149},
\href{http://arxiv.org/abs/1508.04292}{{\tt arXiv:1508.04292 [hep-th]}}.
%%CITATION = ARXIV:1508.04292;%%.

\bibitem{Dolan:2001ih}
L.~Dolan, C.~R. Nappi, and E.~Witten, ``{Conformal operators for partially
  massless states},''
  \href{http://dx.doi.org/10.1088/1126-6708/2001/10/016}{{\em JHEP} {\bf 10}
  (2001)  016}.
arXiv:0109096 [hep-th].
%%CITATION = HEP-TH/0109096;%%.

\bibitem{Noether:1918zz}
E.~Noether, ``{Invariant Variation Problems},''
  \href{http://dx.doi.org/10.1080/00411457108231446}{{\em Gott. Nachr.} {\bf
  1918} (1918)  235--257}, \href{http://arxiv.org/abs/physics/0503066}{{\tt
  arXiv:physics/0503066 [physics]}}.
[Transp. Theory Statist. Phys.1,186(1971)].
%%CITATION = PHYSICS/0503066;%%.

\bibitem{Eastwood:2002su}
M.~G. Eastwood, ``{Higher symmetries of the Laplacian},''
  \href{http://dx.doi.org/10.4007/annals.2005.161.1645}{{\em Annals Math.} {\bf
  161} (2005)  1645--1665}.
arXiv:0206233 [hep-th].
%%CITATION = HEP-TH/0206233;%%.

\bibitem{2006math.....10610E}
M.~{Eastwood} and T.~{Leistner}, ``{Higher Symmetries of the Square of the
  Laplacian},''{\em ArXiv Mathematics e-prints} (Oct., 2006)  ,
  \href{http://arxiv.org/abs/math/0610610}{{\tt math/0610610}}.

\bibitem{Griffin:2013dfa}
T.~Griffin, K.~T. Grosvenor, P.~Horava, and Z.~Yan, ``{Multicritical Symmetry
  Breaking and Naturalness of Slow Nambu-Goldstone Bosons},''
  \href{http://dx.doi.org/10.1103/PhysRevD.88.101701}{{\em Phys. Rev.} {\bf
  D88} (2013)  101701},
\href{http://arxiv.org/abs/1308.5967}{{\tt arXiv:1308.5967 [hep-th]}}.
%%CITATION = ARXIV:1308.5967;%%.

\bibitem{Hinterbichler:2014cwa}
K.~Hinterbichler and A.~Joyce, ``{Goldstones with Extended Shift Symmetries},''
  \href{http://dx.doi.org/10.1142/S0218271814430019}{{\em Int. J. Mod. Phys.}
  {\bf D23} (2014) no.~13, 1443001},
\href{http://arxiv.org/abs/1404.4047}{{\tt arXiv:1404.4047 [hep-th]}}.
%%CITATION = ARXIV:1404.4047;%%.

\bibitem{Griffin:2014bta}
T.~Griffin, K.~T. Grosvenor, P.~Horava, and Z.~Yan, ``{Scalar Field Theories
  with Polynomial Shift Symmetries},''
  \href{http://dx.doi.org/10.1007/s00220-015-2461-2}{{\em Commun. Math. Phys.}
  {\bf 340} (2015) no.~3, 985--1048},
\href{http://arxiv.org/abs/1412.1046}{{\tt arXiv:1412.1046 [hep-th]}}.
%%CITATION = ARXIV:1412.1046;%%.

\bibitem{Griffin:2015hxa}
T.~Griffin, K.~T. Grosvenor, P.~Horava, and Z.~Yan, ``{Cascading
  Multicriticality in Nonrelativistic Spontaneous Symmetry Breaking},''
  \href{http://dx.doi.org/10.1103/PhysRevLett.115.241601}{{\em Phys. Rev.
  Lett.} {\bf 115} (2015) no.~24, 241601},
\href{http://arxiv.org/abs/1507.06992}{{\tt arXiv:1507.06992 [hep-th]}}.
%%CITATION = ARXIV:1507.06992;%%.

\bibitem{Golkar:2014mwa}
S.~Golkar and D.~T. Son, ``{Operator Product Expansion and Conservation Laws in
  Non-Relativistic Conformal Field Theories},''
  \href{http://dx.doi.org/10.1007/JHEP12(2014)063}{{\em JHEP} {\bf 12} (2014)
  063},
\href{http://arxiv.org/abs/1408.3629}{{\tt arXiv:1408.3629 [hep-th]}}.
%%CITATION = ARXIV:1408.3629;%%.

\bibitem{Rychkov:2016iqz}
S.~Rychkov, ``{EPFL Lectures on Conformal Field Theory in $D \ge 3$
  Dimensions},''
\href{http://arxiv.org/abs/1601.05000}{{\tt arXiv:1601.05000 [hep-th]}}.
%%CITATION = ARXIV:1601.05000;%%.

\bibitem{Polchinski:1998rq}
J.~Polchinski, {\em {String theory. Vol. 1: An introduction to the bosonic
  string}}.
\newblock Cambridge University Press,
2007.
\newblock
%%CITATION = INSPIRE-487240;%%.

\bibitem{2009arXiv0911.5265G}
A.~R. {Gover} and J.~{Silhan}, ``{Higher symmetries of the conformal powers of
  the Laplacian on conformally flat manifolds},''{\em ArXiv e-prints} (Nov.,
  2009)  , \href{http://arxiv.org/abs/0911.5265}{{\tt arXiv:0911.5265
  [math.DG]}}.

\bibitem{Anselmi:1998bh}
D.~Anselmi, ``{Theory of higher spin tensor currents and central charges},''
  \href{http://dx.doi.org/10.1016/S0550-3213(98)00783-4}{{\em Nucl. Phys.} {\bf
  B541} (1999)  323--368},
\href{http://arxiv.org/abs/hep-th/9808004}{{\tt arXiv:hep-th/9808004
  [hep-th]}}.
%%CITATION = HEP-TH/9808004;%%.

\bibitem{Karananas:2015ioa}
G.~K. Karananas and A.~Monin, ``{Weyl vs. Conformal},''
  \href{http://dx.doi.org/10.1016/j.physletb.2016.04.001}{{\em Phys. Lett.}
  {\bf B757} (2016)  257--260},
\href{http://arxiv.org/abs/1510.08042}{{\tt arXiv:1510.08042 [hep-th]}}.
%%CITATION = ARXIV:1510.08042;%%.

\bibitem{Shaynkman:2004vu}
O.~V. Shaynkman, I.~{\relax Yu}. Tipunin, and M.~A. Vasiliev, ``{Unfolded form
  of conformal equations in M dimensions and o(M + 2) modules},''
  \href{http://dx.doi.org/10.1142/S0129055X06002814}{{\em Rev. Math. Phys.}
  {\bf 18} (2006)  823--886},
\href{http://arxiv.org/abs/hep-th/0401086}{{\tt arXiv:hep-th/0401086
  [hep-th]}}.
%%CITATION = HEP-TH/0401086;%%.

\bibitem{Graham01121992}
C.~R. Graham, R.~Jenne, L.~J. Mason, and G.~A.~J. Sparling, ``Conformally
  invariant powers of the laplacian, i: Existence,''
  \href{http://dx.doi.org/10.1112/jlms/s2-46.3.557}{{\em Journal of the London
  Mathematical Society} {\bf s2-46} (1992) no.~3, 557--565},
  \href{http://arxiv.org/abs/http://jlms.oxfordjournals.org/content/s2-46/3/557.full.pdf+html}{{\tt
  http://jlms.oxfordjournals.org/content/s2-46/3/557.full.pdf+html}}.
  \url{http://jlms.oxfordjournals.org/content/s2-46/3/557.abstract}.

\bibitem{Farnsworth:2017tbz}
K.~Farnsworth, M.~A. Luty, and V.~Prilepina, ``{Weyl versus Conformal
  Invariance in Quantum Field Theory},''
\href{http://arxiv.org/abs/1702.07079}{{\tt arXiv:1702.07079 [hep-th]}}.
%%CITATION = ARXIV:1702.07079;%%.

\bibitem{yamabe1960}
H.~Yamabe, ``On a deformation of riemannian structures on compact manifolds,''
  {\em Osaka Math. J.} {\bf 12} (1960) no.~1, 21--37.
  \url{http://projecteuclid.org/euclid.ojm/1200689814}.

\bibitem{2008SIGMA...4..036P}
S.~M. {Paneitz}, \href{http://dx.doi.org/10.3842/SIGMA.2008.036}{``{A Quartic
  Conformally Covariant Differential Operator for Arbitrary Pseudo-Riemannian
  Manifolds (Summary)},''{\em SIGMA} {\bf 4} (Mar., 2008)  036},
  \href{http://arxiv.org/abs/0803.4331}{{\tt arXiv:0803.4331 [math.DG]}}.

\bibitem{Fradkin:1981jc}
E.~S. Fradkin and A.~A. Tseytlin, ``{One Loop Beta Function in Conformal
  Supergravities},''
\href{http://dx.doi.org/10.1016/0550-3213(82)90481-3}{{\em Nucl. Phys.} {\bf
  B203} (1982)  157--178}.
%%CITATION = NUPHA,B203,157;%%.

\bibitem{Fradkin:1981iu}
E.~S. Fradkin and A.~A. Tseytlin, ``{Renormalizable asymptotically free quantum
  theory of gravity},''
\href{http://dx.doi.org/10.1016/0550-3213(82)90444-8}{{\em Nucl. Phys.} {\bf
  B201} (1982)  469--491}.
%%CITATION = NUPHA,B201,469;%%.

\bibitem{Fradkin:1982xc}
E.~S. Fradkin and A.~A. Tseytlin, ``{Asymptotic Freedom in Extended Conformal
  Supergravities},''
\href{http://dx.doi.org/10.1016/0370-2693(82)91018-8}{{\em Phys. Lett.} {\bf
  B110} (1982)  117--122}.
%%CITATION = PHLTA,B110,117;%%.

\bibitem{Riegert:1984kt}
R.~J. Riegert, ``{A Nonlocal Action for the Trace Anomaly},''
\href{http://dx.doi.org/10.1016/0370-2693(84)90983-3}{{\em Phys. Lett.} {\bf
  B134} (1984)  56--60}.
%%CITATION = PHLTA,B134,56;%%.

\bibitem{2009arXiv0905.3992J}
A.~{Juhl}, ``{On conformally covariant powers of the Laplacian},''{\em ArXiv
  e-prints} (May, 2009)  , \href{http://arxiv.org/abs/0905.3992}{{\tt
  arXiv:0905.3992 [math.DG]}}.

\bibitem{Juhl:2011ua}
A.~Juhl, ``{Explicit formulas for GJMS-operators and $Q$-curvatures},''
\href{http://arxiv.org/abs/1108.0273}{{\tt arXiv:1108.0273 [math.DG]}}.
%%CITATION = ARXIV:1108.0273;%%.

\bibitem{2012arXiv1203.0360F}
C.~{Fefferman} and C.~R. {Graham}, ``{Juhl's Formulae for GJMS Operators and
  Q-Curvatures},''{\em ArXiv e-prints} (Mar., 2012)  ,
  \href{http://arxiv.org/abs/1203.0360}{{\tt arXiv:1203.0360 [math.DG]}}.

\bibitem{Beccaria:2015vaa}
M.~Beccaria and A.~A. Tseytlin, ``{On higher spin partition functions},''
  \href{http://dx.doi.org/10.1088/1751-8113/48/27/275401}{{\em J. Phys.} {\bf
  A48} (2015) no.~27, 275401},
\href{http://arxiv.org/abs/1503.08143}{{\tt arXiv:1503.08143 [hep-th]}}.
%%CITATION = ARXIV:1503.08143;%%.

\bibitem{Gover:2005mn}
A.~R. Gover, ``{Laplacian operators and Q-curvature on conformally Einstein
  manifolds},''
\href{http://arxiv.org/abs/math/0506037}{{\tt arXiv:math/0506037 [math-dg]}}.
%%CITATION = MATH/0506037;%%.

\bibitem{Manvelyan:2006bk}
R.~Manvelyan and D.~H. Tchrakian, ``{Conformal coupling of the scalar field
  with gravity in higher dimensions and invariant powers of the Laplacian},''
  \href{http://dx.doi.org/10.1016/j.physletb.2006.12.027}{{\em Phys. Lett.}
  {\bf B644} (2007)  370--374},
\href{http://arxiv.org/abs/hep-th/0611077}{{\tt arXiv:hep-th/0611077
  [hep-th]}}.
%%CITATION = HEP-TH/0611077;%%.

\bibitem{Dowker:2010qy}
J.~S. Dowker, ``{Determinants and conformal anomalies of GJMS operators on
  spheres},'' \href{http://dx.doi.org/10.1088/1751-8113/44/11/115402}{{\em J.
  Phys.} {\bf A44} (2011)  115402},
\href{http://arxiv.org/abs/1010.0566}{{\tt arXiv:1010.0566 [hep-th]}}.
%%CITATION = ARXIV:1010.0566;%%.

\bibitem{Dowker:2013oqa}
J.~S. Dowker, ``{Numerical evaluation of spherical GJMS determinants},''
\href{http://arxiv.org/abs/1309.2873}{{\tt arXiv:1309.2873 [math-ph]}}.
%%CITATION = ARXIV:1309.2873;%%.

\bibitem{Dowker:2013ysa}
J.~S. Dowker, ``{Numerical evaluation of spherical GJMS determinants for even
  dimensions},''
\href{http://arxiv.org/abs/1310.0759}{{\tt arXiv:1310.0759 [hep-th]}}.
%%CITATION = ARXIV:1310.0759;%%.

\bibitem{Dowker:2015xya}
J.~S. Dowker and T.~Mansour, ``{Evaluation of spherical GJMS determinants},''
\href{http://dx.doi.org/10.1016/j.geomphys.2015.07.001}{{\em J. Geom. Phys.}
  {\bf 97} (2015)  51--60}.
%%CITATION = JGPHE,97,51;%%.

\bibitem{Beccaria:2015uta}
M.~Beccaria and A.~A. Tseytlin, ``{Conformal a-anomaly of some non-unitary 6d
  superconformal theories},''
  \href{http://dx.doi.org/10.1007/JHEP09(2015)017}{{\em JHEP} {\bf 09} (2015)
  017},
\href{http://arxiv.org/abs/1506.08727}{{\tt arXiv:1506.08727 [hep-th]}}.
%%CITATION = ARXIV:1506.08727;%%.

\bibitem{Brust:2016xif}
C.~Brust and K.~Hinterbichler, ``{Partially Massless Higher-Spin Theory II:
  One-Loop Effective Actions},''
\href{http://arxiv.org/abs/1610.08522}{{\tt arXiv:1610.08522 [hep-th]}}.
%%CITATION = ARXIV:1610.08522;%%.

\bibitem{Gunaydin:2016amv}
M.~GŸnaydin, E.~D. Skvortsov, and T.~Tran, ``{Exceptional $F(4)$ higher-spin
  theory in AdS$_{6}$ at one-loop and other tests of duality},''
  \href{http://dx.doi.org/10.1007/JHEP11(2016)168}{{\em JHEP} {\bf 11} (2016)
  168},
\href{http://arxiv.org/abs/1608.07582}{{\tt arXiv:1608.07582 [hep-th]}}.
%%CITATION = ARXIV:1608.07582;%%.

\bibitem{Chodos:1985hk}
A.~Chodos and E.~Myers, ``{Testing the Surrogate Zeta Function Method},''
\href{http://dx.doi.org/10.1139/p86-117}{{\em Can. J. Phys.} {\bf 64} (1986)
  633--636}.
%%CITATION = CJPHA,64,633;%%.

\bibitem{Wodzicki1987}
M.~Wodzicki, {\em Noncommutative residue Chapter I. Fundamentals},
  \href{http://dx.doi.org/10.1007/BFb0078372}{pp.~320--399}.
\newblock Springer Berlin Heidelberg, Berlin, Heidelberg, 1987.
\newblock \url{http://dx.doi.org/10.1007/BFb0078372}.

\bibitem{GUILLEMIN1985131}
V.~Guillemin, ``A new proof of weyl's formula on the asymptotic distribution of
  eigenvalues,''
  \href{http://dx.doi.org/http://dx.doi.org/10.1016/0001-8708(85)90018-0}{{\em
  Advances in Mathematics} {\bf 55} (1985) no.~2, 131 -- 160}.
  \url{http://www.sciencedirect.com/science/article/pii/0001870885900180}.

\bibitem{zbMATH04149222}
C.~{Kassel}, ``{Le r\'esidu non commutatif. (Noncommutative residue)}.''
  {S\'emin. Bourbaki, Vol. 1988/89, 41e ann\'ee, Exp. No.708, Ast\'erisque
  177-178, 199-229 (1989).}, 1989.

\bibitem{Kontsevich:1994nc}
M.~Kontsevich and S.~Vishik, ``{Determinants of elliptic pseudodifferential
  operators},''
\href{http://arxiv.org/abs/hep-th/9404046}{{\tt arXiv:hep-th/9404046
  [hep-th]}}.
%%CITATION = HEP-TH/9404046;%%.

\bibitem{Kontsevich:1994xe}
M.~Kontsevich and S.~Vishik, ``{Geometry of determinants of elliptic
  operators},''
\href{http://arxiv.org/abs/hep-th/9406140}{{\tt arXiv:hep-th/9406140
  [hep-th]}}.
%%CITATION = HEP-TH/9406140;%%.

\bibitem{Dolan:2005wy}
F.~A. Dolan, ``{Character formulae and partition functions in higher
  dimensional conformal field theory},''
  \href{http://dx.doi.org/10.1063/1.2196241}{{\em J. Math. Phys.} {\bf 47}
  (2006)  062303},
\href{http://arxiv.org/abs/hep-th/0508031}{{\tt arXiv:hep-th/0508031
  [hep-th]}}.
%%CITATION = HEP-TH/0508031;%%.

\end{thebibliography}\endgroup

\end{document}